\documentclass[gmd]{copernicus}
\usepackage{amsmath, amssymb, amsthm, eucal, multicol}
\usepackage{subfig}
\usepackage{tabularx}
\usepackage{booktabs}
\usepackage{comment}
\newcommand{\de}{\mathrm{d}}
\newcommand{\Dt}{\textup{D}_t}

{\left\lbrace\begin{array}{@{}l@{}}}%
{\end{array}\right.}

\newcount\colveccount
\newcommand*\colvec[1]{
        \global\colveccount#1
        \begin{pmatrix}
        \colvecnext
}
\def\colvecnext#1{
        #1
        \global\advance\colveccount-1
        \ifnum\colveccount>0
                \\
                \expandafter\colvecnext
        \else
                \end{pmatrix}
        \fi
}

\DeclareMathOperator{\Tr}{Tr}

\newcommand{\ddt}{\partial_t}
\newcommand{\Div}{\nabla\cdot}

\newcommand{\Rey}{\textup{Re}}
\newcommand{\Reys}{\textup{Re}_\textup{s}}
\newcommand{\Pra}{\textup{Pr}}
\newcommand{\Prat}{\textup{Pr}_\textup{t}}
\newcommand{\St}{\textup{St}}

\newcommand{\Ma}{\textup{Ma}}

\newcommand{\Cs}{C_\textup{s}}

\newcommand{\Cm}{C_\textup{m}}

\newcommand{\Cmt}{\tilde{C}_\textup{m}}

\newcommand{\Rm}{R_\textup{m}}
\newcommand{\gm}{\gamma_\textup{m}}

\newcommand{\rhog}{\rho_\textup{g}}
\newcommand{\rhogh}{\hat{\rho}_\textup{g}}
\newcommand{\yg}{y_\textup{g}}
\newcommand{\rhos}{\rho_\textup{s}}
\newcommand{\rhosh}{\hat{\rho}_\textup{s}}

\newcommand{\rhom}{\rho_\textup{m}}
\newcommand{\rhomo}{\rho_{\textup{m},0}}
\newcommand{\rhombar}{\bar{\rho}_\textup{m}}

\newcommand{\Kg}{K_\textup{g}}
\newcommand{\Kb}{K_\textup{m}}
\newcommand{\Kbt}{\tilde{K}_\textup{m}}

\newcommand{\hg}{h_\textup{g}}
\newcommand{\hm}{h_\textup{m}}
\newcommand{\hmt}{\tilde{h}_\textup{m}}

\newcommand{\ug}{\vec{u}_\textup{g}}
\newcommand{\ugt}{\tilde{\vec{u}}_\textup{g}}
\newcommand{\ub}{\vec{u}_\textup{m}}
\newcommand{\ubt}{\tilde{\vec{u}}_\textup{m}}

\newcommand{\us}{\vec{u}_\textup{s}}
\newcommand{\uj}{\vec{u}_j}

\newcommand{\ur}{\vec{u}_\textup{r}}
\newcommand{\wrr}{\vec{w}_\textup{r}}
\newcommand{\wrrt}{\tilde{\vec{w}}_\textup{r}}
\newcommand{\vr}{\vec{v}_C}
\newcommand{\vrt}{\tilde{\vec{v}}_C}
\newcommand{\vk}{\vec{v}_K}
\newcommand{\vkt}{\tilde{\vec{v}}_K}
\newcommand{\tr}{\mathbb{T}_\textup{r}}
\newcommand{\trt}{\tilde{\mathbb{T}}_\textup{r}}
\newcommand{\Tg}{T_\textup{g}}
\newcommand{\Ts}{T_\textup{s}}

\newcommand{\epss}{\epsilon_\textup{s}}

\newcommand{\taus}{\tau_\textup{s}}
\newcommand{\etaK}{\eta_\textup{K}}
\newcommand{\tauK}{\tau_\textup{K}}

\newcommand{\mut}{\mu_\textup{t}}

\newcommand{\ds}{d_\textup{s}}
\begin{document}
%
\title{ASHEE: a compressible, equilibrium-Eulerian model for volcanic ash plumes}
\author[1,2]{Matteo Cerminara}
\author[2]{Tomaso Esposti Ongaro}
\author[3]{Luigi C. Berselli}
\affil[1]{Scuola Normale Superiore, Pisa, Italy}
\affil[2]{Istituto Nazionale di Geofisica e Vulcanologia, Sezione di Pisa, Italy}
\affil[3]{Dipartimento di Matematica, Universit\`a degli Studi di Pisa, Italy}

\runningtitle{An equilibrium Eulerian model for volcanic plumes}
\runningauthor{M. Cerminara et al.}
\correspondence{Matteo Cerminara \\Via della Faggiola 32, 56126, Pisa -- Italy\\email: matteo.cerminara@ingv.it}

\received{}
\pubdiscuss{} 
\revised{}
\accepted{}
\published{}
\firstpage{1}
\maketitle  

\begin{abstract}
A new fluid-dynamic model is developed to numerically simulate the non-equilibrium dynamics of polydisperse gas-particle mixtures forming volcanic plumes.
Starting from the three-dimensional N-phase Eulerian transport equations \citep{Neri2003} for a mixture of gases and solid dispersed particles, we adopt an asymptotic expansion strategy to derive a compressible version of the first-order non-equilibrium model \citep{Ferry2001}, valid for low concentration regimes (particle volume fraction less than $10^{-3}$) and particles Stokes number (St, i.e., the ratio between their relaxation time and flow characteristic time) not exceeding about 0.2. The new model, which is called ASHEE (ASH Equilibrium Eulerian), is significantly faster than the N-phase Eulerian model while retaining the capability to describe gas-particle non-equilibrium effects.
Direct numerical simulation accurately reproduce the dynamics of isotropic, compressible turbulence in subsonic regime.  For gas-particle mixtures, it describes the main features of density fluctuations and the preferential concentration and clustering of particles by turbulence, thus verifying the model reliability and suitability for the numerical simulation of high-Reynolds number and high-temperature regimes in presence of a dispersed phase.
On the other hand, Large-Eddy Numerical Simulations of forced plumes are able to reproduce their observed averaged and instantaneous flow properties. In particular, the self-similar Gaussian radial profile and the development of large-scale coherent structures are reproduced, including the rate of turbulent mixing and entrainment of atmospheric air.
Application to the Large-Eddy Simulation of the injection of the eruptive mixture in a stratified atmosphere describes some of important features of turbulent volcanic plumes, including air entrainment, buoyancy reversal, and maximum plume height. 
For very fine particles (St~$\to 0$, when non-equilibrium effects are negligible) the model reduces to the so-called dusty-gas model.
However, coarse particles partially decouple from the gas phase within eddies (thus modifying the turbulent structure) and preferentially concentrate at the eddy periphery, eventually being lost from the plume margins due to the concurrent effect of gravity.
By these mechanisms, gas-particle non-equilibrium processes are able to influence the large-scale behavior of volcanic plumes.
\end{abstract}
\introduction
\label{introduction} 
\indent
Explosive volcanic eruptions are characterized by the injection from a vent into the atmosphere of a mixture of gases, liquid droplets and solid particles, at high velocity and temperature. 
In typical magmatic eruptions, solid particles constitute more than 95\% of the erupted mass and are mostly produced by the brittle fragmentation of a highly viscous magma during its rapid ascent in a narrow conduit \citep{wilson:1976,sparks1978dynamics}, with particle sizes and densities spanning over a wide range, depending on the overall character and intensity of the eruption \citep{Kaminski1998,kueppers2006explosive}. The order of magnitude of the plume mixture volumetric concentration very rarely exceed $ \epsilon_\textup{s} \sim 3*10^{-3}$, because the order of magnitude of the ejected fragments density is $\hat{\rho}_\textup{s} \sim 10^3$ kg/m$^3$. Thus, the plume mixture con be considered mainly as a dilute suspension in the sense of \cite{elghobashi1991, elghobashi1994}. This threshold for $\epsilon_\textup{s}$ is overcome in the dense layer forming in presence of pyroclastic density currents~\citep[see e.g.][]{Orsucci2014}. In the literature, collisions between ash particles are usually disregarded when looking at the dynamics of volcanic ash plume, because this dilute character of the plume mixture~\citep[cf.][]{Morton1956,Woods2010}.

After injection in the atmosphere, this multiphase eruptive mixture can rise convectively in the atmosphere, forming either a buoyant volcanic plume or collapse catastrophically forming pyroclastic density currents.
Since these two end-members have different spatial and temporal scales and different impacts on the surrounding of a volcano, understanding the dynamics of volcanic columns and the mechanism of this bifurcation is one of the topical aims of volcanology and one of the main motivations for this work.

The term {\em volcanic column} will be adopted in this paper to generically indicate the eruptive character (e.g. convective/collapsing column).
Following the fluid-dynamic nomenclature, we will term {\em jet} the inertial regime of the volcanic column and {\em plume} the buoyancy-driven regime. A {\em forced plume} is characterized by an initial momentum-driven jet stage, transitioning into a plume. 

In this work, we present a new computational fluid-dynamic model to simulate turbulent gas-particle forced plumes in the atmosphere. 
Although the focus of the paper is on multiphase turbulence in subsonic regimes, the model is also suited to be applied to transonic and supersonic flows.
In many cases, indeed, the eruptive mixture is injected in the atmosphere at pressure higher than atmospheric, so that the flow is initially driven by a rapid, transonic decompression stage. This is suggested by numerical models predicting choked flow conditions at the volcanic vent \citep{wilson1980relationships,wilson1980explosive}, implying a supersonic transition above the vent or in the crater \citep{kieffer:1984,woods1995decompression,koyaguchi2010effects} and it is supported by field evidences of the emission of shock waves during the initial stages of an eruptions \citep{morrissey1997burst}. Despite the importance of the decompression stage for the subsequent development of the volcanic plume \citep{Pelanti2006,ogden:2008b,orescanin:2010,Carcano2013} and for the stability of the eruptive column \citep{ogden:2008a}, our analysis is limited to the plume region where flow pressure is equilibrated to the atmospheric pressure. From laboratory experiments, this is expected to occur within less than 20 inlet diameters above the ground \citep{yuceil2002scaling}.

\subsection{Dusty gas modeling of volcanic plumes}
Starting from the assumption that the large-scale behavior of volcanic columns is controlled by the {\em bulk} properties of the eruptive mixture, most of the previous models of volcanic plumes have considered the eruptive mixture as homogeneous (i.e., they assume that particles are perfectly coupled to the gas phase). Under such hypothesis, the multiphase transport equations can be largely simplified and reduce to a set of mass, momentum and energy balance equations for a single fluid (named {\em dusty-gas} or {\em pseudo-gas}) having average thermo-fluid dynamic properties (mixture density, velocity and temperature) and equation of states accounting for the incompressibility of the particulate phase and gas covolume \citep{Marble1970a}. 

By adopting the dusty gas approximation, volcanic plumes have been studied in the framework of jet \citep{prandtl1963essentials} and plume theory \citep{Morton1956,Morton1959}. One-dimensional, steady-state pseudo-gas models of volcanic plumes have thus had a formidable role in volcanology to identify the main processes controlling their dynamics and scaling properties \citep{wilson:1976,Woods1988a,sparks:1997}. 

Accordingly, volcanic plume dynamics is schematically subdivided into two main stages. The lower, jet phase is driven by the initial flow momentum. Mixture buoyancy is initially negative (the bulk density is larger than atmospheric) but the mixture progressively expands adiabatically thanks to atmospheric air entrainment and heating, eventually undergoing a buoyancy reversal. 
When buoyancy reversal does not occur, partial or total collapse of the jet from its maximum thrust height (where the jet has lost all its initial momentum) and generation of pyroclastic density currents are expected.

Above the jet thrust region, the rise of volcanic plumes is driven by buoyancy and it is controlled by turbulent mixing until, in the stratified atmosphere, a level of neutral buoyancy is reached. Above that height, the plume starts to spread out achieving its maximum height and forming an umbrella ash cloud, dispersing in the atmosphere and slowly falling-out.

In one-dimensional, time-averaged models, entrainment of atmospheric air is described by one empirical coefficient (the entrainment coefficient) relating the influx of atmospheric air to the local, vertical plume velocity. The entrainment coefficient also determines the plume shape \citep{ishimine2006sensitivity} and can be empirically determined by means of direct field observations or ad-hoc laboratory measurements.

Further development of volcanic plume models have included the influence of atmospheric stratification and humidity \citep{woods1993moist,glaze1996sensitivity}, the effect of cross wind \citep{Bursik2001}, loss and reentrainment of solid particles from plume margins \citep{Woods1991,veitch2002particle}, and transient effects \citep{scase2009evolution,WoodhouseUnsteady}.
However, one-dimensional models strongly rely on the self-similarity hypothesis, whose validity cannot be experimentally ascertained for volcanic eruptions.

To overcome the limitations of one-dimensional models, three-dimensional dusty-gas models have been developed to simulate volcanic plumes. 
\citet{Suzuki2005a} have developed a three-dimensional dusty gas model (SK-3D) able to accurately resolve the relevant turbulent scales of a volcanic plume, allowing a first, theoretical determination of the entrainment coefficient \citep{suzuki:2010}, without the need of an empirical calibration.

To simulate the three-dimensional large-scale dynamics of volcanic plumes including particle settling and the complex microphysics of water in volcanic plumes, the ATHAM (Active Tracer High Resolution Atmospheric Model) code has been designed \citep{Oberhuber1998,graf:1999,VanEaton2015}.
ATHAM describes the dynamics of gas-particle mixtures by assuming that particles are in kinetic equilibrium with the gas phase only in the horizontal component, whereas along the vertical direction they are allowed to have a differential velocity. Thermal equilibrium is assumed.
In this sense, ATHAM relaxes the dusty-gas approximation (while maintaining its fundamental structure and the same momentum transport equations) by describing the settling of particles with respect to the gas.

\subsection{Multiphase flow models of volcanic plumes}
Notwithstanding all the above advantages, dusty-gas models are still limited by the equilibrium assumption, which can be questionable at least for the coarsest part of the granulometric spectrum in a plume.
Turbulence is indeed a non-linear, multiscale process and the time and space scales of gas-particle interaction may be comparable with some relevant turbulent scales, thus influencing the large-scale behavior of volcanic plumes. 

To model non-equilibrium processes, Eulerian multiphase flow models have been developed,  which solve the full set of mass, momentum, and energy transport equations for a mixture of gas and dispersed particles, treated as interpenetrating fluids. \citet{Valentine1989} and \citet{Dobran1993,Neri1994} have first analyzed the influence of erupting parameters on the column behavior to identify: By means of two-dimensional numerical simulations, they individuate a threshold from collapsing to convective columns. Lately, two-dimensional \citep{DiMuro2004,Dartevelle2004} and three-dimensional numerical simulations \citep{Esposti2008} has contributed to modify the view of a sharp transition between convecting and collapsing columns in favor of that of a transitional regime, characterized by a progressively increasing fraction of mass collapsing.
However, previous works could not investigate in detail the non-equilibrium effects in volcanic plumes, mainly because of their averaged description of turbulence: a detailed resolution of the relevant turbulent scales in three dimensions would indeed be computationally prohibitive for N-phase systems.

The main objective of the present work is therefore to develop a new physical model and a fast three-dimensional numerical code able to resolve the spatial and temporal scales of the interaction between gas and particles in turbulent regime and to describe the kinetic non-equilibrium dynamics and their influence on the {\em observable} features of volcanic plumes. To this aim, a development of the so-called {\em equilibrium-Eulerian} approach \citep{Ferry2001,Balachandar2010} has been adopted. It is a generalization of the dusty-gas model keeping the kinematic non-equilibrium as a first order correction of the \citet{Marble1970a} model with respect to the Stokes number of the solid particles/bubbles in the mixture.

The derivation of the fluid dynamic model describing the non-equilibrium gas-particle mixture is described in detail in Section \ref{model}. The computational solution procedure and the numerical code development are reported in Section \ref{numerics}. Section \ref{validation} focuses on verification and validation issues in the context of applications to turbulent volcanic plumes. In particular, here we discuss: three-dimensional numerical simulations of compressible, isotropic turbulence (with and without particles); experimental scale forced plumes; Sod's shock tube problem.
Finally, Section \ref{results} presents numerical simulations of volcanic plumes and discusses some aspects related to numerical grid resolution in practical cases.

\section{The multiphase flow model}
\label{model}
To derive an appropriate multiphase flow model to describe gas-particle volcanic plumes, we here introduce the
non-dimensional scaling parameters characterizing gas particle and particle particle interactions.

The drag force between gas and particles introduces in the system a time scale $\taus$, the {\em
particle relaxation time}, which is the time needed to a particle to equilibrate to a change of gas
velocity.
Gas-particle drag is a non-linear function of the local flow variables and, in
particular, it depends strongly on the relative Reynolds number, defined as:
\begin{equation}\label{eq:Re_s}
\Reys=\frac{\rhogh |\us - \ug| \ds}{\mu}
\end{equation}
here $\ds$ is the particle diameter, $\rhogh$ is the gas density, $\mu$ is the gas dynamic viscosity coefficient and $\vec{u}_{g(s)}$ is the gas (solid) phase velocity field. Being $\hat{\rho}_{\textup{g}(\textup{s})}$ the gaseous (solid) phase density and $\epss = V_\textup{s}/V$ the volumetric concentration of the solid phase, it is useful to define the gas
bulk density $\rhog \equiv (1 - \epss)\rhogh \simeq \rhogh$ and the solid bulk
density $\rhos\equiv \epss \rhosh$ (even though in our applications $\epss$ is order $10^{-3}$, 
$\rhos$ is non-negligible since $\rhosh / \rhogh $ is of order $10^3$).

For an individual point-like particle (i.e., having diameter $\ds$ much smaller than the scale of the problem under analysis), 
at $\textrm{Re}_s < 1000$, the drag force per volume unity can be given by the Stokes' law: 
\begin{equation}\label{eq:Stokes}
 \vec{f}_d = \frac{\rhos}{\taus}(\ug - \us)\,,
\end{equation}
where 
\begin{equation} 
\label{eq:taus}
\taus \equiv \frac{\rhosh}{\rhogh} \frac{\ds^2}{18\nu\,\phi_\textup{c}(\Reys)}
\end{equation}
is the characteristic time of particle velocity relaxation with respect the gas, $\rhosh$ is the particle density, $\nu$ is the gas kinematic viscosity and \mbox{$\phi_\textup{c} = 1 + 0.15\,\Reys^{0.687}$} is a correction factor (obtained from the Schiller--Naumann correlation) for finite particle Reynolds number~\citep[cf.][]{Clift1978, Balachandar2009, Balachandar2010, cerminara2015phd}. In Eq.~\eqref{eq:Stokes} we disregard all the effects due to the pressure gradient, the added mass, the Basset history and the Saffman terms, because we are considering {\em heavy particles}: \mbox{$\rhosh/\rhogh \gg 1$} \citep[cf.][]{Ferry2001, Bagheri2013}. 
Equation \eqref{eq:Stokes} has a linear dependence on the fluid-particle relative velocity only when $\Reys \ll 1$, so that $\phi_\textup{c} \simeq 1$ and the classic Stokes drag expression is recovered. On the other hand, if the relative Reynolds number $\Reys$ grows, non-linear effects become much more important in Eq.~\eqref{eq:taus}. The \cite{Clift1978} empirical relationship used in this work has been used and tested in a number of papers \citep[e.g.,][]{Balachandar2010, Wang1993, Bonadonna2002}, and it is equivalent to assuming the following gas-particle drag coefficient:
\begin{equation}\label{eq:CD}
C_\textup{D}(\Reys) = \frac{24}{\Reys}(1 + 0.15\,\Reys^{0.687})\,.
\end{equation}
\citet{Wang1993} discussed nonlinear effects due to this correction on the dynamics of point-like particles falling under gravity in an homogeneous and isotropic turbulent surrounding.
We recall here the terminal velocity that can be found by setting $\ug = 0$ in Eq.~\eqref{eq:Stokes} is:
\begin{equation}\label{eq:settling_w}
\vec{w}_s = \sqrt{\frac{4 \ds \rhosh}{3 C_\textup{D} \rhog g}}\,\vec{g} = \taus \,\vec{g}\,.
\end{equation}
As previously pointed out, correction used in Eq.~\eqref{eq:CD} is valid if $\Reys < 10^3$, the regime addressed in this work for ash particles  much denser then the surrounding fluid and smaller than $1\,\textup{mm}$. 
As shown by \citet{Balachandar2009}, maximum values of $\Reys$ are associated to particle gravitational settling (not to turbulence). Using formula~\eqref{eq:CD} and~\eqref{eq:settling_w}, it is thus possible to estimate $\Reys$ of a falling particle with diameter $\ds$. We obtain that $\Reys$ is always smaller than $10^3$ for ash particles finer than 1~mm in air.
If regimes with a bigger decoupling needs to be explored, more complex empirical correction has to be used for $\phi_\textup{c}$ \citep[][]{Neri2003,Burger2001}.

The same reasoning can be applied to estimate the {\em thermal relaxation time} between gas and particles.
In terms of the solid phase specific heat capacity $\Cs$ and its
thermal conductivity $k_\textup{g}$, we have:
\begin{equation}
 \tau_T = \frac{2}{\textup{Nu}_\textup{s}}\frac{\rhosh \Cs}{k_\textup{g}} \frac{\ds^2}{12}\,,
\end{equation}
where $\textup{Nu}_\textup{s} = \textup{Nu}_\textup{s}(\Reys, \textup{Pr})$ is the Nusselt number, usually function of the relative Reynolds number and of the Prandtl number of the carrier fluid \citep[][]{Neri2003}. In terms of $\tau_T$, the heat exchange between a particle at temperature $\Ts$ and the surrounding gas at temperature $\Tg$ per unit volume is:
\begin{equation}
 \mathrm{Q}_T = \frac{\rhos \Cs}{\tau_T} (\Ts -\Tg)\,.
\end{equation}

Comparing the kinetic and thermal relaxation times we get:
\begin{equation}
\label{eq:tauTtaus}
 \frac{\tau_T}{\taus} = \frac{3}{2} \frac{2 \phi_\textup{c}}{\textup{Nu}_\textup{s}} \frac{\Cs \mu}{k_\textup{g}}\,.
\end{equation}
In order to estimate this number, firstly we notice that factor $2\phi_\textup{c}/\textup{Nu}_\textup{s}$ tends to 1 if $\Reys \to 0$, and it remains smaller than $\simeq 2$ if $\Reys < 10^3$ \citep[][]{Neri2003,cerminara2015phd}. Then, in the case of ash particles in air, we have (in SI units) $\mu \simeq 10^{-5}$, $\Cs \simeq 10^3$, $k_\textup{g} \simeq 10^{-2}$. Thus we have that $\tau_T/\taus \simeq 1$, meaning that the thermal equilibrium time is typically of the same order of magnitude of the kinematic one. This bound is very useful when we write the equilibrium-Eulerian and the dusty gas models, because it tells us that the thermal Stokes number is of the same order of the kinematic one, at least for volcanic ash finer than 1 mm.

The non-dimensional Stokes number (St) is defined as the ratio between the kinetic (or thermal) relaxation time and a characteristic time of the flow under investigation $\tau_F$, namely $\displaystyle \textup{St}_\textup{s}=\taus/\tau_F$.
The definition of the flow time-scale can be problematic for high-Reynolds number flows (typical of volcanic plumes), which are characterized by a wide range of interacting length- and time-scales, a distinctive feature of the turbulent regime. 
For volcanic plumes, the more energetic time-scale would be of the order $\displaystyle \tau_L= L/U$, where $L$ and $U$ are the plume diameter and velocity at the vent, which gives the characteristic turnover time of the largest eddies in a turbulent plume \citep[e.g.,][]{Zhou2001}.
On the other hand, the smallest time-scale (largest $\textup{St}_\textup{s}$) can be defined by the Kolmogorov similarity law by $\displaystyle \tau_\eta \sim \tau_L\,\Rey_L^{-1/2}$, where the macroscopic Reynolds number is defined, at a first instance, by $\displaystyle \textrm{Re}_L=U L/\nu$, $\nu$ being the kinematic viscosity of the gas phase alone.
For numerical models, it is also useful introducing the Large-Eddy Simulation (LES) time-scale $\tau_{\xi}$, with respect to the resolved scales $\xi$, related to the numerical grid resolution, size of the explicit filter, and discretization accuracy \citep{lesieur2005,garnier2009large,Balachandar2010,Cerminara2013}. At LES scale $\xi$, $\textup{St}_\textup{s}$ is not as large as at the Kolmogorov scale, thus the decoupling between particles and the carrier fluid is mitigated by the LES filtering operation. We found that $\textup{St}_\textup{s} \lesssim 0.2$ for LES of volcanic ash finer than 1 mm.

The model presented here is conceived for resolving {\em dilute} suspensions, namely mixtures of gases and particles with volumetric concentration $\frac{V_\textup{s}}{V}\equiv\epss \lesssim 10^{-3}$. We here use the definition of dilute suspension by \citet{elghobashi1991, elghobashi1994} and \citet{Balachandar2009}, corresponding to regimes in which particle-particle collisions can be disregarded. 
This can also be justified by analyzing the time scale of particle--particle collisions.
In the dilute regime, in which we can assume an equilibrium Maxwell distribution of particle velocities, the mean free path of solid particles is given by \citep{gidaspow1994multiphase}:
\begin{equation}
\lambda_\textup{p-p} = \frac{1}{6\sqrt{2}}\frac{\ds}{\epss}\,.
\end{equation}
Consequently, particle-particle collisions are relatively infrequent ($\lambda_\textup{p-p} \sim 0.1$ m $\ll \ds$), so that we can neglect, as a first approximation, particle-particle collisions and consider the particulate fluid as pressure-less, inviscid and non-conductive.

In volcanic plumes the particle volumetric concentration can exceed of one order of magnitude the threshold \mbox{$\epss \simeq 10^{-3}$} only near the vent \citep[see, e.g.,][]{sparks:1997,esposti2008transient}. However, the region of the plume where the dilute suspension requirement is not fulfilled remains small with respect the size of the entire plume, weakly influencing its global dynamics. Indeed, as we will show in Sec.~\ref{results}, air entrainment and particle fallout induce a rapid decrease of the volumetric concentration. On the contrary, the mass fraction of the solid particles can not be considered small, because particles are heavy: $\epss \ast \rhosh \equiv \rhos \simeq \rhog$. Thus, particles inertia will be considered in the present model: in other words, we will consider the {\em two way} coupling between dispersed particles and the carrier gas phase. 

Summarizing, our multiphase model focuses and carefully takes advantage of the hypotheses characterizing the following regimes: heavy particles (\mbox{$\rhosh/\rhogh \gg 1$}) in dilute suspension ($\epss \lesssim 10^{-3}$) with dynamical length scales much larger than the particles diameter (point-particle approach) and relative Reynolds number smaller than $10^3$.

\subsection{Eulerian-Eulerian multiphase flow model}

When the Stokes number is smaller than one, and the number of particles is very large, it is convenient to use an
Eulerian approach, where the carrier and the dispersed phase are modeled as interpenetrating continua, 
and their dynamics is described by the laws of fluid mechanics \citep{Balachandar2010}.

Here we want to model a polydisperse mixture of \mbox{$i \in [1,2,\dots,I] \equiv \mathcal{I}$} gaseous phases and \mbox{$j \in [1,2,\dots,J] \equiv \mathcal{J}$} solid phases. From now on, we will use the subscript $(\cdot)_j$ instead of $(\cdot)_\textup{s}$ for the jth solid phase. Solid phases represent the discretization of a virtually continuous grain-size distribution into discrete bins, as usually done in volcanological studies ~\citep[cf.][]{Cioni2003, Neri2003}. Another possible approach is the method of moments, in which the evolution of the {\em moments} of the grain size distribution is described. This has recently been applied in volcanology to integral plume models by \citet{Vitturi2015}. In the present work we opted for the classical discretization of the grain size distribution ~\citep[cf.][]{Neri2003}. In~\citep{cerminara2015phd} we analyze the Eulerian-Eulerian model under the barotropic regime to show the existence of weak solutions of the corresponding partial differential equations problem. 
%

%

In the regime described above, the Eulerian-Eulerian equations for a mixture of a gas and a solid dispersed phase are \citep{feireisl2004dynamics, Marble1970a, Neri2003, gidaspow1994multiphase, garnier2009large, Berselli2014, esposti2008transient}:

\begin{equation}\label{eq:EulerianEulerian}
\begin{cases}
 \ddt\rho_i + \Div (\rho_i \vec{u}_i) = 0\,, \qquad i \in \mathcal{I}\,; \\
  \ddt\rho_j + \Div (\rho_j \uj) = S_j\,, \qquad j \in \mathcal{J}\,; \\
 \ddt (\rhog \ug) + \Div (\rhog \ug \otimes \ug) + \nabla p = \\
 \qquad\qquad\qquad\qquad\qquad = \Div \mathbb{T} + \rho \vec{g} - \displaystyle\sum_{j\in\mathcal{J}} \vec{f}_j\,;\\
 \ddt (\rho_j \uj) + \Div(\rho_j \uj\otimes\uj) = \\
 \quad\qquad\qquad\qquad =  \rho_j \vec{g} + \vec{f}_j + S_j \uj\,, \qquad j \in \mathcal{J}\,; \\
 \ddt (\rhog \hg) + \Div ( \rhog \hg \ug) + \Div( \vec{q} - \mathbb{T}\cdot \ug) = \\
 \qquad = \ddt p - \ddt(\rhog \Kg) - \Div(\rhog \Kg \ug) + \\
 \qquad\qquad + \rhog (\vec{g}\cdot\ug) - \displaystyle\sum_{j\in\mathcal{J}} ( \uj\cdot\vec{f}_j + \mathrm{Q}_j)\,; \\
 \ddt (\rho_j h_j) + \Div (\rho_j h_j \uj ) = \mathrm{Q}_j + S_j h_j\,, \quad j \in \mathcal{J}\,;
\end{cases}
\end{equation}
with the following constitutive equations (\vec{g} is the gravitational acceleration):
\begin{itemize}
\item Given $y_{i(j)}$ the mass fractions of the gaseous (solid) phases and $\rhom$ the bulk density of the mixture, the bulk density of the gas phase is $\rhog = \sum_\mathcal{I} \rho_i = \sum_\mathcal{I} y_i \rhom$, while the mass fraction of the solid phases \mbox{$\rhos = \sum_\mathcal{J} \rho_j = \sum_\mathcal{J} y_j \rhom$}. Consequently, $\rhom = \rhog + \rhos$.

\item The volumetric concentration of the ith(jth) phase is given by $\epsilon_i = \rho_i/\hat{\rho}_i$.

 \item Perfect gas: $p = \sum_\mathcal{I} \hat{\rho}_i R_i \Tg$, with $R_i$ the gas constant of the ith gas phase. This law can be simplified by nothing that $\epss \ll 1$, thus $\epsilon_i \simeq 1$ and $\hat{\rho}_i \simeq \rho_i$ \citep[cf.][]{Suzuki2005a}. Anyway, in this work we use the complete version of the perfect gas law. It can be written in convenient form for a poly-disperse mixture as:
 \begin{equation}\label{eq:perfectGas}
 \frac{1}{\rhom} = \sum_{j\in\mathcal{J}} \frac{y_j}{\hat{\rho}_j} + \sum_{i\in\mathcal{I}} \frac{y_i R_i \Tg}{p}\,.
 \end{equation}
 \item Newtonian gas stress tensor:
 \begin{equation}
 \mathbb{T} = 2\mu(\Tg)\,\big(\mathrm{sym}(\nabla
\ug)-\frac{1}{3}\Div\ug\mathbb{I}\big)\,,
\end{equation}
where \mbox{$\mu(T) = \sum_\mathcal{I} \epsilon_i \mu_i(T)$} is the gas dynamic viscosity and $\mu_i$ is that of the ith gas component.
 \item Enthalpy per unit of mass of the gas (solid) phase:\\ $\hg = \sum_\mathcal{I} \rho_i C_{p,i}\Tg/\rhog$ $\big(h_j = C_j T_j\big)$, with $C_{p,i}$ $\big(C_j\big)$ the specific heat at constant pressure of the ith (jth) phase.
 \item The Fourier law for the heat transfer in the gas: \mbox{$\vec{q} = -k_\textup{g}\,\nabla T$}, where $k_\textup{g} = \sum_\mathcal{I} \epsilon_i k_i$ and $k_i$ is the conductivity of the ith gas component.
 \item $\mathrm{Q}_j$ and $\vec{f}_j$ refer to $\mathrm{Q}_T$ and $\vec{f}_d$ of the jth solid phase; $S_j$ is the source or sink term (when needed) of the jth phase. $K_{i} = |\vec{u}_{i}|^2/2$ is the kinetic energy per unit of mass of the ith gas phase ($K_j$ for the jth solid phase).
\end{itemize}

%
\subsection{Equilibrium-Eulerian model}
In the limit $\St_j \ll 1$, the drag terms $\vec{f}_j$ and the thermal exchange terms $\mathrm{Q}_j$ can be calculated by knowing $\ug$ and $\Tg$, and the Eulerian-Eulerian model can be largely simplified by considering the dusty-gas (also known as pseudo-gas)
approximation \citep{Marble1970a}. A refinement of this approximation (valid if $\textup{St}_j \lesssim 0.2$), has been developed by \citet{maxey:1987}, as a first-order approximation of the Lagrangian particle momentum balance (see Eq.~\eqref{eq:EulerianEulerian}d):
\begin{equation}\label{eq:lagrangian}
 \ddt \uj + \uj\cdot \nabla\,\uj = \frac{1}{\tau_j}(\ug -\uj) +
\vec{g}\,.
\end{equation}
By using the Stokes law and a perturbation method, and by defining 
$\vec{a} \equiv \Dt \ug$
(with $\Dt = \frac{\partial}{\partial t} + \vec{u}\cdot \nabla$ ), we obtain a correction to particle velocity up to first order
\begin{equation}
\uj =\ug + \vec{w}_j - \tau_j (\vec{a} + \vec{w}_j\cdot\nabla\ug) + O(\tau_j^2)\label{eq:uEqEu}
\end{equation}
leading to the so-called equilibrium-Eulerian model developed 
by \citet{Ferry2001} and \citet{Balachandar2010} for incompressible multiphase flows.
It is worth noting that at the zeroth order we recover \mbox{$\uj = \ug + \vec{w}_j$}, where $\vec{w}_j$ is the settling velocity defined in Eq.~\eqref{eq:settling_w}.

To write the compressible version of that model, we define the relative jth particle velocity field $\vec{v}_j$ so that $\uj = \ug + \vec{v}_j$. Recalling the definitions of the mass fraction and the mixture density given above, we define:
\begin{align}
& \ur = -\sum_{j\in\mathcal{J}} y_j \vec{v}_j\\
& \ub = \ug - \ur\\
& \tr = \sum_{j\in\mathcal{J}}(y_j \vec{v}_j\otimes\vec{v}_j) - \ur \otimes \ur\,,
\end{align}
By summing up the gas momentum equation with the solid momentum equations in Eq.~\eqref{eq:EulerianEulerian}, we thus obtain:
\begin{multline}
\ddt (\rhom \ub) + \Div (\rhom \ub\otimes \ub + \rhom \tr) = \\
= - \nabla p + \Div\mathbb{T} + \rhom \vec{g} + \displaystyle\sum_{j\in\mathcal{J}} S_j \uj\,.
\end{multline}
This momentum balance equation is equivalent to the compressible Navier-Stokes equation with the substitution $\ug \to \ub$ and the addition of the term $\Div (\rhom\tr)$ which takes into account the first order effects of particle decoupling on momentum (two-way coupling). We keep this term because of the presence of the settling velocity $\vec{w}_j$ in $\vec{v}_j$ which is at the leading order.

Moving to the mass conservation, summing up over $i$ and $j$ the continuity equations in \eqref{eq:EulerianEulerian}, we obtain the continuity equation for the mixture:
\begin{equation}\label{eq:mass_beta_eqEu_1}
\ddt \rhom + \Div( \rhom \ub) = \sum_{j\in\mathcal{J}} S_j\,,
\end{equation}
while for the phases we have:
\begin{align}
&\ddt (\rhom y_i) + \Div(\rhom\ug y_i) = 0\,,\quad i \in \mathcal{I}\\
&\ddt (\rhom y_j) + \Div[\rhom(\ug + \vec{v}_j) y_j] = S_j\,,\quad j \in \mathcal{J}\,.
\end{align}
It is worth noting that the mixture density follows the classical continuity equation with velocity field $\ub$. We refer to $\ub$ as the mixture velocity field.

As pointed out in Eq.~\eqref{eq:tauTtaus} and below, in our physical regime the thermal Stokes time is of the same order of magnitude of the kinematic one. However, this regime has been thoroughly analyzed in the incompressible case by \cite{ferry2005equilibrium}, demonstrating that the error made by assuming thermal equilibrium is at least one order of magnitude smaller than that on the momentum equation (at equal Stokes number), thus justifying the limit $T_j \to \Tg = T$ as done for the thermal equation in the dusty gas model. 

By summing up all enthalpy equations in \eqref{eq:EulerianEulerian}, and by defining $\hm = \sum_\mathcal{I} y_i h_i + \sum_\mathcal{J} y_j h_j = \Cm T$ and $\Kb = \sum_\mathcal{I} y_i K_i + \sum_\mathcal{J} y_j K_j$, we obtain:
\begin{multline}
\ddt(\rhom \hm) + \Div\left[\rhom \hm(\ub + \vr)\right] = \\
= \ddt p -\ddt(\rhom \Kb) - \Div\left[\rhom \Kb(\ub + \vk)\right] +\\
 + \Div(\mathbb{T}\cdot\ug - \vec{q}) + \rhom (\vec{g}\cdot\ub) + \sum_{j\in\mathcal{J}} S_j (h_j + K_j)\,.
\end{multline}

The terms
\begin{align}\label{eq:vr}
& \vr = \ur + \dfrac{\sum_\mathcal{J} y_j C_j \vec{v}_j}{\Cm} = \dfrac{\sum_\mathcal{J} y_j (C_j - \Cm)\vec{v}_j}{\Cm}\\
& \vk = \ur + \dfrac{\sum_\mathcal{J} y_j K_j \vec{v}_j}{\Kb} = \dfrac{\sum_\mathcal{J} y_j (K_j - \Kb)\vec{v}_j}{\Kb}\,,
\end{align}
take into account the combined effect of the kinematic decoupling and the difference between the specific heat ($\vr$) and kinetic energy ($\vk$) of the mixture and of the jth specie.

Summarizing, the compressible equilibrium-Eulerian model is:
\begin{equation}\label{eq:equilibriumEulerian}
\begin{cases}
 \ddt\rhom + \Div(\rhom \ub) = \displaystyle\sum_{j\in\mathcal{J}} S_j\,;\\
 \ddt(\rhom y_i) + \Div(\rhom \ug y_i) = 0\,, \quad i \in \mathcal{I}\,;\\
  \ddt(\rhom y_j) + \Div(\rhom \uj y_j) = S_j\,, \quad j \in \mathcal{J}\,;\\
 \ddt (\rhom \ub) + \Div (\rhom \ub\otimes \ub + \rhom \tr) = \\
 \qquad\qquad = - \nabla p + \Div\mathbb{T} + \rhom \vec{g} + \displaystyle\sum_{j\in\mathcal{J}} S_j \uj\,;\\
 \ddt (\rhom \hm) + \Div\left[\rhom \hm(\ub + \vr)\right] = \\
 \quad = \ddt p - \ddt(\rhom \Kb) -\Div\left[\rhom \Kb (\ub + \vk)\right] + \\
 \quad\quad + \Div(\mathbb{T}\cdot \ug - \vec{q}) + \rhom(\vec{g}\cdot\ub) + \displaystyle\sum_{j\in\mathcal{J}} S_j (h_j + K_j)\,.
\end{cases}
\end{equation}
The first equation is redundant, because it is contained in the second and third set of continuity equations. Note that we have not used the explicit form of $\vec{v}_j$ for deriving Eqs.~\eqref{eq:equilibriumEulerian}, which therefore can be used for any multiphase flow model with $I$ phases moving with velocity $\ug$ and temperature $T$, and $J$ phases each moving with velocity $\uj = \ug + \vec{v}_j$ and temperature $T$. However, in what follows we will use Eq.~\eqref{eq:uEqEu} when referring to the compressible Equilibrium-Eulerian model.

It is also worth noting that in the Navier-Stokes equations it is critical to accurately take into account the non-linear terms contained by the conservative derivative $\ddt \psi + \Div(\psi \vec{u})$ because they are the origin of the major difficulties in turbulence modeling. A large advantage of the dusty gas and equilibrium-Eulerian models is that in both models the the most relevant part of the drag ($\sum_\mathcal{J} \vec{f}_j$) and heat exchange ($\sum_\mathcal{J} \mathrm{Q}_j$) terms have been absorbed into the conservative derivatives for the mixture. This fact allows the numerical solver to implicitly and accurately solve the particles contribution on mixture momentum and energy (two-way coupling), using the same numerical techniques developed in Computational Fluid Dynamics for the Navier-Stokes equations. The dusty gas and Equilibrium-Eulerian models are best suited for solving multiphase system in which the particles are strongly coupled with the carrier fluid and the bulk density of the particles is not negligible with respect to that of the fluid.


The equilibrium-Eulerian model thus reduces to a set of mass, momentum, and energy balance equations
for the gas-particle mixture plus one equation for the mass transport of the particulate phase.
In this respect, it is similar to the dusty-gas equations, to which it reduces for $\taus \equiv 0$.
With respect to the dusty-gas model, here we solve for the mixture velocity $\ub$, which is slightly different from the carrier gas velocity $\ug$. Moreover, here kinematic decoupling is taken into account by moving the $I$ gas phases and the $J$ solid phases with different velocity fields, respectively $\ug$ and $\uj$. Thus, we are accounting for the imperfect coupling of the particles to the gas flow, leading to 
preferential concentration and settling phenomena (the vector $\vec{v}_j$ includes a convective and a gravity
accelerations terms).

The equilibrium-Eulerian method becomes even more efficient (relative to the standard Eulerian)
for the polydisperse case ($J > 1$). For each species of particle tracked, the standard Eulerian method requires
four scalar fields, the fast method require one. Furthermore, the computation of the
correction to $\vec{v}_j$ needs only to be done for one particle species. The correction has the
form $-\tau_j\vec{a}$, so once the term $\vec{a}$ is computed, velocities for all species of
particles may be obtained simply by scaling the correction factor based on the species' response
times $\tau_{j}$.
To be more precise, the standard Eulerian method needs $I + 5J +4$ scalar partial differential equations, while the equilibrium-Eulerian model needs just $I + J + 4$, i.e. $4 J$ equations less.
%
%
%
%
%
%
%
\subsection{Sub-grid scale models}

The spectrum of the density, velocity and temperature fluctuations of turbulent flows at high Reynolds number typically span over many orders of magnitude. In the cases where the turbulent spectrum extend beyond the numerical grid resolution, it is necessary to model the effects of the high-frequency fluctuations (those that cannot be resolved by the numerical grid) on the resolved flow. This leads to the so-called Large-Eddy Simulation (LES) techniques, in which a low-pass filter is applied to the model equations to filter out the small scales of the solution. In the incompressible case the theory is well-developed~\citep[see][]{berselli2005mathematics,sagaut2006large}, but LES for compressible flows is still a new and open research field.
In our case, we apply a spatial filter, denoted by an overbar ($\delta$ is the filter scale):
\begin{equation}
\bar{\psi} = \int_\Omega G(\vec{x} - \vec{x}'; \delta) \psi(\vec{x}') \de \vec{x}'\,.
\end{equation}
Some example of LES filters $G(\vec{x}; \delta)$ used in compressible turbulence are reviewed in \cite{garnier2009large}. 
In compressible turbulence it is also useful to introduce the so-called Favre filter:
\begin{equation}\label{eq:favre_filter}
\tilde{\psi} = \frac{\overline{\rhom \psi}}{\rhombar}\,.
\end{equation}

First, we apply this filter to the Equilibrium-Eulerian model fundamental equation~\eqref{eq:uEqEu} modified as follows:
\begin{multline}
\uj = \ug + \vec{w}_j + \\
 -\tau_j ( \ddt \ub + \ub\cdot\nabla\ub + (\wrr + \vec{w}_j)\cdot\nabla\ub) + O(\tau_j^2)
\end{multline}
moving all the new second order terms into $O(\tau_j^2)$, using $\ddt y_j + \uj\cdot\nabla y_j = 0$ and defining:
\begin{equation}
\wrr = - \sum_j y_j \vec{w}_j\,.
\end{equation}
Multiplying the new expression for $\uj$ by $\rhom$ and Favre-filtering, at the first order we obtain:
\begin{multline}\label{eq:umEqEu_filt}
\rhombar \tilde{\vec{u}}_j = \rhombar (\ugt + \vec{w}_j) + \\
 - \tau_j \left(\ddt (\rhombar\ubt) + \Div (\rhombar \ubt\otimes \ubt) + \rhombar(\wrrt + \vec{w}_j)\cdot \nabla\ubt\right) + \\
  -\tau_j \Div \mathbb{B}\,,
\end{multline}
where we have used $\tilde\tau_j = \tau_j$ and consequently $\tilde{\vec{w}}_j = \vec{w}_j$ because the Stokes time changes only at the large scale and it can be considered constant at the filter scale. Moreover, we have defined the subgrid-scale Reynolds stress tensor:
\begin{equation}
{\mathbb{B}} = \rhombar(\widetilde{\ub\otimes\ub} - \ubt\otimes\ubt)\,.
\end{equation}
As discussed and tested in~\cite{Shotorban2007}, the subgrid terms can be considered $O(\tau_j)$ and neglected when multiplied by first order terms. 
Another form of Eq.~\eqref{eq:umEqEu_filt} can be recovered by noting that at the leading order $\ubt \simeq \ugt - \wrrt$:
\begin{equation}\label{eq:ugEqEu_filt}
\tilde{\vec{u}}_j = \ugt + \vec{w}_j - \tau_j \left(\ddt \ugt + \ugt\cdot\nabla \ugt + \vec{w}_j\cdot \nabla\ugt + \Div \mathbb{B}/\rhombar\right)\,.
\end{equation}

We recall here the Boussinesq eddy viscosity hypothesis:
\begin{equation}\label{eq:sgsB}
\mathbb{B} = \frac{2}{\mathsf{d}} \rhombar K_\textup{t} \mathbb{I} - 2 \mut \tilde{\mathbb{S}}_\textup{m}\,,
\end{equation}
 where the deviatoric part of the subgrid stress tensor can be modeled with an eddy viscosity $\mut$ times the rate-of-shear tensor $\tilde{\mathbb{S}}_\textup{m} = \mathrm{sym}(\nabla\ubt) - \frac{1}{3}\Div\ubt \mathbb{I}$. The first term on the right hand side of Eq.~\eqref{eq:sgsB} is the isotropic part of the subgrid-scale tensor, proportional to the subgrid-scale kinetic energy $K_\textup{t}$. While in incompressible turbulence the latter term is absorbed into the pressure, it must be modeled for compressible flows (cf.~\cite{Moin1991} and \cite{Yoshizawa1986}). \cite{ducros1995} showed another way to treat this term by absorbing it into a new {\em macro-pressure} and {\em macro-temperature} (cf. also \cite{lesieur2005} and~\cite{Lodato2009}).
 We recall here also the eddy diffusivity viscosity model (cf.~\cite{Moin1991}): any scalar $\psi$ transported by $\ub$ generates a subgrid-scale vector that can be modeled with the large eddy variables. We have:
\begin{equation}\label{eq:eddy_diffusivity}
\rhombar(\widetilde{\ub\psi} - \ubt \tilde\psi) = -\frac{\mut}{\Prat} \nabla \tilde\psi\,,
\end{equation}
where $\Prat$ is the subgrid-scale turbulent Prandtl number.

We apply the Favre filter defined in Eq.~\eqref{eq:favre_filter} to Eqs.~\eqref{eq:equilibriumEulerian} (for the application of the Favre filter to the compressible Navier-Stokes equations cf.~\cite{garnier2009large}, \cite{Moin1991} and~\cite{Erlebacher1990}), obtaining:
\begin{align}
& \ddt \rhombar + \Div( \rhombar \ubt) = \tilde{S}_\textup{m}\,; 
\label{eq:mass_beta_eqEu_filt}\\
&\ddt (\rhombar \tilde y_i) + \Div(\rhombar \ugt \tilde y_i) = - \Div \mathcal{Y}_i\,,\quad i \in \mathcal{I}\,;
\label{eq:mass_i_eqEu_filt}\\
&\ddt (\rhombar \tilde y_j) + \Div[\rhombar \tilde{\vec{u}}_j \tilde y_j] = \tilde{S}_j - \Div \mathcal{Y}_j\,,\quad j \in \mathcal{J}\,;
\label{eq:mass_j_eqEu_filt}\\
& \ddt (\rhombar \ubt) + \Div(\rhombar\ubt\otimes\ubt + \rhombar\trt) + \nabla \bar p = \nonumber\\
& \qquad\qquad  = \Div\tilde{\mathbb{T}} + \sum_{j\in\mathcal{J}} \tilde{S}_j \tilde{\vec{u}}_j + \rhombar \vec{g} - \Div \mathbb{B}
\label{eq:mom_eqEu_filt}\\
& \ddt (\rhombar \hmt) + \Div[\rhombar(\ubt + \vrt) \hmt] = \nonumber\\
&\qquad = \ddt \bar p - \ddt (\rhombar \Kbt) - \Div[\rhombar(\ubt + \vkt) \Kbt] + \nonumber\\
&\qquad\qquad + \Div(\tilde{\mathbb{T}} \cdot \ugt - \tilde{\vec{q}}) + \rhombar(\vec{g}\cdot\ubt) + \nonumber\\
&\qquad\qquad\qquad + \sum_{j\in\mathcal{J}} \tilde{S}_j (\tilde{h}_j + \tilde{K}_j) -\Div\mathcal{Q}\,,
\label{eq:en_eqEu_filt}
\end{align}
where
\begin{align}
& \mathcal{Y}_i = \rhombar(\widetilde{y_i \ug} - \tilde y_i \ugt) = - \frac{\mut}{\Prat}\nabla \tilde y_i \\
& \mathcal{Y}_j = \rhombar(\widetilde{y_j \uj} - \tilde y_i \tilde{\vec{u}}_j) = - \frac{\mut}{\Prat}\nabla \tilde y_j \\
& \mathbb{B} = \rhombar(\widetilde{\ub\otimes\ub} - \ubt\otimes\ubt) = \frac{2}{\mathsf{d}}\rhombar K_\textup{t} \mathbb{I} - 2 \mut \tilde{\mathbb{S}}_\textup{m} \\
& \mathcal{Q} = \rhombar(\widetilde{\hm \ub} - \hmt \ubt) = -\frac{\mut}{\Prat} \nabla \hmt\,,
\label{eq:sgs_Q}
\end{align}
are: the subgrid eddy diffusivity vector of the ith phase; of the jth phase; the subgrid-scale stress tensor; the diffusivity vector of the temperature; respectively.
Other approximations have been used to derive the former LES model: the viscous terms in momentum and energy equations, and the pressure-dilatation and conduction terms in the energy equations are all non-linear terms and we here treat them as done by \cite{Erlebacher1990} and \cite{Moin1991}. The subgrid terms corresponding to the former non-linear terms could be neglected so that, for example, $\overline{p\,\Div \ug} \simeq \bar{p}\,\Div\ugt$. In particular, this term has been neglected also in presence of shocks (cf. \cite{Garnier2002}). We refer to \cite{Vreman1995} for an {\em a priori} and {\em a posteriori} analysis of all the neglected terms of the compressible Navier-Stokes equations. Moreover, in our model the mixture specific heat $\Cm$ and the mixture gas constant $\Rm$ vary in the domain because $y_i$ and $y_j$ vary. Thus, also the following approximations should be done, coherently with the other approximations used: $\hmt = \widetilde{\Cm T} \simeq \Cmt \tilde T$ and $\widetilde{\Rm T} \simeq \tilde R_\textup{m} \tilde T$.

In order to close the system, terms $\mut$, $K_\textup{t}$ and $\Prat$ must be chosen on the basis of LES models, either static or dynamic \citep[see][]{Moin1991, bardina1980, germano1991}. In the present model, we implemented several sub-grid scale (SGS) models to compute the SGS viscosity, kinetic energy and Prandtl number \citep{cerminara2015phd}. Currently, the code offers the possibility of choosing between: 1) the compressible Smagorinsky model, both static and dynamic \citep[see][]{Fureby1996, Yoshizawa1993, pope2000, Chai2012, garnier2009large}; 2) the sub-grid scale K-equation model, both static and dynamic \citep[see][]{chacon2013, Fureby1996,Yoshizawa1993,Chai2012}; 3) the dynamical Smagorinsky model in the form by \citet{Moin1991}; 4) the WALE model, both static and dynamic \citep[see][]{Nicoud1999, Lodato2009, Piscaglia2013}.

All through this paper, we present results obtained with the dynamic WALE model (see Fig.~\ref{fig:ek} and the corresponding section for a study on the accuracy of this LES model). A detailed analysis of the influence of subgrid-scale models to simulation results is beyond the scopes of this paper and will be addressed in future works.
%
\section{Numerical solver}
\label{numerics}
The Eulerian model described in Section \ref{model}, is solved numerically to obtain a time-dependent description of all independent flow fields in a three-dimensional domain with prescribed initial and boundary conditions.
%
We have chosen to adopt an open-source approach to the code development in order to guarantee control on the numerical solution procedure and to share scientific knowledge. We hope that this will help building a wider computational volcanology community.
As a platform for developing our solver, we have chosen the unstructured, finite volume (FV) method based open source C++ library OpenFOAM (version 2.1.1). OpenFOAM, released under the Gnu Public License (GPL), has gained a vast popularity during the recent years. The readily existing solvers and tutorials provide a quick start to using the code also to inexperienced users. Thanks to a high level of abstraction in the programming of the source code, the existing solvers can be freely and easily modified in order to create new solvers (e.g., to solve a different set of equations) and/or to implement new numerical schemes. OpenFOAM is well integrated with advanced tools for pre-processing (including meshing) and post-processing (including visualization). The support of the OpenCFD Ltd, of the OpenFOAM foundation and of a wide developers and users community guarantees ease of implementation, maintenance and extension, suited for satisfying the needs of both volcanology researchers and of potential users, e.g. in volcano observatories. Finally, all solvers can be run in parallel on distributed memory architectures, which makes OpenFOAM suited for envisaging large-scale, three-dimensional volcanological problems.

The new computational model, called ASHEE (ASH Equilibrium Eulerian model) is documented in the VMSG (Volcano Modeling and Simulation Gateway) at Istituto Nazionale di Geofisica e Vulcanologia (http://vmsg.pi.ingv.it) and is made available through the VHub portal (https://vhub.org). 
\subsection{Finite Volume discretization strategy}
In the FV method \citep{ferziger1996computational}, the governing partial differential equations are integrated over a computational cell, and the Gauss theorem is applied to convert the volume integrals into surface integrals, involving surface fluxes. Reconstruction of scalar and vector fields (which are defined in the cell centroid) on the cell interface is a key step in the FV method, controlling both the accuracy and the stability properties of the numerical method.
 
OpenFOAM implements a wide choice of discretization schemes.
In all our test cases, the temporal discretization is based on the second-order Crank-Nicolson scheme \citep{ferziger1996computational}, with a blending factor of 0.5 (0 meaning a first-order Euler scheme, 1 a second-order, bounded implicit scheme) and an adaptive time stepping based on the maximum initial residual of the previous time step \citep{kay2010adaptive}, and on a threshold that depends on the Courant number ($C < 0.2$). 
All advection terms of the model are treated implicitly to enforce stability. Diffusion terms are also discretized implicit in time, with the exception of those representing subgrid turbulence.
The pressure and gravity terms in the momentum equations and the continuity equations are solved explicitly. However, as discussed below, the PISO (Pressure Implicit with Splitting of Operators, \citet{issa1986solution}) solution procedure based on a pressure correction algorithm makes such a coupling implicit.
Similarly, the pressure advection terms in the enthalpy equation and the relative velocity $\vec{v}_j$ are made implicit when the PIMPLE (mixed SIMPLE and PISO algorithm, \citet{ferziger1996computational}) procedure is adopted.
The same PIMPLE scheme is applied treating all source terms and the additional terms deriving from the equilibrium-Eulerian expansion.

In all described test cases, the spatial gradients are discretized by adopting an unlimited centered linear scheme \citep[which is second-order accurate and has low \emph{numerical diffusion} --][]{ferziger1996computational}. Analogously, implicit advective fluxes at the control volume interfaces are reconstructed by using a centered linear interpolation scheme (also second order accurate). The only exception is for pressure fluxes in the pressure correction equation, for which we adopt a TVD (Total Variation Diminishing) limited linear scheme (in the subsonic regimes) to enforce stability and non-oscillatory behavior of the solution. We refer to \citet{jasak1996phd} for a complete description of the discretization strategy adopted in OpenFOAM.
\subsection{Solution procedure}
Instead of solving the set of algebraic equations deriving from the discretization procedure as a whole, most of the existing solvers in OpenFOAM are based on a segregated solution strategy, in which partial differential equations are solved sequentially and their coupling is resolved by iterating the solution procedure. 
In particular, for Eulerian fluid equations, momentum and continuity equation (coupled through the pressure gradient term and the gas equation of state) are solved by adopting the PISO algorithm. The PISO algorithm consists of one predictor step, where an intermediate velocity field is solved using pressure from the previous time-step, and of a number of PISO corrector steps, where intermediate and final velocity and pressure fields are obtained iteratively. 
The number of corrector steps used affects the solution accuracy and usually at least two steps are used. Additionally, coupling of the energy (or enthalpy) equation can be achieved in OpenFOAM through additional PIMPLE iterations \citep[which derives from the SIMPLE algorithm by ][]{patankar:1980}.
For each transport equation, the linearized system deriving from the implicit treatment of the advection-diffusion terms is solved by using the PbiCG solver (Preconditioned bi-Conjugate Gradient solver for asymmetric matrices) and the PCG (Preconditioned Conjugate Gradient solver for symmetric matrices), respectively, preconditioned by a Diagonal Incomplete Lower Upper decomposition (DILU) and a Diagonal Incomplete Cholesky (DIC) decomposition. The segregated system is iteratively solved until a global tolerance threshold $\epsilon_\textup{PIMPLE}$ is achieved. In our simulations, we typically use $\epsilon_\textup{PIMPLE} < 10^{-7}$ for this threshold.

The numerical solution algorithm is designed as follows:
\begin{enumerate}
\item Solve the (explicit) continuity equation \eqref{eq:mass_beta_eqEu_filt} for mixture density $\rhom$ (predictor stage: uses fluxes from previous iteration).
\item \label{pimplel} Solve the (implicit) transport equation for all gaseous and particulate mass fractions: $y_i, \quad i=1,...,I$ and $y_j, \quad j=1,...,J$.
\item Solve the (semi-implicit) momentum equation to obtain $\ub$ (predictor stage: uses the pressure field from previous iteration).
\item Solve the (semi-implicit) enthalpy equation to update the temperature field $T$ and the compressibility $\rhom/p$ (pressure from previous iteration).
\item \label{pisol} Solve the (implicit) pressure equation to update the pressure (uses predicted values of fluxes).
\item Correct density, velocity and fluxes with the new pressure field (keeping $T$ and $\rhom/p$ fixed).
\item Iterate from \ref{pisol} evaluating the continuity error as the difference between the kinematic and thermodynamic calculation of the density (PISO loop).
\item Compute explicit terms (transport coefficients and decoupling).
\item Evaluate the numerical error $\epsilon_\textup{PIMPLE}$ and iterate from \ref{pimplel} if prescribed (PIMPLE loop).
\item Compute LES subgrid terms.
\end{enumerate}

With respect to the standard solvers implemented in OpenFOAM (v2.1.1) for compressible fluid flows (e.g. {\tt sonicFoam} or {\tt rhoPimpleFoam}), the main modification required are the following:
\begin{enumerate}
\item The mixture density and velocity replaces the fluid ones.
\item A new scalar transport equation is introduced for the mass fraction of each particulate and gas species.
\item The equations of state are modified as described in Eq.\eqref{eq:perfectGas}.
\item First-order terms from the equilibrium-Eulerian model are added in the mass, momentum and enthalpy equations.
\item \label{dvlim} Equations are added to compute flow acceleration and velocity disequilibrium.
\item Gravity terms and ambient fluid stratification are added.
\item New SGS models are implemented.
\end{enumerate}
Concerning point \ref{dvlim}, it is worth remarking that, accordingly to \citet{Ferry2003}, the first-order term in $\tau_j$ in Eq.\eqref{eq:uEqEu} must be limited to avoid the divergence of preferential concentration in a turbulent flow field (and to keep the effective Stokes number below 0.2). In other word, we impose at each time step that $|\tau_j (\vec{a} + \vec{w}_j\cdot\nabla\ug)| \leq 0.2 |\ug+\vec{w}_j|$. We tested the effect of this limiter on preferential concentration in Sec.~\ref{MHIT} below.

\section{Verification and validation study}
\label{validation}
A wide set of numerical tests has been performed to assess the adequacy of the ASHEE model for the intended volcanological application and the reliability of the numerical solution method. 
Validation tests are focused on the dynamics of gas (Section \ref{DHIT}) and multiphase (Section \ref{MHIT}) turbulence and on the mixing properties of buoyant plumes (Section \ref{ZHOU}).
Compressibility likely exerts a controlling role to the near-vent dynamics during explosive eruptions \citep[e.g.,][]{Carcano2013}. Although this is not the focus of this work, we briefly discuss in Section \ref{SOD} the performance of the model on a standard one-dimensional shock wave numerical test.

\subsection{Compressible decaying homogeneous and isotropic turbulence}
\label{DHIT}
Turbulence is a key process controlling the dynamics of volcanic plumes since it controls the rate of mixing and air entrainment.
To assess the capability of the developed model to resolve turbulence \citep[which requires low numerical diffusion and controlled numerical error][]{geurts2002framework}, we have tested the numerical algorithm against different configurations of decaying homogeneous and isotropic turbulence (DHIT). 

In this configuration, the flow is initialized in a domain $\Omega$ which is a box with side $L = 2\pi$ with periodic boundary conditions. As described in \citet{lesieur2005,Honein2004,Liao2009,Pirozzoli2004,blaisdell1991numerical}, we chose the initial velocity field so that its energy spectrum is
\begin{equation}
\mathcal{E}(k) = \frac{16}{3}\sqrt{\frac{2}{\pi}} \frac{u_\textup{rms}}{k_0} \left(\frac{k}{k_0}\right)^4\,e^{-\frac{2 k^2}{k_0^2}}\,,
\end{equation}
with peak initially in $k = k_0$ and so that the initial kinetic energy and enstrophy are:
\begin{align}
& K_0 = \int_0^\infty \mathcal{E}(k) \de k = \frac{1}{2} u_\textup{rms}^2 \\
& \mathcal{H}_0 = \int_0^\infty k^2\,\mathcal{E}(k) \de k = \frac{5}{8} u_\textup{rms}^2 k_0^2\,.
\end{align}
As reviewed by \citet{pope2000}, the Taylor microscale can be written as a function of the dissipation $\epsilon = 2 \nu \mathcal{H}$:
\begin{equation}
\label{taylormic}
\lambda_\textup{T}^2 \equiv \frac{5 \nu u_\textup{rms}^2}{\epsilon} = \frac{5 K}{\mathcal{H}}\,,
\end{equation}
thus in our configuration, the initial Taylor micro scale is:
\begin{equation}
\lambda_{\textup{T}, 0} = \sqrt{\frac{5 K_0}{\mathcal{H}_0}} = \frac{2}{k_0}\,.
\end{equation}
We have chosen the non-dimensionalization keeping the root mean square of the magnitude of velocity fluctuations ($\vec{u}'$) equal to $u_\textup{rms}$:
\begin{equation}
u_\textup{rms} \equiv \frac{1}{(2\pi)^3}\int_\Omega \sqrt{\vec{u}'\cdot \vec{u}'} \de \vec{x} = 2 \int_0^\infty \mathcal{E}(k)\, \de k\,.
\end{equation}
We also chose to make the system dimensionless by fixing $\rhomo = 1$, $T_0 = 1$, $\Pra = 1$, so that the ideal gas law becomes:
\begin{equation}
p = \rhom \Rm T = \Rm\,,
\end{equation}
and the initial Mach number of the mixture based on the velocity fluctuations reads:
\begin{equation}
\Ma_{\textup{rms}} = \sqrt{\frac{u_\textup{rms}^2}{c_\textup{m}^2}} = \sqrt{\frac{2 K_0 \rhom}{\gm p}} = u_\textup{rms} (\gm\,p)^{-\frac{1}{2}}\,.
\end{equation}
This means that $\Ma_\textup{rms}$ can be modified keeping fixed $u_\textup{rms}$ and modifying $p$. Following \citet{Honein2004}, we define the eddy turnover time:
\begin{equation}
\tau_\textup{e} = \frac{\sqrt{3}\lambda_T}{u_\textup{rms}}\,.
\end{equation}

The initial compressibility ratio $C_0$ is defined as the ratio between the kinetic energy and its compressible component $K_c$:
\begin{equation}
C_0 = \frac{K_{c,0}}{K_0} = \frac{1}{2(2\pi)^3K_0} \int_\Omega \sqrt{\vec{u}_c'\cdot\vec{u}_c'} \de \vec{x}\,.
\end{equation}
Here, $\vec{u}_c'$ is the compressible part of the velocity fluctuations, so that $\Div\vec{u}' = \Div \vec{u}_c'$ and $\nabla\wedge \vec{u}_c' = 0$.

The last parameter, i.e. the dynamical viscosity, can be given both by fixing the Reynolds number based on $\lambda_{\textup{T},0}$ or $k_0$:
\begin{align}
& \Rey_{\lambda} = \frac{\rhom u_\textup{rms} \lambda_{T, 0}}{\sqrt{3}\,\mu}\\
& \Rey_{k_0} = \frac{\rhom u_\textup{rms}}{k_0\, \mu}\,.
\end{align}
It is useful to define the maximum resolved wavenumber on the selected $N$-cells grid and the Kolmogorov length scale based on $\Rey_{k_0}$. They are, respectively:
\begin{align}
&k_\textup{max} = \left(\frac{N}{2} - 1\right)\frac{2\pi}{L}\frac{N}{N - 1}\,,\\
&\etaK = \frac{2\pi}{k_0} \Rey_{k_0}^{-\frac{3}{4}}\,.
\end{align}
In order to have a DNS, the smallest spatial scale $\delta$ should be chosen in order to have $k_\textup{max}\etaK > 2$ \citep{Pirozzoli2004}.
%

We compare the DNS of compressible decaying homogeneous and isotropic turbulence with a reference, well tested numerical solver for Direct Numerical Simulations of compressible turbulence by \citet{Pirozzoli2004,Bernardini2009}. For this comparison we fix the following initial parameters: $p = \Rm = 1$, $\gm = 1.4$, $\Pra = 1$, $\Ma_\textup{rms} = 0.2$, $C_0 = 0$, $u_\textup{rms}^2 = 2K_0 = 0.056$, $k_0 = 4$, $\lambda_\textup{T} = 0.5$, $\tau_\textup{e} \simeq 3.6596$, $\mu = 5.885846*10^{-4}$, $\Rey_\lambda \simeq 116$, $\Rey_{k_0} \simeq 100$. 
Thus a grid with $N = 256^3$ cells gives $k_\textup{max} \simeq 127$ and $k_\textup{max} \etaK = 2\pi$, big enough to have a DNS. 
The simulation has been performed on 1024 cores on the {\em Fermi Blue Gene/Q} infrastructure at Italian CINECA super-computing center (http://www.cineca.it), on which about $5$ h are needed to complete the highest-resolution runs ($256^3$ of cells) up to time  $t/\tau_\textup{e}=5.465$ (about 3500 time-steps). The average required total CPU time on 1024 Fermi cores is about 1--3 millions of cells per second, with the variability associated with the number of solid phases described by the model. This value is confirmed in all benchmark cases presented in this paper.

\begin{figure}[h]
\centering
\includegraphics[width=0.8\columnwidth]{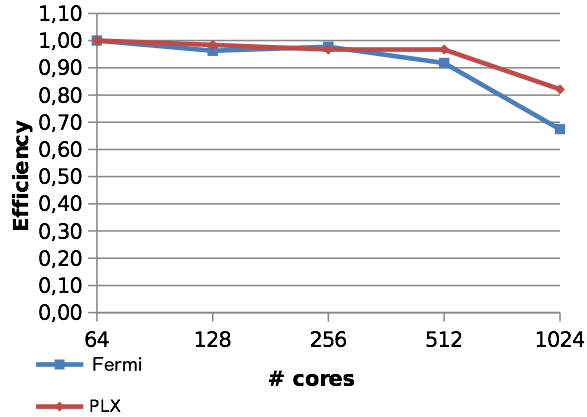}
\caption{ASHEE parallel efficiency on Fermi and PLX supercomputers at CINECA (www.cineca.it).}
\label{fig:efficiency}
\end{figure}

Fig. \ref{fig:efficiency} reports the parallel efficiency on both the Fermi and the PLX (a Linux cluster based on Intel Xeon esa- and quad-core processors at 2.4 GHz) machines at CINECA. The ASHEE code efficiency is very good (above 0.9) up to 512 cores (i.e., up to about 30000 cells per core), but it is overall satisfactory for 1024 cores, with efficiency larger than 0.8 on PLX and slightly lower (about 0.7) on Fermi probably due to the limited level of cache optimization and input/output scalability \citep{Culpo2011}. The code was run also on 2048 cores on Fermi with parallel efficiency of 0.45 \citep{dagna2013}. 

\begin{figure}[h]
\centering
\includegraphics[width=\columnwidth]{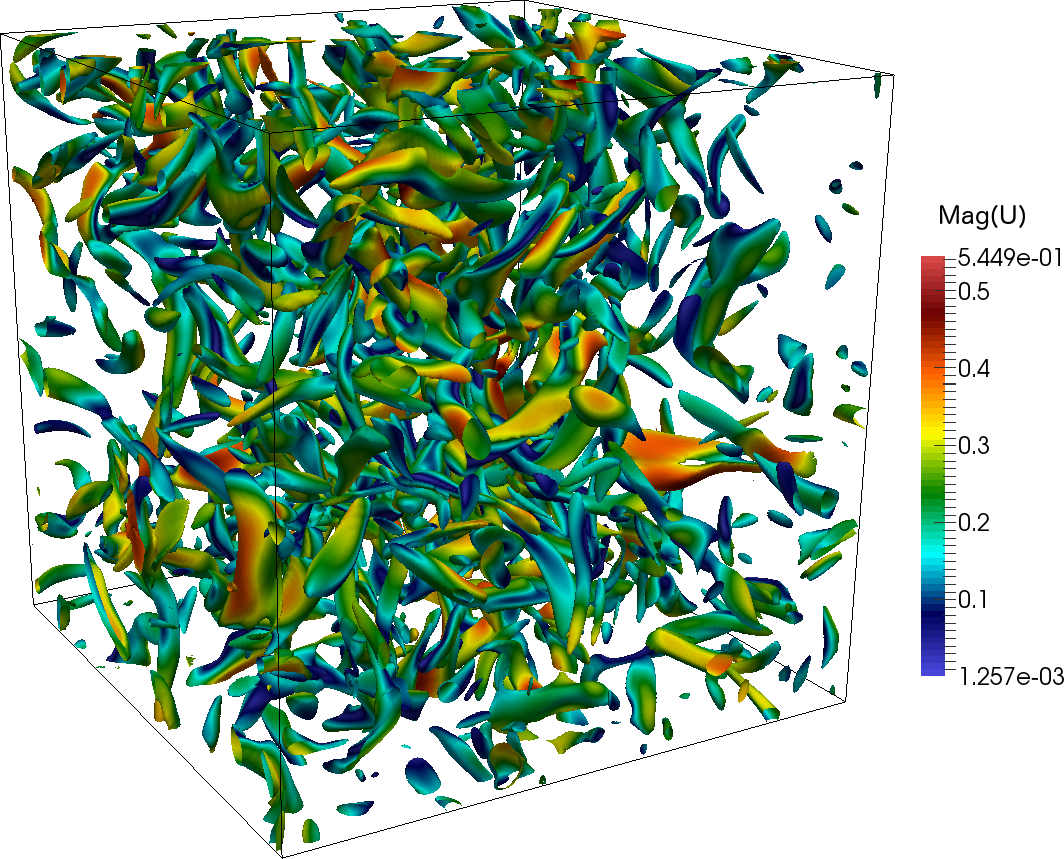}
\caption{Isosurface at $Q_{\vec{u}} \simeq 19$ Hz$^2$ and $t/\tau_\textup{e} \simeq 2.2$, representing zones with coherent vortices.}
\label{fig:dhit_Q}
\end{figure}

Fig. \ref{fig:dhit_Q} shows an isosurface of the second invariant of the velocity gradient, defined as:
\begin{equation}
Q_{\vec{u}} = \frac{1}{2}\left(\Tr(\nabla\vec{u})^2 - (\nabla\vec{u}\cdot \nabla \vec{u})\right)\,,
\label{eq:Q_u}
\end{equation}
The so called Q-criterion \citep{garnier2009large} allows indeed the identification of coherent vortices inside a three dimensional velocity field.

\begin{figure}[th!]
\centering
\includegraphics[width=0.95\columnwidth]{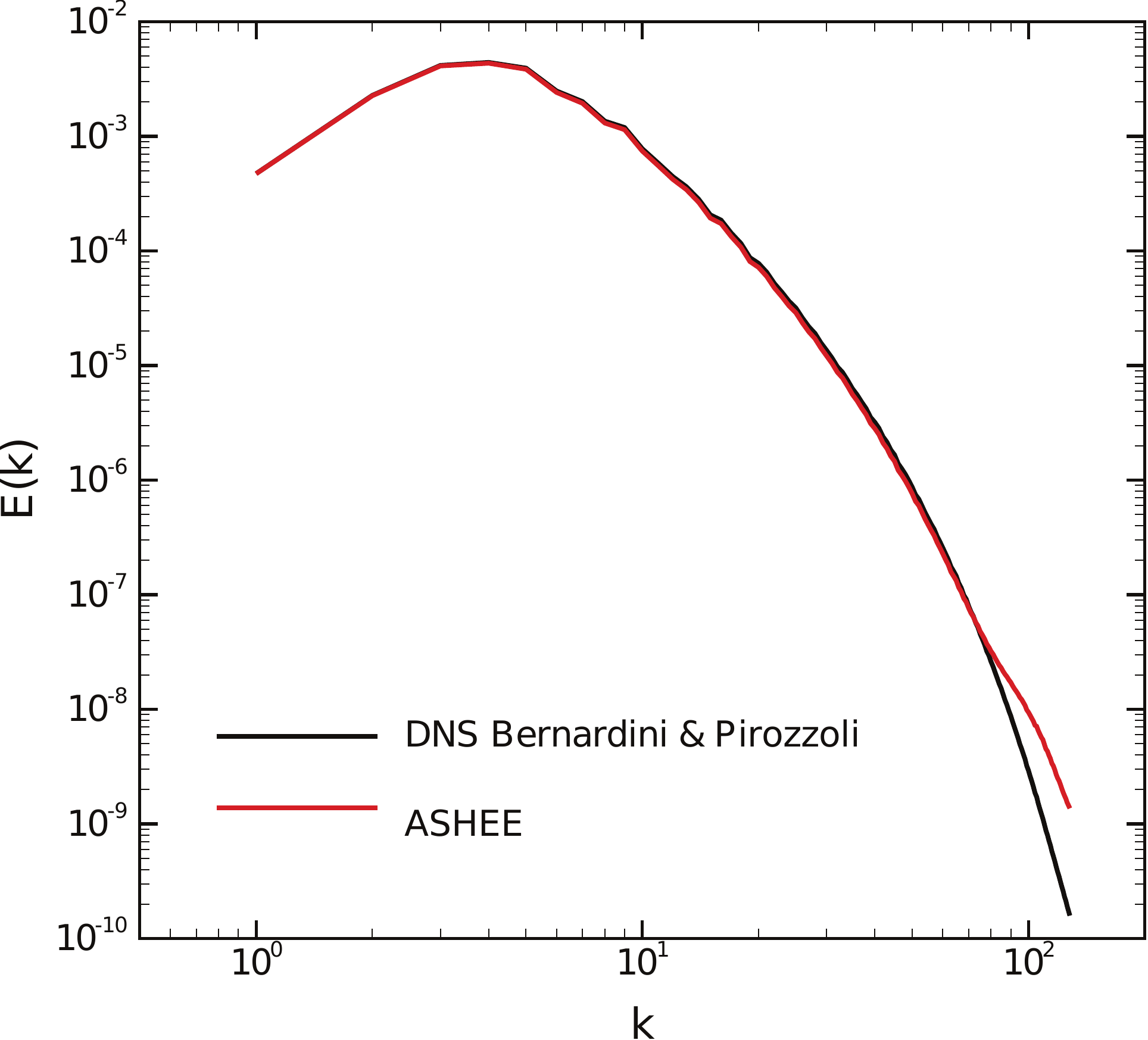}
\caption{Comparison of a DNS executed with the eight order scheme by \cite{Pirozzoli2004} and our code implemented using the \texttt{C++} libraries of OpenFOAM at $t/\tau_\textup{e} = 1.093$. The $L^2$ norm between the two spectra is $4.0*10^{-4}$. The main parameters are $\Rey_\lambda \simeq 116$, $\Ma_\textup{rms} = 0.2$.}
\label{fig:bernardini}
\end{figure}

In Fig.~\ref{fig:bernardini} we present a comparison of the energy spectrum $\mathcal{E}(k)$ obtained with the ASHEE model and the model by \citet{Bernardini2009} after approximatively 1 eddy turnover time; the $L^2$ norm of the difference between the two spectra is $4.0*10^{-4}$. This validates the accuracy of our numerical code in the single-phase and shock-free case. 

\begin{figure*}[h]
\centering
\subfloat[$\rho_\textup{rms}$, $C$, $\rho_\textup{max}/\rho_\textup{min}$][]{\fbox{\includegraphics[width=0.8\columnwidth]{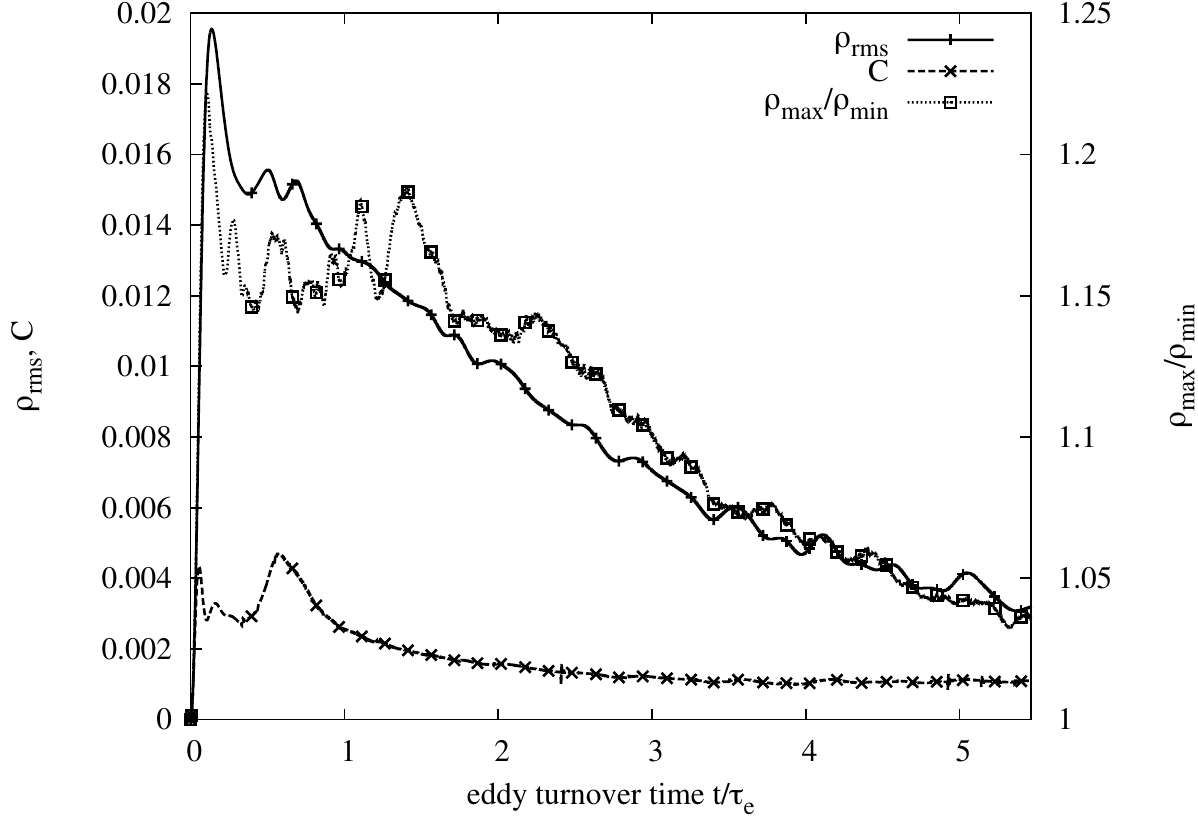}\label{fig:rhoRMS}}}\quad
\subfloat[$\mathcal{E}(k)$][]{\fbox{\includegraphics[width=0.8\columnwidth]{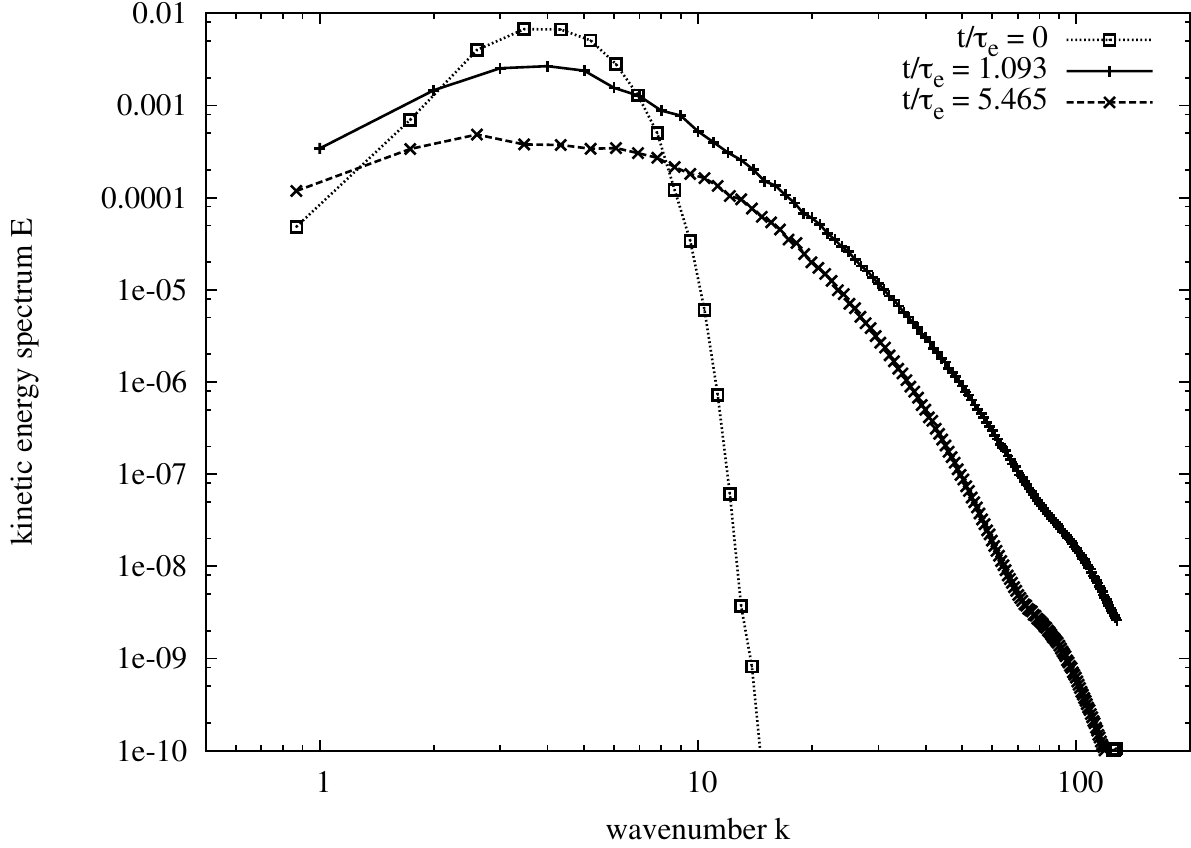}\label{fig:spectrum_evolution}}}\\
\subfloat[$K$, $\mathcal{H}$ and $\lambda_\textup{T}$][]{\fbox{\includegraphics[width=0.8\columnwidth]{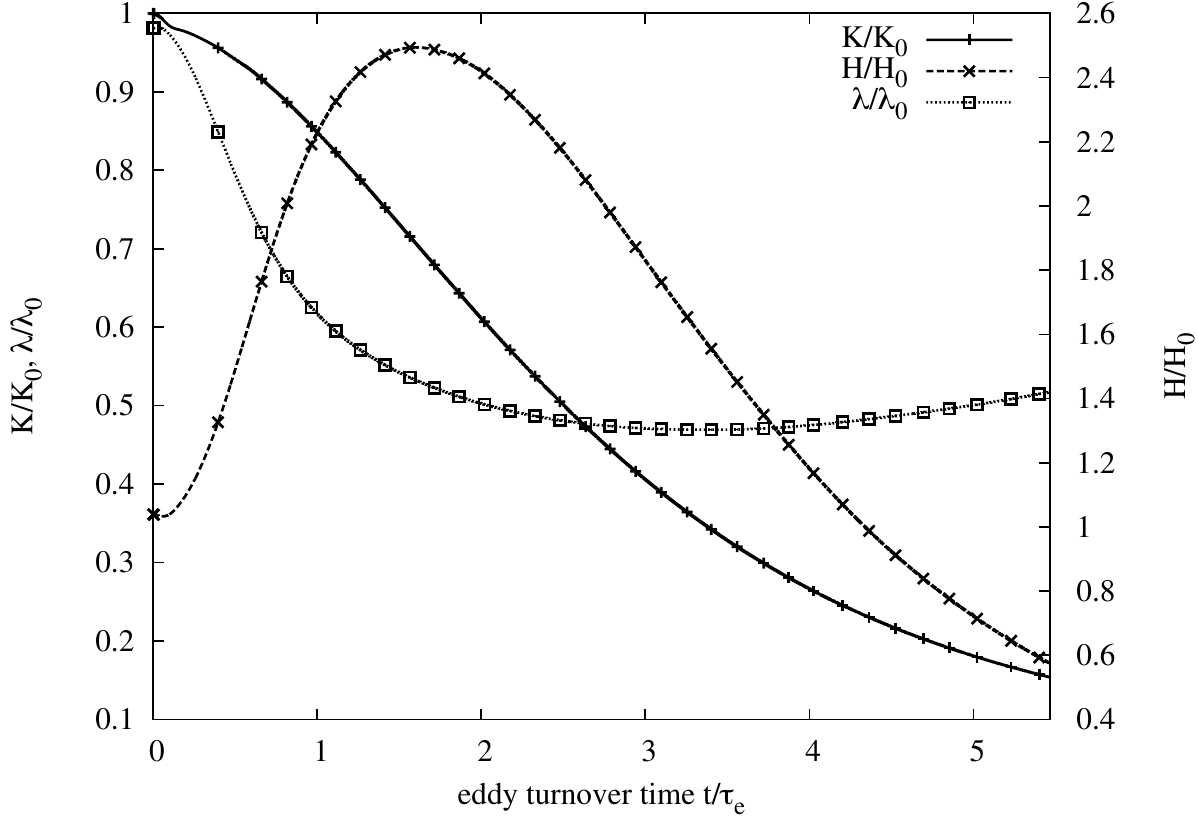}\label{fig:kin}}}\quad
\subfloat[Evolution of the Kolmogorov time microscale $\tauK$.][]{\fbox{\includegraphics[width=0.8\columnwidth]{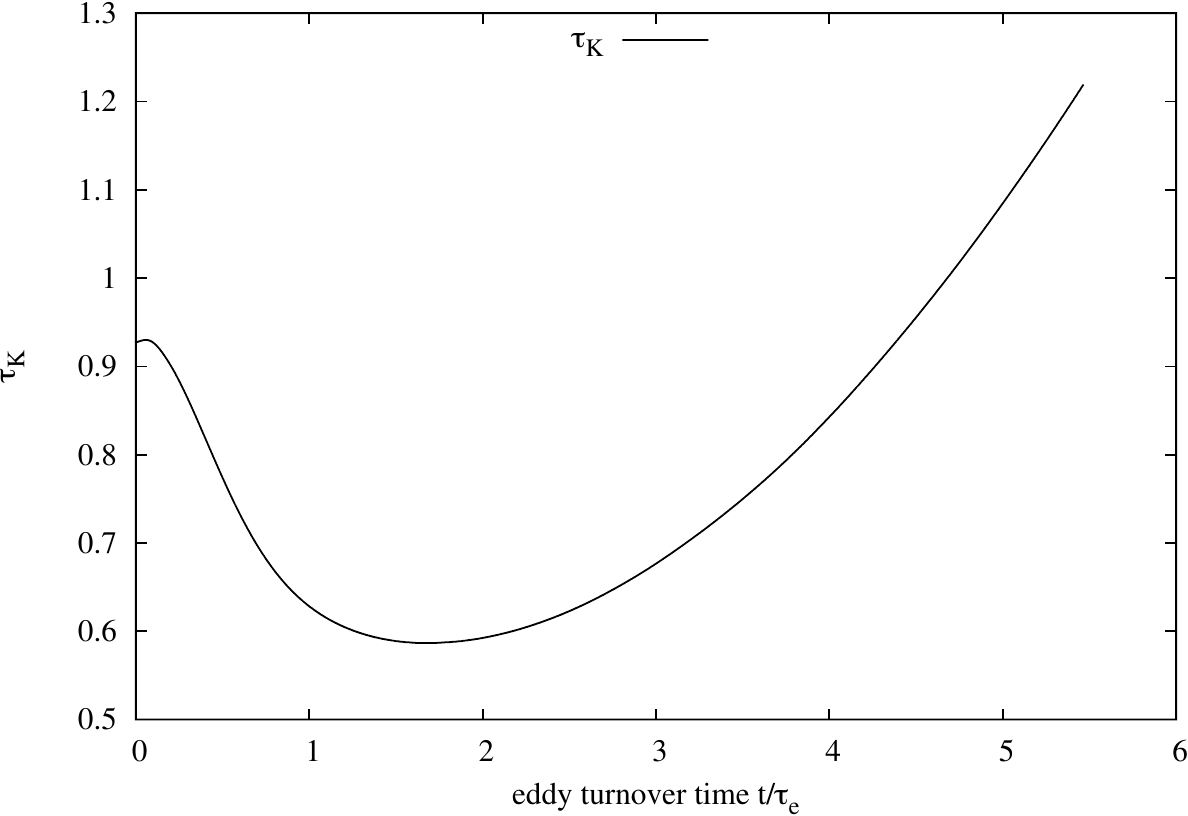}\label{fig:tauEta}}}
\caption{Evolution of some dynamical quantities in DHIT with $\Rey_\lambda \simeq 116$ and $\Ma_\textup{rms} = 0.2$ at $t/\tau_\textup{e} = 5.465$.}
\label{fig:evolution}
\end{figure*}

Fig.~\ref{fig:evolution} shows the evolution of several integral parameters describing the dynamics of the decaying homogeneous and isotropic turbulence.
Fig.~\ref{fig:rhoRMS} displays the density fluctuations $\rho_\textup{rms} = \sqrt{\langle(\rho - \langle \rho \rangle_\Omega)^2\rangle_\Omega}$, the density contrast $\rho_\textup{max}/\rho_\textup{min}$ and the standard measure of compressibility $C = \langle|\Div\vec{u}|^2\rangle_\Omega/\langle|\nabla\vec{u}|^2\rangle_\Omega$ which takes value between 0 (incompressible flow) and 1 (potential flow) \citep{Boffetta2007}.
All the quantities showed in Fig.~\ref{fig:rhoRMS} depend on the initial Mach number end compressibility. For the case shown, $\Ma_\textup{rms} = 0.2$ and we obtain very similar result to those ported in Fig. 18 and 19 by \citet{Garnier1999}.

Fig.~\ref{fig:spectrum_evolution} shows the kinetic energy spectrum at $t/\tau_\textup{e} = 0,\,1.093,\,5.465$. We notice that the energy spectrum widens from the initial condition until its tail reach $k \simeq k_\textup{max} \simeq 127$. Then system becomes to dissipate and the maximum of the energy spectrum decreases. The largest scales tend to lose energy slower than the other scales and the spectrum widens also in the larger scale direction.

Fig.~\ref{fig:kin} presents the evolution of $K$ (total turbulent kinetic energy), $\mathcal{H}$ (enstrophy), $\lambda_\textup{T}$ (Taylor microscale). 
We notice that the total kinetic energy decreases monotonically and at $t \simeq 5.5 \tau_\textup{e}$ just $\simeq 15\%$ of its initial value is conserved. On the other hand, enstrophy increases until it reaches a maximum at $1.5<t/\tau_\textup{e}<2$. It then starts to decrease monotonically. 
This behavior is related to the two different stages we have highlighted in the analysis of the energy spectrum evolution. In the first stage, viscous effects are negligible and enstrophy increases due to vortex stretching. During the second stage, viscous diffusion starts to have an important role and distorted dissipative structures are created \citep{Garnier1999}. 
Also the Taylor microscale reflects this behavior, reaching a minimum at the end of the first stage and increasing monotonically during the second stage of the evolution. It is a characteristic of the magnitude of the velocity gradients in the inertial range: by comparing it with $\delta$ we can have an idea of the broadness of the range of wave numbers where the flow is dissipative. 
In this DNS, we have $\lambda_\textup{T} \simeq 10.2 \delta$ at $t \simeq 5.5 \tau_\textup{e}$. 

In Fig.~\ref{fig:tauEta} we show the evolution of the Kolmogorov time scale $\tauK$ during the evolution of the decaying turbulence.

\begin{figure}[t]
\centering
\includegraphics[width=\columnwidth]{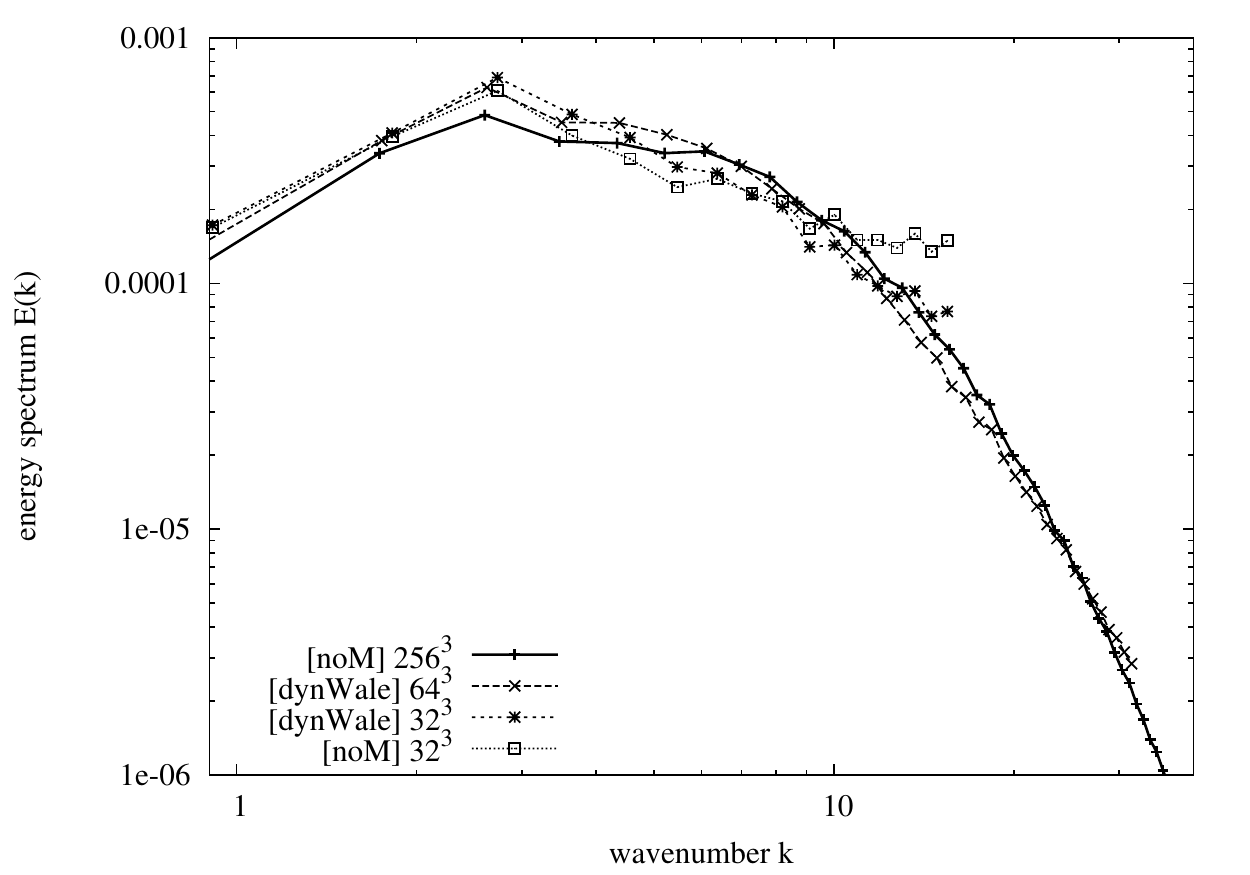}
\caption{Energy spectrum $\mathcal{E}(k)$ at $t/\tau_\textup{e} = 5.465$ obtained with different spatial resolutions and with/without subgrid scale LES model.}\label{fig:ek}
\end{figure}

We finally compare in Fig.~\ref{fig:ek} the DNS described with simulations at lower resolution with $N=32^3$ and $N = 64^3$ cells. In this case, it is expected that the spectra diverge from the DNS, unless an appropriate {\em subgrid model} is introduced to simulate the effects of the unresolved to the resolved scales. Several subgrid models have been tested \citep{cerminara2015phd}, both static and dynamic. Fig. \ref{fig:ek} presents the resulting spectrum using the dynamic WALE model \citep{Nicoud1999,Lodato2009}.
In this figure, we notice how the dynamic WALE model works pretty well for both the $32^3$ and $64^3$ LES, avoiding the smallest scales to accumulate unphysical energy.
%
%
%
\subsection{Two-phase isotropic turbulence}
\label{MHIT}
In this section we test the capability of our numerical code to correctly describe the decoupling between solid and gaseous phases when $\St_j < 0.2$ and to explore its behavior when the equilibrium-Eulerian hypothesis $\St_j < 0.2$ is not fulfilled so that a limiter to the relative velocity $\ug - \uj$ is applied.

To this aim, we performed a numerical simulation of homogeneous and isotropic turbulence with a gas phase initialized with the same initial and geometric conditions described in Sec.~\ref{DHIT}. 
We added to that configuration 5 solid particle classes ($j = 2\div6$) chosen in such a way that $\St_j \in [0.03,1]$, homogeneously distributed and with zero relative velocity: \mbox{$\vec{v}_j(\vec{x},0) = 0$}.
From Fig.~\ref{fig:tauEta}, we see that, during turbulence decay, approximately $\tauK \in [0.6,1.2]$. Therefore, for a given particle class with $\tau_j$ fixed, during the time interval $t/\tau_\textup{e}\in [0,5.5]$ we have  $\St_\textup{max}/\St_\textup{min} \simeq 2$. 
In Tab.~\ref{tab:HIT_Stokes} we report the main properties of the particles inserted in the turbulent box. To evaluate the Stokes time here we used $\tau_j = \hat{\rho}_j d_j^2/(18\mu)$ 
because in absence of settling $\Rey_j < 1$ when $\St_j < 1$ \citep{Balachandar2009}.
We set the material density of all the particles to $\hat{\rho}_j = 10^3$.
In order to have a small contribution of the particle phases to the fluid dynamics -- one way coupling -- here we set the solid particles mass fraction to a small value, $y_j = 0.002$, so that $\yg = 0.99$.
\begin{table}[t]
\centering
\begin{tabular}{ccc}
\hline
$\tau_j$ & $\textup{St}_\textup{max} = \tau_j/0.6$ & $d_j\;(\hat{\rho}_j =10^3)$\\
\hline
0.60 & 1.0 & $2.521*10^{-3}$\\
0.30 & 0.5 & $1.783*10^{-3}$\\
0.15 & 0.25 & $1.261*10^{-3}$\\
0.075 & 0.125 & $8.914*10^{-4}$\\
0.0375 & 0.0625 & $6.303*10^{-4}$\\
\hline
\end{tabular}
\caption{Stokes time, maximum Stokes number and diameter of the solid particles inserted in the turbulent box.}\label{tab:HIT_Stokes}
\end{table}

\begin{figure}[t]
\centering
\subfloat[][Mass fraction]{\includegraphics[width=0.48\columnwidth]{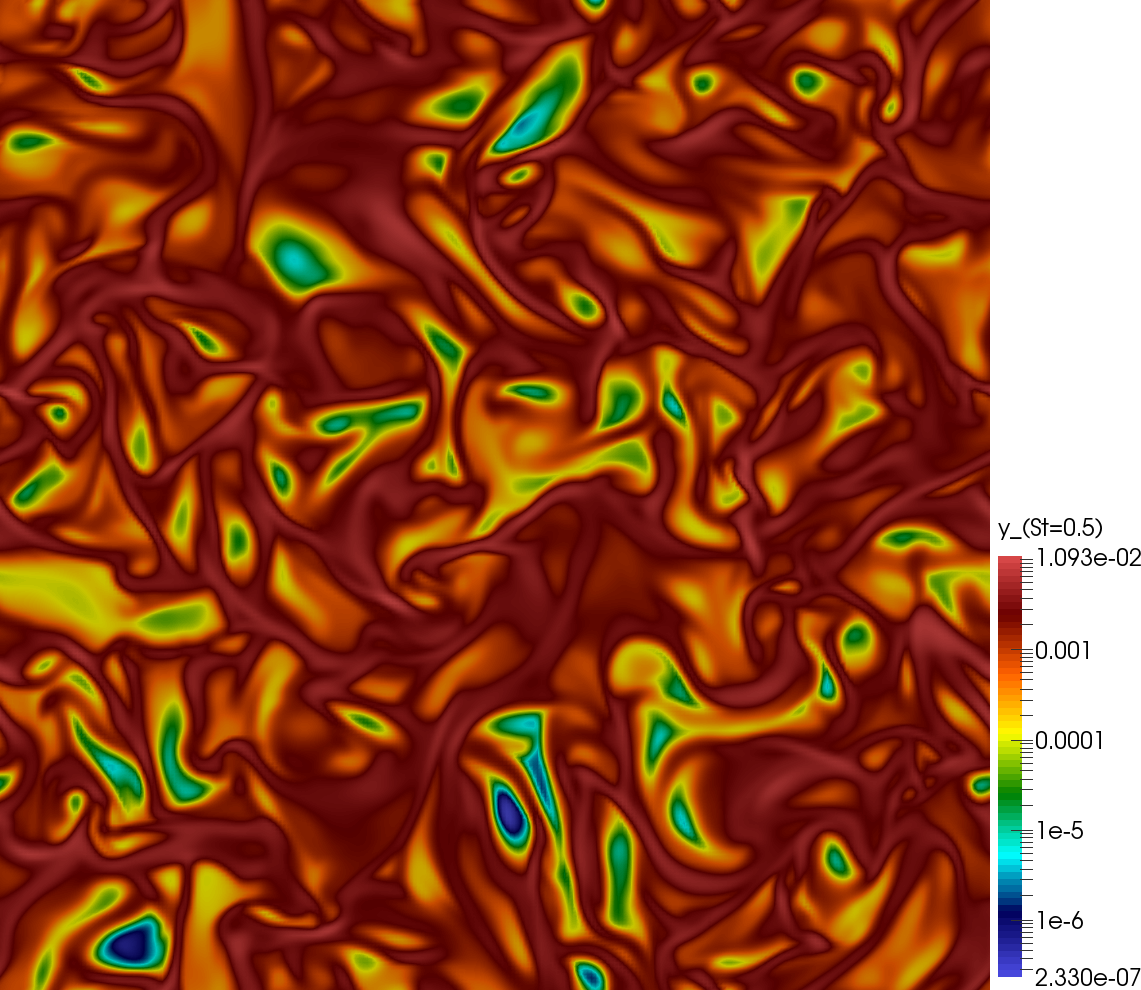}}\quad
\subfloat[][Acceleration]{\includegraphics[width=0.48\columnwidth]{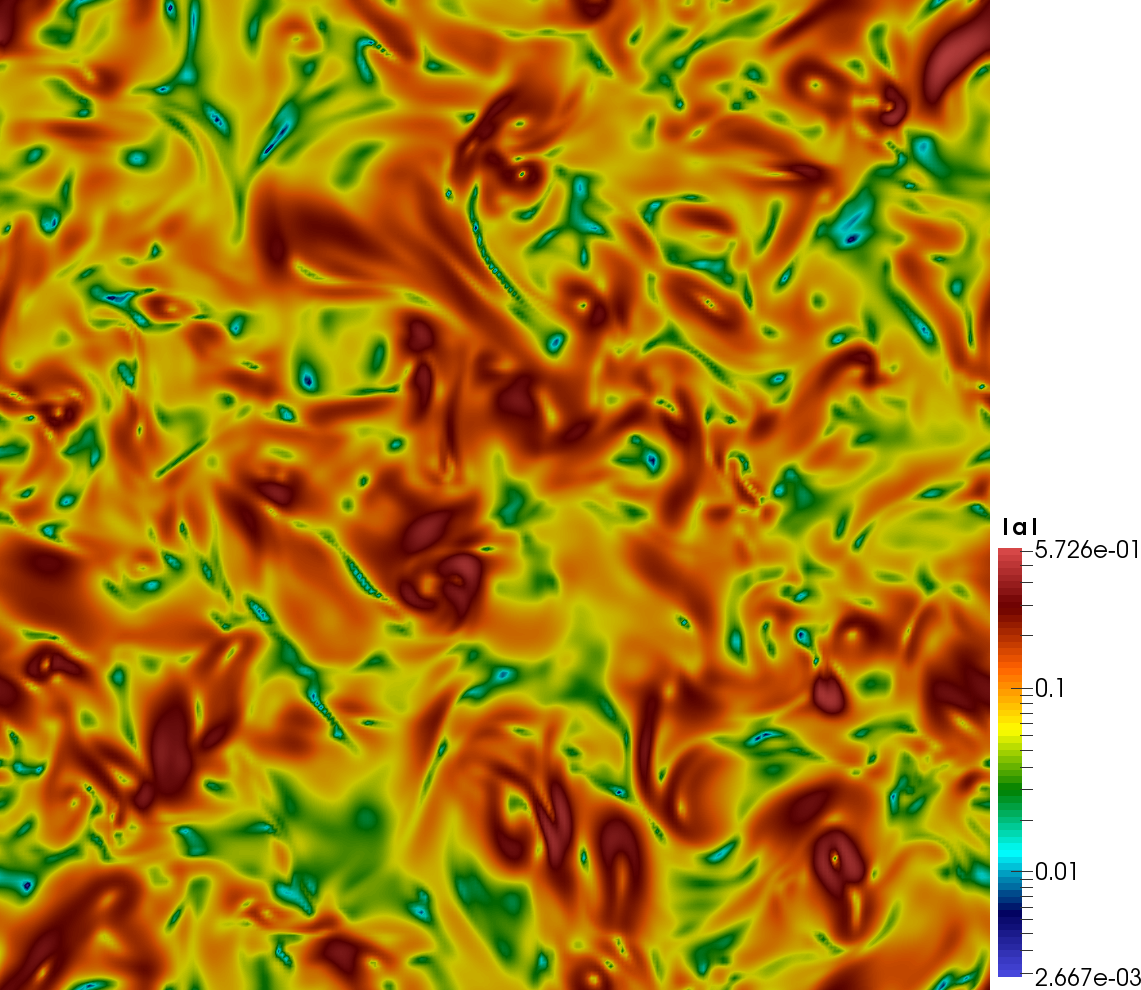}}
\caption{Slice of the turbulent box at $t/\tau_\textup{e} \simeq 2.2$. The two panels represent respectively a logaritmic color map of $y_3$ ($\St_\textup{max} = 0.5$) and of $|\vec{a}_g|$.}\label{fig:HIT_oneWay}
\end{figure}

In Fig.~\ref{fig:HIT_oneWay} we show a slice of the turbulent box at $t/\tau_\textup{e} \simeq 2.2$.
Panel a) displays the solid mass fraction, highlighting the preferential concentration and clustering of particles in response to the development of the acceleration field (panel b) associated with turbulent eddies.

As described in~\cite{maxey:1987,Rani2003}, a good measure for the degree of preferential concentration in incompressible flows is the weighted average on the particle mass fraction of the quantity $(|\mathcal{D}|^2 - |\mathcal{S}|^2)$, where $\mathcal{S}$ is the vorticity tensor, i.e. the skew symmetric part of the gas velocity gradient and $\mathcal{D}$ is its symmetrical part. For compressible flows, we choose to consider
\begin{multline}\label{eq:preferentialConcentration}
\langle \mathcal{P}\rangle_j \equiv \langle \left(|\mathcal{D}|^2 - |\mathcal{S}|^2 - |\Tr(\mathcal{D})|^2\right)\rangle_j \equiv \\
\equiv  \frac{\langle y_j \left(\mathcal{P} - \langle \mathcal{P}\rangle_\Omega\right)\rangle_\Omega}{\langle y_j \rangle_\Omega}\,.
\end{multline}
This is a good measure because (use integration by parts, Gauss theorem and Eq.~\eqref{eq:uEqEu} with $\vec{w}_j = 0$):
\begin{multline}
\langle \Div\uj \rangle_\Omega = - \tau_j \left\langle \sum_{l,m}\left( \partial_l u_m \partial_m u_l - \partial_l u_l \partial_m u_m\right)\right\rangle_\Omega = \\
 = - \tau_j \left\langle \left(|\mathcal{D}|^2 - |\mathcal{S}|^2 - |\Tr(\mathcal{D})|^2\right)\right\rangle_\Omega\,.
\end{multline}
Moreover, it is worth noting that $\langle \mathcal{P}\rangle_j$ vanishes in absence of preferential concentration.
By dimensional analysis, preferential concentration is expected to behave as:
\begin{equation}\label{eq:preferentialConcentrationDim}
\langle \mathcal{P}\rangle_j \propto
\begin{cases}
 \tau_j/\tauK^{3}\quad  \mbox{DNS}\\
  \tau_j/\tau_\xi^{3}\quad  \mbox{LES}\,,
 \end{cases}
\end{equation}
because it must be proportional to $\tau_j$ and have a dimension of [s$^{-2}$].
As described by [Pope 2001], the typical time-scale corresponding to an eddy length-scale $\xi$ in the inertial sub-range, can be evaluated by means of the Kolmogorov's theory as:

\begin{equation}
\tau_\xi = \tau_\lambda \left(\frac{\xi}{\lambda_\textup{T}}\right)^\frac{2}{3}\,,
\end{equation}
where the Taylor microscale $\lambda_\textup{T}$ is defined by Eq.~\ref{taylormic}.
Since the time based on the Taylor microscale is defined as
\begin{equation}
\tau_\lambda = \frac{\sqrt{3}\lambda_\textup{T}}{u_\textup{rms}}\,,
\end{equation}
we can evaluate the typical time at the smallest resolved LES scale $\xi$ knowing the kinetic energy $K(t)$ and $\lambda_\textup{T}(t)$:
\begin{equation}\label{eq:tau_xi}
\tau_\xi(t) = \sqrt{\frac{3}{2 K(t)}}\, \xi^\frac{2}{3} \lambda_\textup{T}(t)^\frac{1}{3}\,.
\end{equation}

\begin{figure}[t]
\centering
\includegraphics[width=\columnwidth]{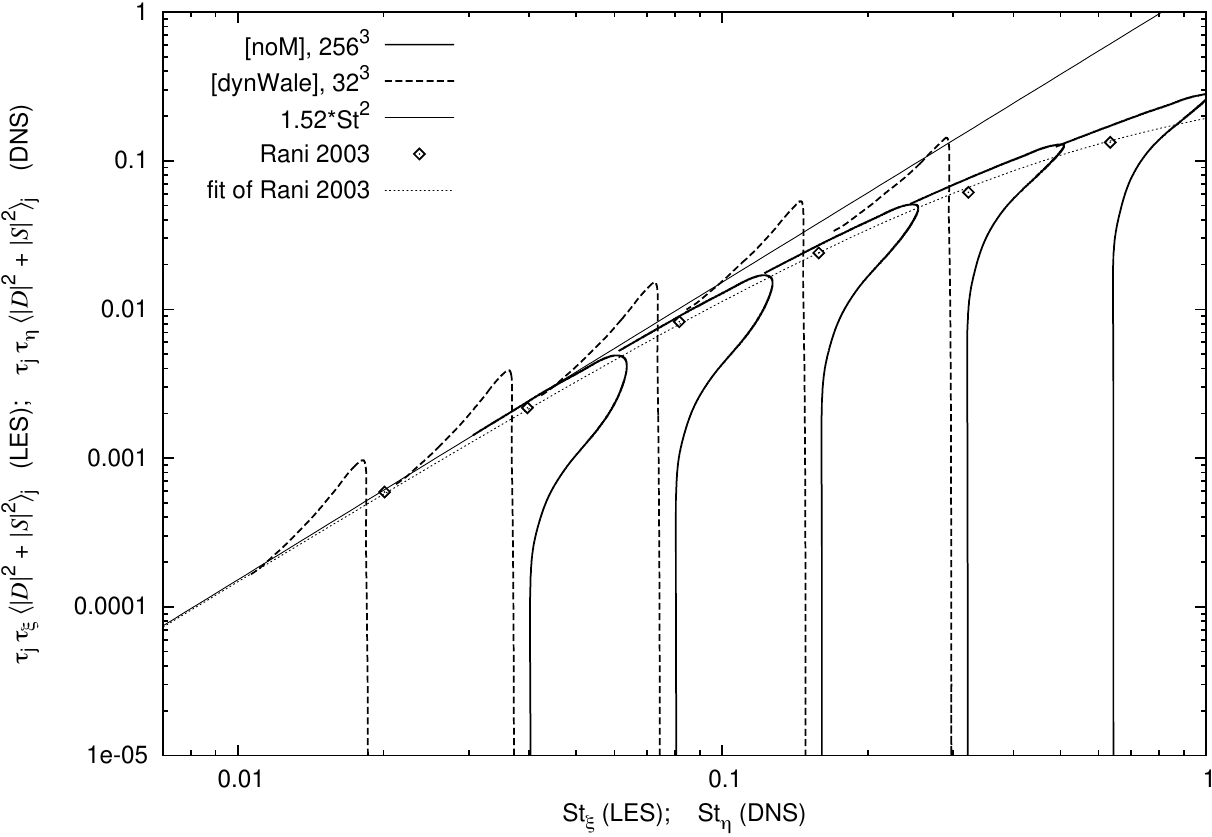}
\caption{Evolution of the degree of preferential concentration with $\St_\xi$ (LES) or $\St_\eta$ (DNS). We obtain a good agreement between equilibrium-Eulerian LES/DNS and Lagrangian DNS simulations. The fit for the data by~\cite{Rani2003} is found in Eq.~\eqref{eq:preferentialConcentrationFit}.}\label{fig:preferentialConcentration}
\end{figure}

In Fig.~\ref{fig:preferentialConcentration}
we show the time-evolution of the degree of preferential concentration as a function of the Stokes number for both DNS with $256^3$ cells and the LES with $32^3$ cells. There, we multiply $\langle \mathcal{P}\rangle_j$ by $\tau_\xi \tau_j$ in order to make it dimensionless and to plot on the same graph all the different particles at different times together. 

At $t=0$ the preferential concentration is zero for all Stokes number. Then, preferential concentration of each particle class increases up to a maximum value and then it decreases because of the decaying of the turbulent energy. 
The maximum degree of preferential concentration is reached by each particle class when $\tauK$ is minimum (at $t/\tau_\textup{e} \simeq 1.7$, cf. Fig.~\ref{fig:tauEta}). Then, $\langle \mathcal{P}\rangle_j$ decreases and merges with the curve relative to the next particle class at the final simulation time, when $\tauK$ is about twice its minimum.
Note that the expected behavior of Eq.~\eqref{eq:preferentialConcentrationDim} is reproduced for $\textup{St}_j < 0.2$ and in particular we find:
\begin{equation}
\langle \mathcal{P}\rangle_j \simeq
\begin{cases}
 1.52\, \St_j\, \tauK^{-2}\quad \mbox{DNS}\\
  1.52\, \St_j\, \tau_\xi^{-2}\quad \mbox{LES}\,.
 \end{cases}
\end{equation}
Moreover, by comparing our results with the Eulerian-Lagrangian simulation described in \cite{Rani2003}, we note that our limiter for the preferential concentration when $\textup{St} > 0.2$ is well behaving.

For the sake of completeness, we found that the best fit in the range $\St < 2.5$ for the data found by \citet{Rani2003} is:
\begin{equation}\label{eq:preferentialConcentrationFit}
\langle \mathcal{P}\rangle_j \simeq 1.52*\frac{\St_j}{1+3.1*\St_j+3.8*\St_j^2}\,\, \tauK^{-2}\,,
\end{equation}
with root mean square of residuals $8.5*10^{-3}$.

For what concerns the $32^3$ LES simulation, Fig.~\ref{fig:preferentialConcentration} shows that the Stokes number of each particle class in the LES case is much smaller than its DNS counterpart. Accordingly with \citet{Balachandar2010}, we have
\begin{equation}
\St_\xi = \St_\eta \left(\frac{\etaK}{\xi}\right)^\frac{2}{3}\,,
\end{equation}
confirming that the equilibrium-Eulerian model widens its applicability under the LES approximation. We also notice that the presented LES is able to reproduce the expected degree of preferential concentration with a satisfactory level of accuracy when $\St < 0.2$. In particular, the LES slightly overestimates preferential concentration and the time needed to reach the equilibrium and to ``forget'' the particle initial condition.
\subsection{Turbulent forced plume}
\label{ZHOU}
As a second benchmark, we discuss high-resolution, three-dimensional numerical simulation of a forced gas plume, produced by the injection of a gas flow from a circular inlet into a stable atmospheric environment at lower temperature (and higher density). Such an experiment allows to test the numerical model behavior against some of the fundamental processes controlling volcanic plumes, namely density variations, non-isotropic turbulence, mixing, air entrainment, and thermal exchange. Our study is mainly aimed at assessing the capability of the numerical model to describe the time-average behavior of a turbulent plume and to reproduce the magnitude of large-scale fluctuations and large-eddy structures.
We will mainly refer to laboratory experiments by \citet{George1977} and \citet{Shabbir1994} and numerical simulations by \citet{Zhou2001} for a quantitative assessment of model results.

\begin{figure*}[tp]
\centering
\subfloat[][Fluctuations at an axial point]{\includegraphics[width=0.48\textwidth]{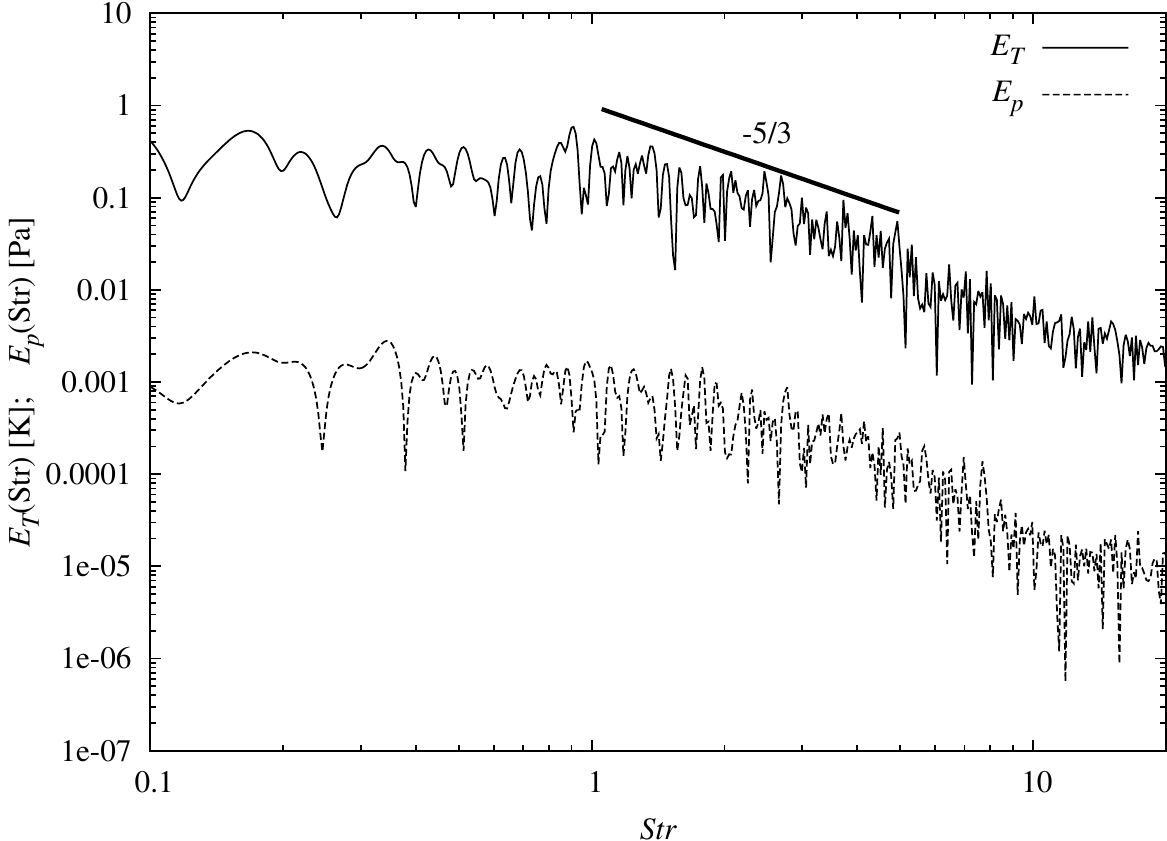}}\quad
\subfloat[][Fluctuations at an outer edge point]{\includegraphics[width=0.48\textwidth]{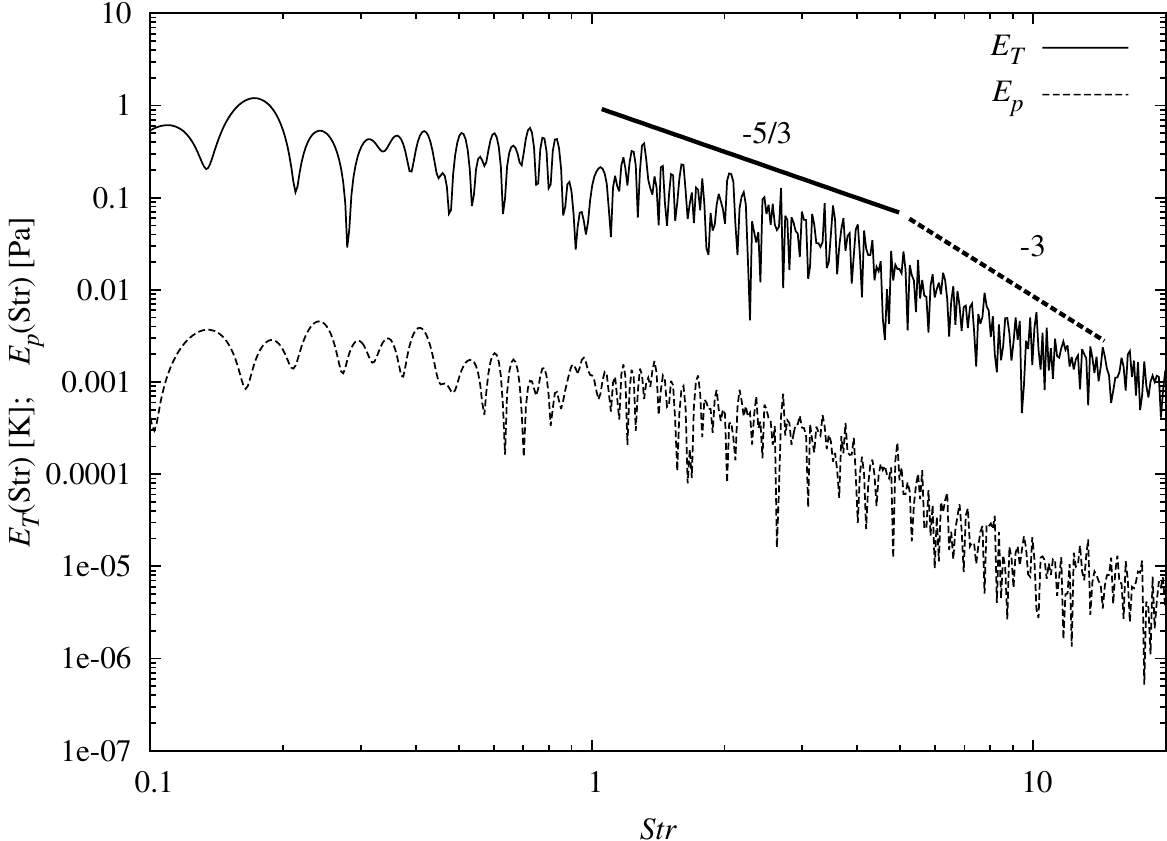}}
\caption{Temperature (solid) and pressure (dashed) fluctuations energy spectra: a) at a point along the plume axis \mbox{(0, 0, 0.5715) [m]}; b) at a point along the plume outer edge \mbox{(0, 0.06858, 0.5715) [m]}. The slopes $\textup{Str}^{-5/3}$ and $\textup{Str}^{-3}$ are represented with a thick solid and dashed line respectively.}\label{fig:spectrum}
\end{figure*}

Numerical simulations describe a vertical round forced plume with heated air as the injection fluid. The plume axis is aligned with the gravity vector and is subjected to a positive buoyancy force. The heat source diameter $2b_0$ is $6.35$ [cm], the exit vertical velocity on the axis $u_0$ is 0.98 [m/s], the inflow temperature $T_0$ is $568$ [K] and the ambient air temperature $T_a$ is 300 [K]. The corresponding Reynolds number is $1273$, based on the inflow mean velocity, viscosity and diameter. Air properties at inlet are $C_p = 1004.5$ [J/(K kg)]; $R=287$ [J/(K kg)]; $\mu=3\times10^{-5}$ [Pa s].
  
As discussed by \citet{Zhou2001} the development of the turbulent plume regime is quite sensitive to the inlet conditions: we therefore tested the model by adding a periodic perturbation and a non-homogeneous inlet profile to anticipate the symmetry breaking, and the transition from a laminar to a turbulent flow. 
The radial profile of vertical velocity has the form:
\begin{equation}
U_0(r) = \frac{1}{2}u_0 \left[ 1-\tanh\left( \frac{b_0}{4\delta_r} \left(\frac{r}{b_0}-\frac{b_0}{r}\right) \right) \right] 
\end{equation}
where $\delta_r$ is the thickness of the turbulent boundary layer at the plume inlet, that we have set at $\delta_r = 0.1 b_0$.
A periodical forcing and a random perturbation of intensity $0.05 U_0(r)$ has been superimposed to mimic a turbulent inlet.

The resulting average mass, momentum and buoyancy flux are $q_0=2.03\times 10^{-3}$ [kg s$^{-1}$], $m_0=1.62\times 10^{-3}$[kg m s$^{-2}$], $f_0=1.81\times 10^{-3}$ [kg s$^{-1}$].

The computational grid is composed of $360 \times 180 \times 180$ uniformly spaced cells (deformed near the bottom plane to conform to the circular inlet) in a box of size $12.8 \times 6.4 \times 6.4$ diameters. In particular, the inlet is discretized with 400 cells. The adaptive time step was set to keep the Courant number less than 0.2. Based on estimates by \citet{Plourde2008}, the selected mesh refinement is coarser than the required grid to fully resolve turbulent scales in a DNS (which would require about $720 \times 360 \times 360$ cells). Nonetheless, this mesh is resolved enough to avoid the use of a subgrid-scale model. This can be verified by analyzing the energy spectra of fluctuations on the plume axis and at the plume outer edges. In Fig.~\ref{fig:spectrum} we show the energy spectra of temperature and pressure as a function of the non-dimensional frequency: the Strouhal number \mbox{$\textup{Str} = f \ast \, 2b_0/u_0$} ($f$ is the frequency in [Hz]). We recover a result similar to \citet{Plourde2008}, where the inertial--convective regime with the decay $-5/3$ and the inertial--diffusive regime with the steeper decay $-3$ are observable \citep[][]{list1982}.

%
%

\begin{figure*}
\begin{center}
\includegraphics[width=\textwidth]{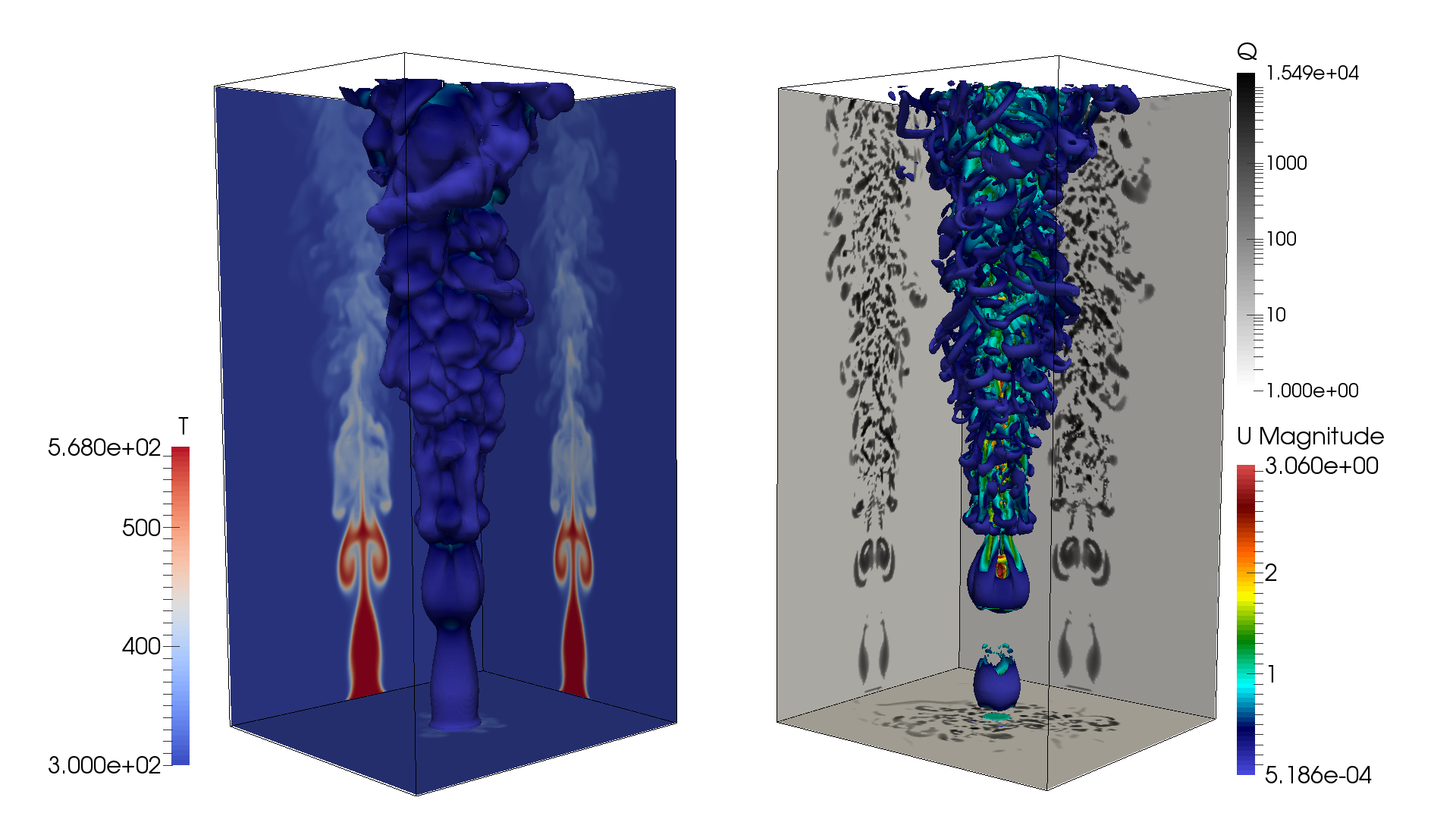}
\caption{Three-dimensional numerical simulation of a forced gas plume at $t=10 s$. a) Isosurface of temperature $T=305$ [K], colored with the magnitude of velocity, and the temperature distribution on two orthogonal slices passing across the inlet center. b) Isosurface of $Q_{\vec{u}}=100$ [s$^{-2}$] colored with the value of the velocity magnitude, and its distribution across two vertical slices passing through the inlet center}\label{fig:Zhou3D}
\end{center}
\end{figure*}

Model results describe the establishment of the turbulent plume through the development of fluid-dynamic instabilities near the vent (puffing is clearly recognized as a toroidal vortex in Fig. \ref{fig:Zhou3D}a). The breaking of large-eddies progressively leads to the onset of the developed turbulence regime, which is responsible of the mixing with the surrounding ambient air, radial spreading of the plume and decrease of the plume average temperature and velocity. Figure \ref{fig:Zhou3D}a displays the spatial distribution of gas temperature. Mixing becomes to be effective above a distance of about 4 diameters. Figure \ref{fig:Zhou3D}b displays the distribution of the vorticity, represented by values of the $Q_{\vec{u}}$ invariant (Eq.~\ref{eq:Q_u}). The figure clearly identifies the toroidal vortex associated to the first instability mode (puffing, dominant at such Reynolds numbers). We have observed the other instability modes \citep[helical and meandering,][]{lesieur2005} only by increasing the forcing intensity (not shown).

\begin{figure*}[t]
\centering
\includegraphics[width=\textwidth]{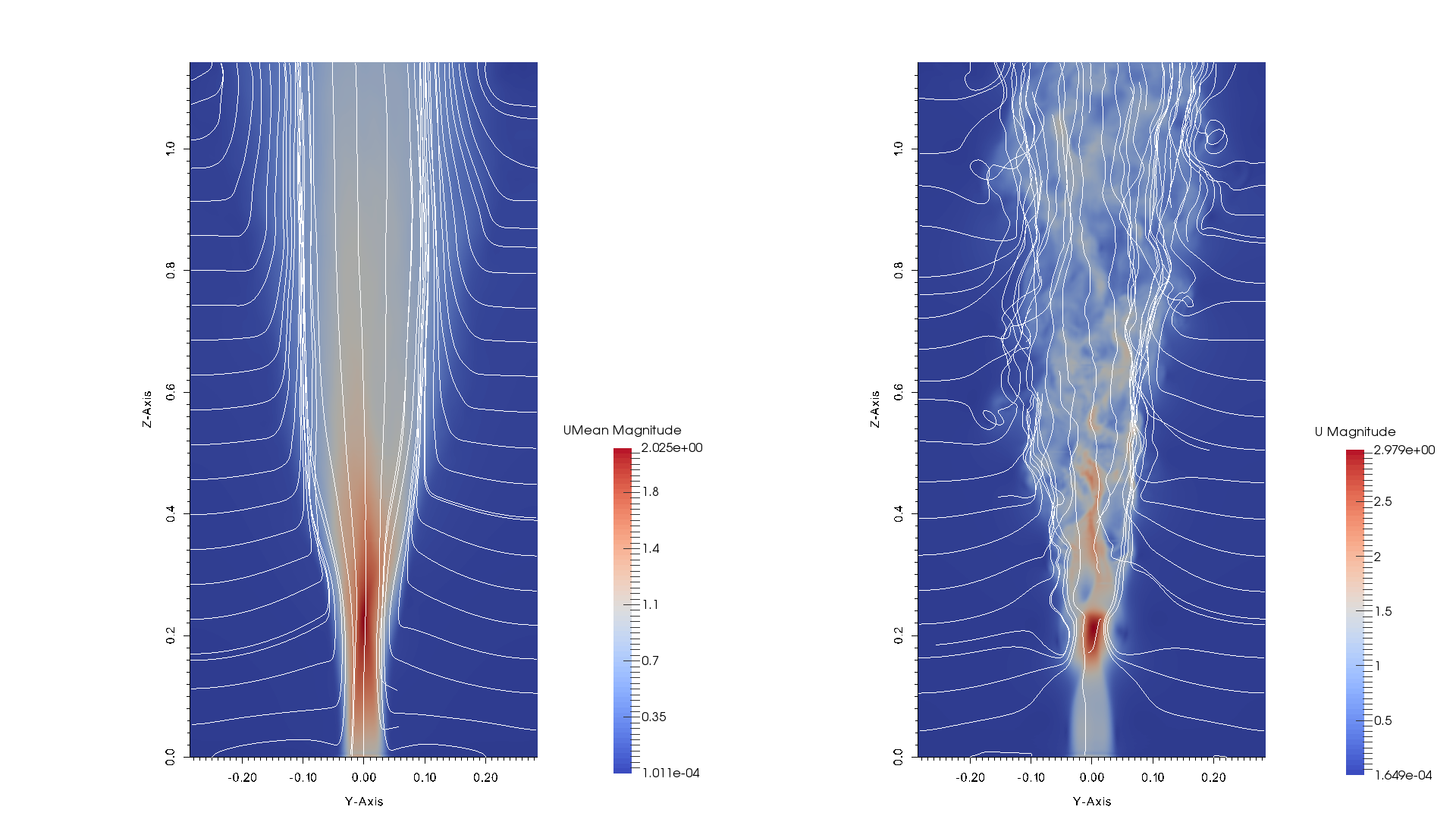}
\caption{Two-dimensional slice and streamlines of the velocity field: a) time-averaged velocity field; b) instantaneous velocity field at $t = 10$ s. The mean velocity field outside the plume is approximatively horizontal while in the plume it is approximately vertical. The region where the mean velocity field change direction is the region where the entrainment of air by the plume occurs.}
\label{fig:ZhouStream}
\end{figure*}

Experimental observations by \citet{George1977} and \citet{Shabbir1994} reveal that the behavior of forced plumes far enough from the inlet can be well described by integral one-dimensional plume models \citep{Morton1956,Morton1959} provided that an adequate empirical entrainment coefficient is used. In the buoyant plume regime at this Reynolds number \citet{George1977} obtained an entrainment coefficient of 0.153.

To compare numerical result with experimental observations and one-dimensional average plume models, we have time-averaged the numerical results between 4 and 10 s (when the turbulent regime was fully developed) and computed the vertical mass $Q(z)$, momentum $M(z)$ and buoyancy $F(z)$ fluxes as a function of the height. To perform this operation, we define the time averaging operation $(\bar{\cdot})$ and the radial domain
\begin{multline}\label{eq:omegaPlume}
\Omega(z) = \{(x,y)\in \mathbb{R}^2 \,|\,\\
| \, (\bar{y}_\textup{tracer}(\vec{x}) > 0.01* y_{\textup{tracer},0})\, \wedge \, (\bar{u}_z (\vec{x}) > 0) \}\,,
\end{multline}
where $(x,y,z) = \vec{x}$ are the spatial coordinates, $y_\textup{tracer}$ is the mass fraction field of a tracer injected from the vent with initial mass fraction $y_{\textup{tracer},0}$ and $u_z$ is the axial component of the velocity field. 
We use this definition for $\Omega(z)$ for coherence with integral plume models, where the mean velocity field is assumed to have the same direction of the plume axis \citep[cf. ][]{Morton1956, Woods1988a, cerminara2015b, cerminara2015phd}. 
This hypothesis is tested in Fig.~\ref{fig:ZhouStream}a, where it can be verified that the time-averaged streamlines inside the plume are parallel to the axis (Fig.~\ref{fig:ZhouStream}b shows the instantaneous streamlines and velocity magnitude field).

The plume fluxes are evaluated as follows \citep[cf.][]{George1977, Shabbir1994, kaminski2005}:
\begin{itemize}
\item mass flux $Q(z) = \displaystyle\int_\Omega \bar{\rho}\,\bar{u}_z \,\de x \de y$
\item momentum flux $M(z) = \displaystyle\int_\Omega  \bar{\rho}\,\bar{u}_z^2\,\de x \de y$
\item buoyancy flux $F(z) = \displaystyle\int_\Omega \bar{u}_z\,\left(\rho_\alpha - \bar{\rho}\right)\,\de x \de y$
\end{itemize}
where $\rho_\alpha = \rho_\alpha(z)$ is the atmospheric density.
From these quantities it is possible to retrieve the main plume parameters
\begin{itemize}
\item plume radius $b(z) = \sqrt{\frac{Q(F+Q)}{\pi \rho_\alpha M}}$
\item plume density $\beta(z) = \rho_\alpha \, \frac{Q}{(F+Q)}$
\item plume temperature $T_\beta(z) = T_\alpha\, \frac{F+Q}{Q}$
\item plume velocity $U(z) = \frac{M}{Q}$
\item entrainment coefficient $k(z) = \frac{Q'}{2\pi \rho_\alpha U b}$
\end{itemize}
where $(\cdot)'$ is the derivative along the plume axis and $T_\alpha$ is the atmospheric temperature profile.
\begin{figure*}
\begin{center}
\includegraphics[width=0.9\textwidth]{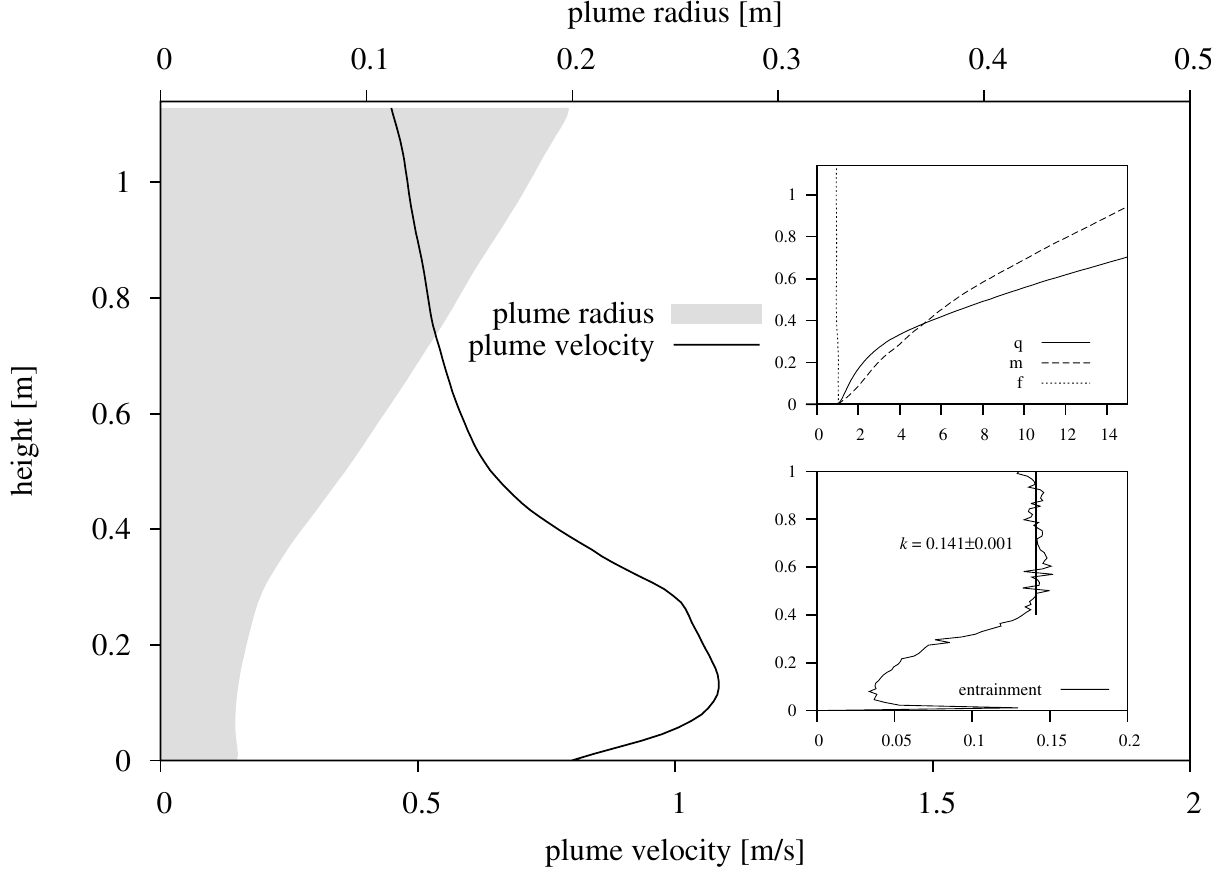}
\end{center}
\caption{Time-averaged plume radius and velocity. The insets display the non-dimensional mass, momentum and buoyancy fluxes (top) and the time-averaged entrainment coefficient. The line in the entrainment panel is a constant fit, from which results $k = 0.141\pm 0.001$.}
\label{fig:Zhou1D}
\end{figure*}
\begin{figure}
\begin{center}
\includegraphics[width=\columnwidth]{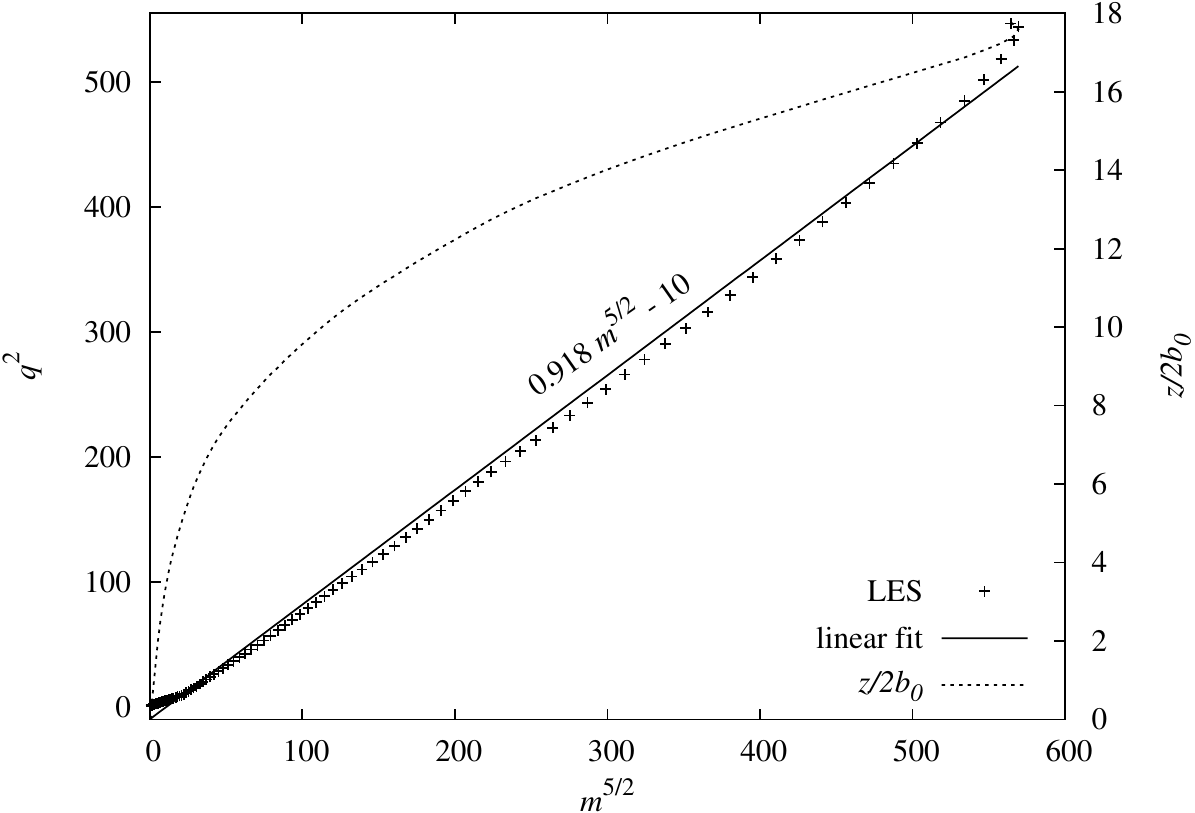}
\end{center}
\caption{Linear regression between $m^{5/2}$ and $q^{2}$ for the plume simulation with azimuthal forcing.}\label{fig:Zhou_integral}
\end{figure}
\begin{figure}
\centering
\includegraphics[width=\columnwidth]{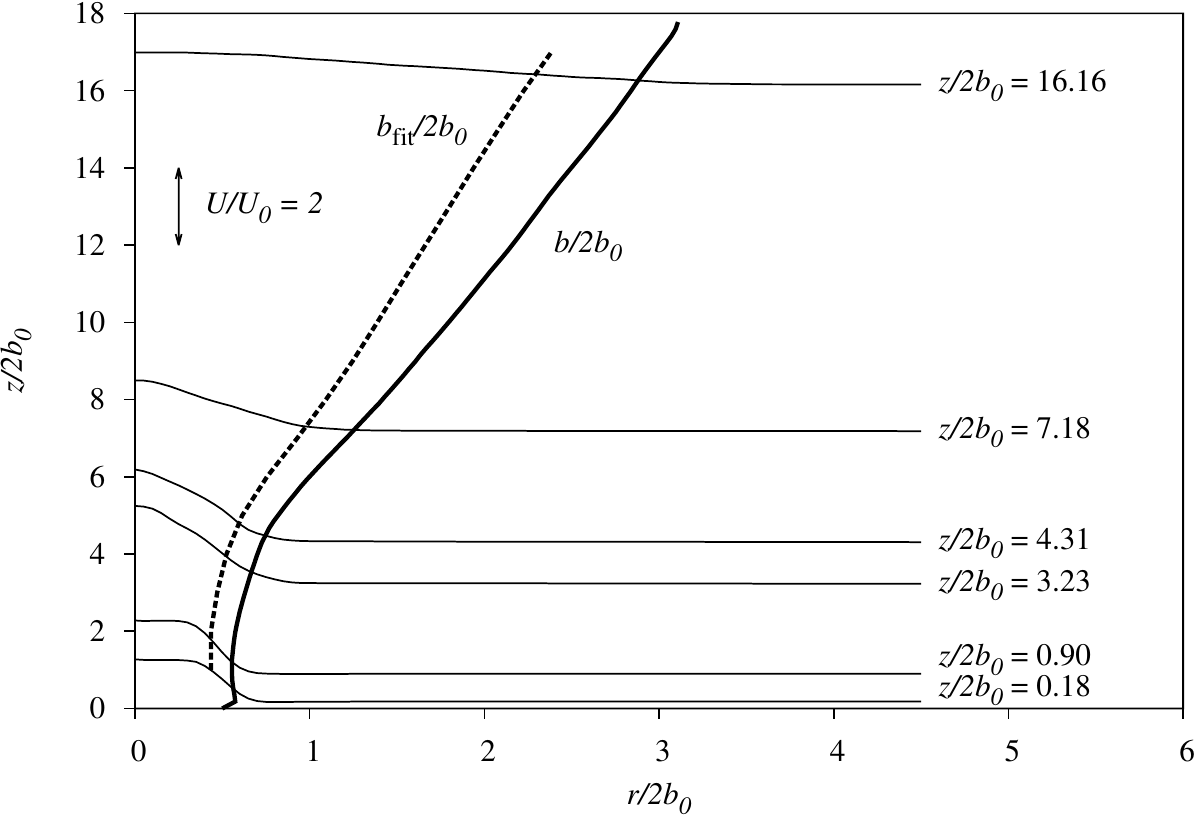}
\caption{Radial profiles of the time-averaged velocity field at various height. The scale for these profiles is indicated by the up-down arrow on the left in the panel. The thick solid line is the plume radius evaluated from the mass, momentum and buoyancy fluxes, while the thick dashed line is the plume radius evaluated from Gaussian fits of horizontal profiles.}\label{fig:profiles}
\end{figure}
\begin{figure*}[t]
\begin{center}
\includegraphics[width=\textwidth]{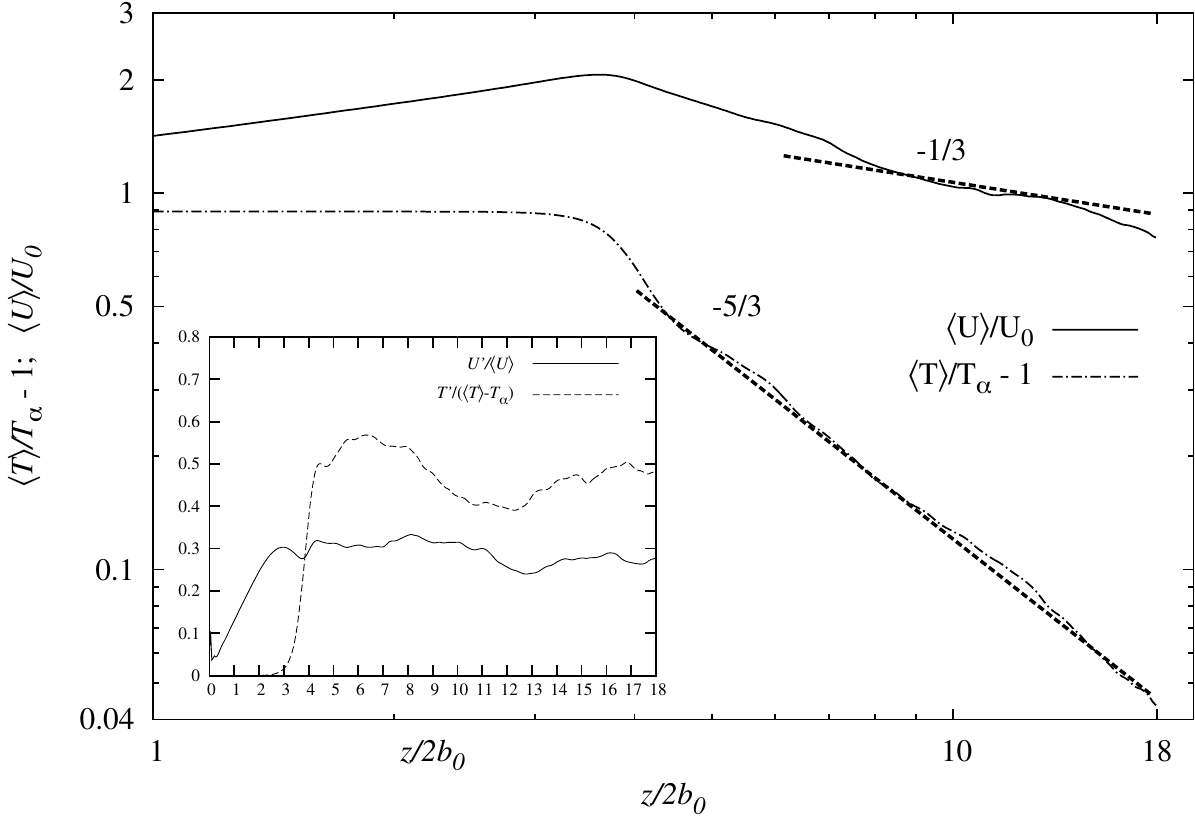}
\end{center}
\caption{Centerline time-average axial velocity, and temperature profiles with azimuthal forcing. Inset) centerline correlations of fluctuating velocity and temperature.}\label{fig:ZhouAxisMean}
\end{figure*}

Figure \ref{fig:Zhou1D} displays the average plume radius and velocity. As previously reported by \citet{Fannelop2003} and \citet{Plourde2008}, the plume radius initially shrinks due to the sudden increase of velocity due to buoyancy (at $z=0.1$ m). Above, turbulent mixing becomes to be effective and increases the plume radius while decreasing the average velocity.
The upper inset in Fig.~\ref{fig:Zhou1D} represents the values of the vertical mass $q=Q/Q_0$, momentum $m=M/M_0$ and buoyancy $f=F/F_0$, normalized with the inlet values. 
All variables have the expected trends and, in particular, the buoyancy flux is constant (as expected for weak ambient stratification) whereas $q$ and $m$ monotonically increase and attain the theoretical asymptotic trends shown also in Fig.~\ref{fig:Zhou_integral}. 
Indeed, \citet{Fannelop2003} have shown that an integral plume model for non-Boussinesq regimes (i.e., large density contrasts) in the approximation of weak ambient stratification and adopting the \citet{ricou1961} formulation for the entrainment coefficient, has a first integral such that $q^2$ is proportional to $m^{5/2}$ at all elevations. Figure \ref{fig:Zhou_integral} demonstrates that this relationship is well reproduced by our numerical simulations, as also observed in DNS by \citet{Plourde2008}. 

The lower inset in Fig.~\ref{fig:Zhou1D} shows the computed entrainment coefficient, which is very close to the value found in experiments \citep{George1977, Shabbir1994} and numerical simulations \citep{Zhou2001} of an analogous forced plume. We found a value around 0.14 in the buoyant plume region ($6.4<z/2b_0<16$). 

The analysis of radial profiles led to a similar conclusions: in Fig.~\ref{fig:profiles}, we show the evolution of the radial profiles for the mean vertical velocity field. In this figure, we also report the plume radius as evaluated from Gaussian fits of these profiles on horizontal slices:
\begin{equation}
\bar{u}_z(x,y) = U_\textup{fit}\,\exp\left(-\frac{x^2+y^2}{b_\textup{fit}^2}\right)\,.
\end{equation}
The slope of the function $b_\textup{fit}(z)$ has been evaluated in the region $6.4<z/2b_0<16$, to obtain $b_\textup{fit}/z = 0.142\pm0.001$ to be compared with the result of \citet{George1977}: $b_\textup{fit}/z = 0.135\pm0.010$.

Finally, figure \ref{fig:ZhouAxisMean} reports the time-average values of the vertical velocity and temperature along the plume axis. As observed in laboratory experiments, velocity is slightly increasing and temperature is almost constant up to above 4 inlet diameters, before the full development of the turbulence. When the turbulent regime is established, the decay of the velocity and temperature follows the trends predicted by the one-dimensional theory and observed in experiments. The insets displays the average value of the vertical velocity and temperature fluctuations along the axis. Coherently with experimental results \citep{George1977}, velocity fluctuations reach their maximum value and a stationary trend (corresponding to about the 30\% of the mean value) at a lower height (about 3 inlet diameters) with respect to temperature fluctuations (which reach a stationary value about the 40\% above 4 inlet diameters).

%
%
%

\subsection{Transonic and supersonic flows}
\label{SOD}
Although not essential in the present application, the ability of solving transonic and supersonic regimes is also required for the full-scale simulation of volcanic processes.	 
We here test the behavior of the ASHEE code in presence of shocks in the classical Sod's shock tube test case \citep{Sod1978} describing the expansion of a compressible, single-phase gas having adiabatic index $\gamma = 1.4$. 
At $t=0$ the domain of length 10 m is subdivided in two symmetric subsets. In the first subset (spatial coordinate $x < 0$) we set $u =0$, $p = 10^5$ Pa, $T = 348.432$ K, so that $\rho = 1$. In the second subset ($x > 0$), we set $u =0$, $p = 10^4$ Pa, $T = 278.746$ K, so that $\rho = 0.125$ kg/m$^3$. We indicate with $c = 374.348$ m/s the speed of sound of the gas in the $x < 0$ part of the domain.
We impose zero gradient boundary conditions ($\partial_x (\cdot) = 0$) for all the variables $u$, $p$, $T$.
As described in \citet{Sod1978}, a reference analytic solution exists for this problem.

\begin{figure}
\centering
\includegraphics[width=\columnwidth]{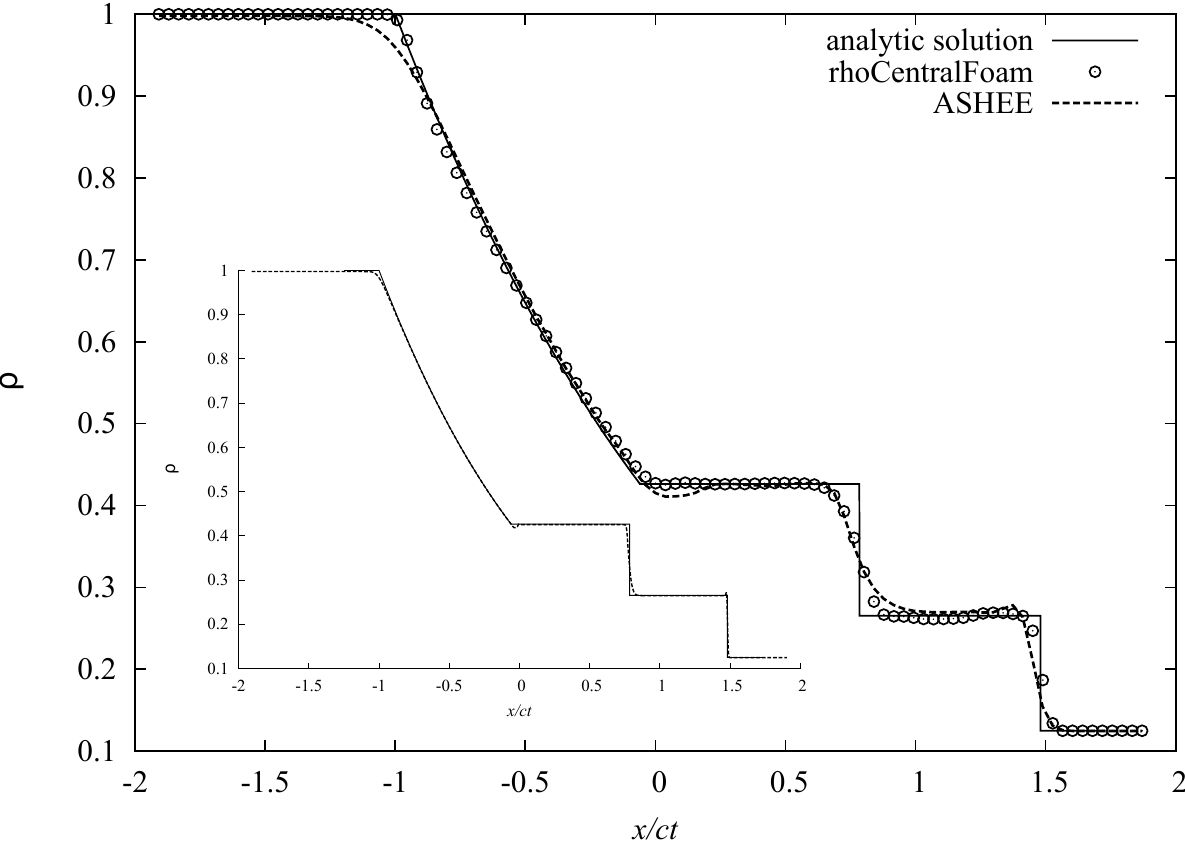}
\caption{The Sod's shock tube density after 0.007 s (here $c = 374.348$ m/s). Here we compare the analytic solution (solid) with two simulations performed with ASHEE model (dashed line) and the OpenFOAM {\tt rhoCentralFoam} solver (circles). The resolution is 100 cells, while in the inset it is reported the solution obtained with the ASHEE model with resolution of 1000 cells.}
\label{fig:shockTube}
\end{figure}
In Fig.~\ref{fig:shockTube} we show the density profile obtained with the ASHEE model after 0.007 s of simulation. We performed two simulations at different resolution. The first has 100 cells and it is compared with the OpenFOAM solver {\tt rhoCentralFoam} with a second order semi-discrete, non staggered central scheme of \citet{Kurganov2001} for the fluxes, and a total variation diminishing limiter \citep{VanLeer1997} for the interpolation. We refer to \citet{Greenshields2009} for a presentation of {\tt rhoCentralFoam} and of the Sod's shock tube test case.
The inset of Fig.~\ref{fig:shockTube} is the simulation with an higher resolution (1000 cells). In this figure, we notice that the code performs satisfactorily both at low and high resolution. It is capable to capture the shocks pretty well, with a diffusion that is comparable with that obtained with {\tt rhoCentralFoam}, a solver conceived for simulating shocks.

\section{3D simulation of a turbulent volcanic plume}
\label{results}
Numerical simulations of volcanic plumes were conducted in the framework of the IAVCEI (International Association of Volcanology and Geochemistry of the Earth Interior) plume model intercomparison initiative \citep{costa_etal_2015}, consisting in performing a set of simulations using a standard set of input parameters so that independent results could be meaningfully compared and evaluated, discuss different approaches, and identify crucial issues of state of the art models. 
We here discuss three-dimensional numerical simulation of a weak volcanic plume in a stratified, calm atmosphere, whose input data were set assuming parameters and meteorological conditions similar to those of the 26 January 2011 Shinmoe-dake eruption \citep[][]{suzuki2013}. Initial conditions and injection parameters are reported in Table \ref{tab:weakplume}.
\begin{table}[h]
\centering
\begin{tabular}{ll} 
\toprule
Parameter &	Value \\ \midrule
Vent elevation	& $1500$ m \\ 
Vent diameter  &  $54$ m\\
Mass eruption rate &	$1.5\times 10^6$ kg/s \\
Exit velocity  &	$135$ m/s \\ 
Exit temperature  &	$1273$ K \\
Exit water fraction &	$3$ wt.\% \\
Mixture density at vent & $4.85$ kg/m$^3$ \\ \bottomrule
\end{tabular}
\caption{Vent conditions for the weak volcanic plume simulation.}
\label{tab:weakplume}
\end{table}

\begin{figure}[h]
\centering
\includegraphics[width=\columnwidth]{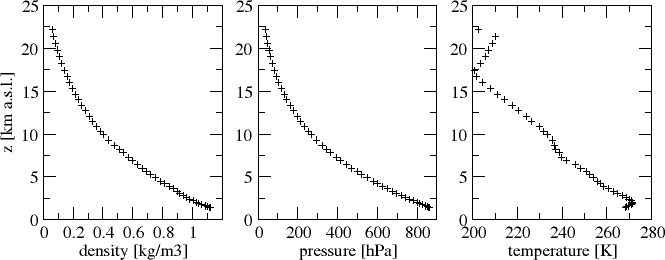}
\caption{Atmospheric profiles as provided by the Japan Meteorological Agency's Non-Hydrostatic Model \citep{hashimoto2012tephra} for Shinmoe-dake volcano at 00 JST of 27 January 2011.}
\label{fig:atmWP}
\end{figure}

The particle size distribution is composed of two individual classes of pyroclasts in equal weight proportion representing, respectively, fine (diameter $d=0.0625$ mm; density $\rho=2700$ kg/m$^3$, volume fraction $\epsilon=0.00086821$) and coarse ash (diameter $d=1.0000$ mm; density $\rho=2200$ kg/m$^3$, volume fraction $\epsilon=0.00106553$).
With respect to the laboratory benchmark case of Section \ref{ZHOU}, volcanic plumes are characterized by non-Boussinesq regimes at the vent and buoyancy reversal (with the initial mixture density about 4 times larger than the atmospheric one) and by a stratified atmosphere (Fig. \ref{fig:atmWP}). However, the most relevant difference is due to the significant temperature contrast (900 K) and to the presence of a high particle content which may strongly affect the mixing properties of the plume.

The Stokes number of the solid particles is, in general, a complex function of time and space, since the turbulent flow is characterized by a wide spectrum of relevant time and length scales.
Generally, the Stokes number is associated with the most energetic turbulent eddy scale which, for laboratory plumes, has a typical turnover time of the order of $\tau_L \sim \textup{Str} \frac{D_v}{U_v} \approx 0.12$ s, where $D_v$ and $U_v$ are the plume diameter and velocity at the vent, respectively, and $\textup{Str}$ is the Strouhal number, of the order $\textup{Str} = 0.3$ \citep{Zhou2001}.
Based on this time scale, and computing the particle relaxation time from Eq. \ref{eq:taus}, the Stokes number for the two adopted particle classes is about $\St_\textup{coarse}\approx 5$ and $\St_\textup{fine}\approx 0.2$, so we expect to see non-equilibrium phenomena for both particles classes, with more evident effects on the coarsest phase. However, the Stokes number, as an average value in all the plume is not as high as calculated above. Indeed, by using Eq.~\eqref{eq:tau_xi} as reference time for the turbulent dynamics, we obtain $\St_\textup{coarse} \approx 0.6$ and $\St_\textup{fine} \approx 0.03$.
It is worth recalling here that the equilibrium-Eulerian approach is accurate and advantageous for particles having $\St\leq 0.2$ and that, in our model, we numerically limit the acceleration field in order to keep the turbulent non-equilibrium within this limit, as explained in Sect. \ref{numerics} and tested in Sect.~\ref{MHIT} Fig.~\ref{fig:preferentialConcentration}. The averaged value of this limit -- measuring the importance of the decoupling limiter for this simulation -- is approximately $0.6$.

\begin{figure*}[h]
\centering
\subfloat[MeshH][]{\includegraphics[width=0.9\columnwidth]{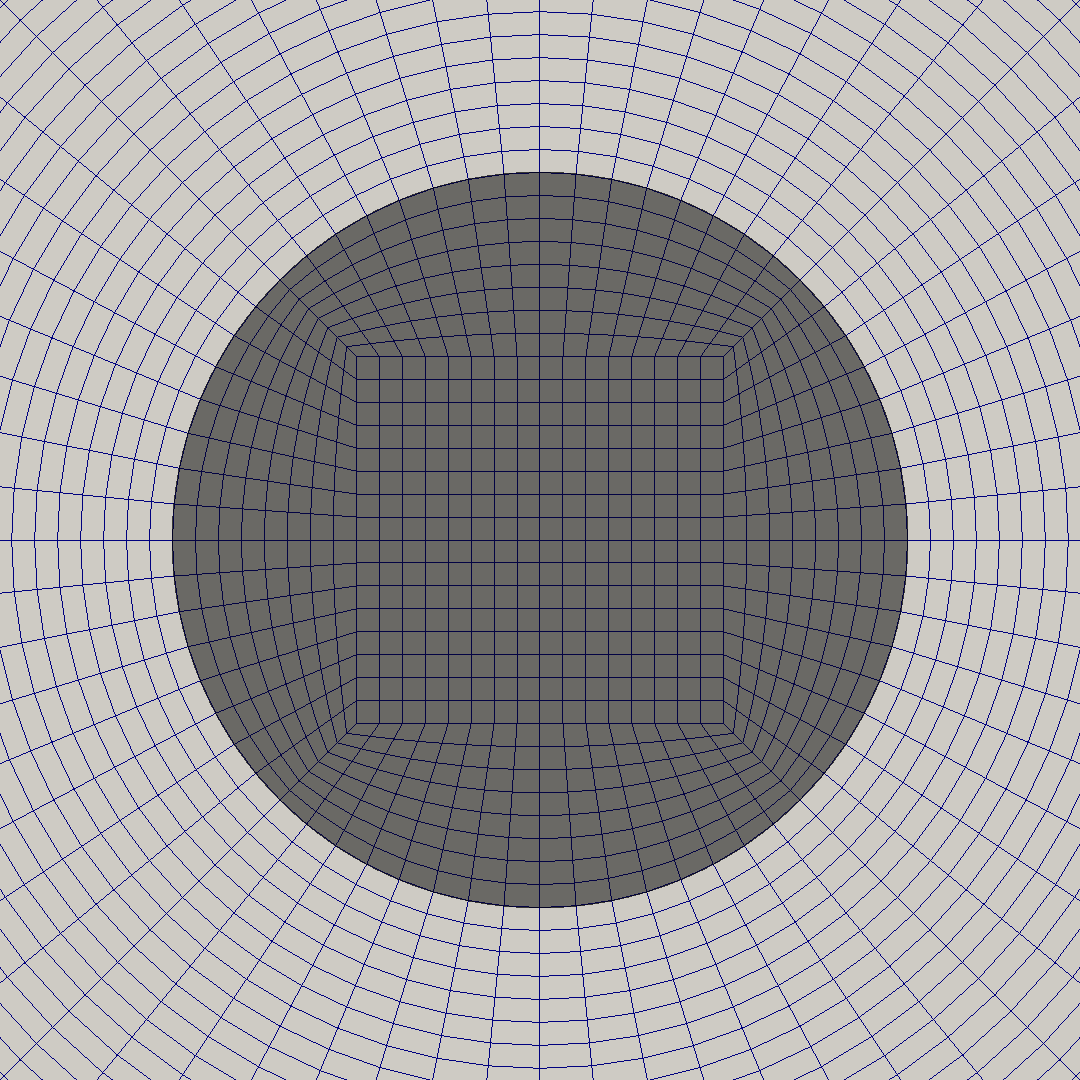}\label{MeshH}}\quad
\subfloat[MeshV][]{\includegraphics[width=0.9\columnwidth]{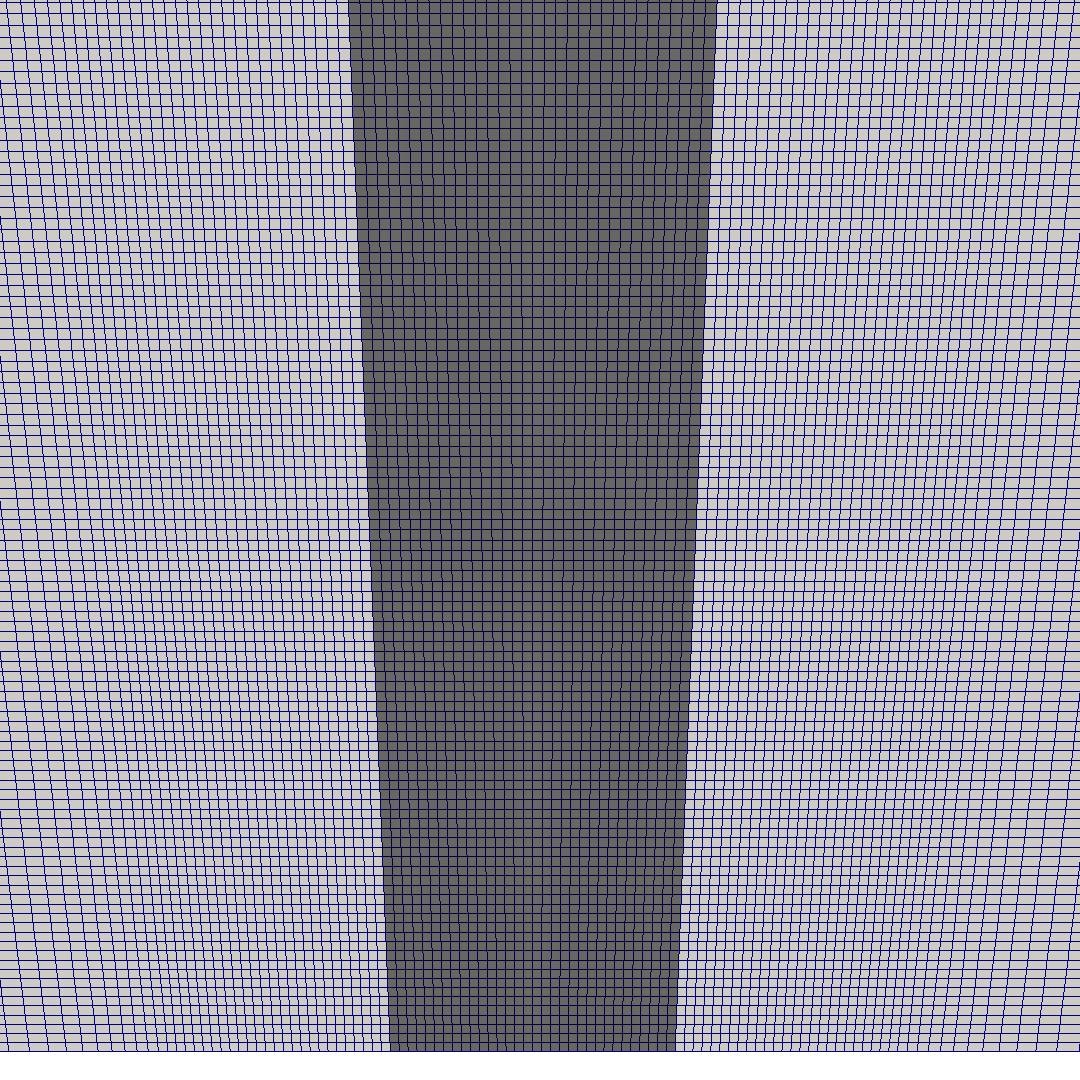}\label{MeshV}}\\
\caption{Zoom of the computational grid used for volcanic plume simulations.}
\label{fig:grid}
\end{figure*}

The computational domain is cylindrical and is extended $483 b_0 \times 765 b_0$ in the radial and vertical directions ($b_0$ being the vent radius). The  numerical grid is non-uniform and non-orthogonal. The discretization of the vent is represented in Fig.~\eqref{MeshH}. For the highest resolution run, the cell size increases from a minimum grid size $\Delta r = 2b_0/32$ with no radial grading factor in the region where the plume is expected to develop (Fig.~\ref{MeshV}), whose initial radius is equal to $2.5 b_0$ and increases linearly with an angle $\theta$ such that $\tan \theta = 0.147$, slightly larger than $\tan \theta = 0.12$ predicted by the Morton's plume theory with entrainment $k=0.1$ \citep{ishimine2006sensitivity}. Outside this region, a radial grading factor of 1.0446 is applied. 
Along $z$, 2048 cells are utilized. The minimum vertical cell size is $\Delta z = 2b_0/32$, and a grading factor of 1.00187 is imposed.
The azimuthal resolution is constant and equal to $\frac{1}{32} \pi$ (5.625 degrees). The resulting total number of cells is $10,747,904$.
This numerical mesh guarantees accuracy of the results: the solution procedure utilizes 2 PISO and 2 PIMPLE loops to achieve an absolute residual $\epsilon_\textup{PIMPLE}=10^{-7}$ (see Sec.~\ref{numerics}).

Simulation of 720 s of eruption required about 490,000 time steps (imposing a CFL constrain of 0.2, resulting in an average time-step $dt\approx 1.5$ ms, with a maximum velocity at the vent of about 150 m/s) for a total run-time of about 25 days on 1024 cores on the Fermi architecture at CINECA (meaning about 2.25 millions of cells per second, consistently with estimates of Sec.~\ref{validation}).

\begin{figure*}
\centering
\includegraphics[width=0.9\textwidth]{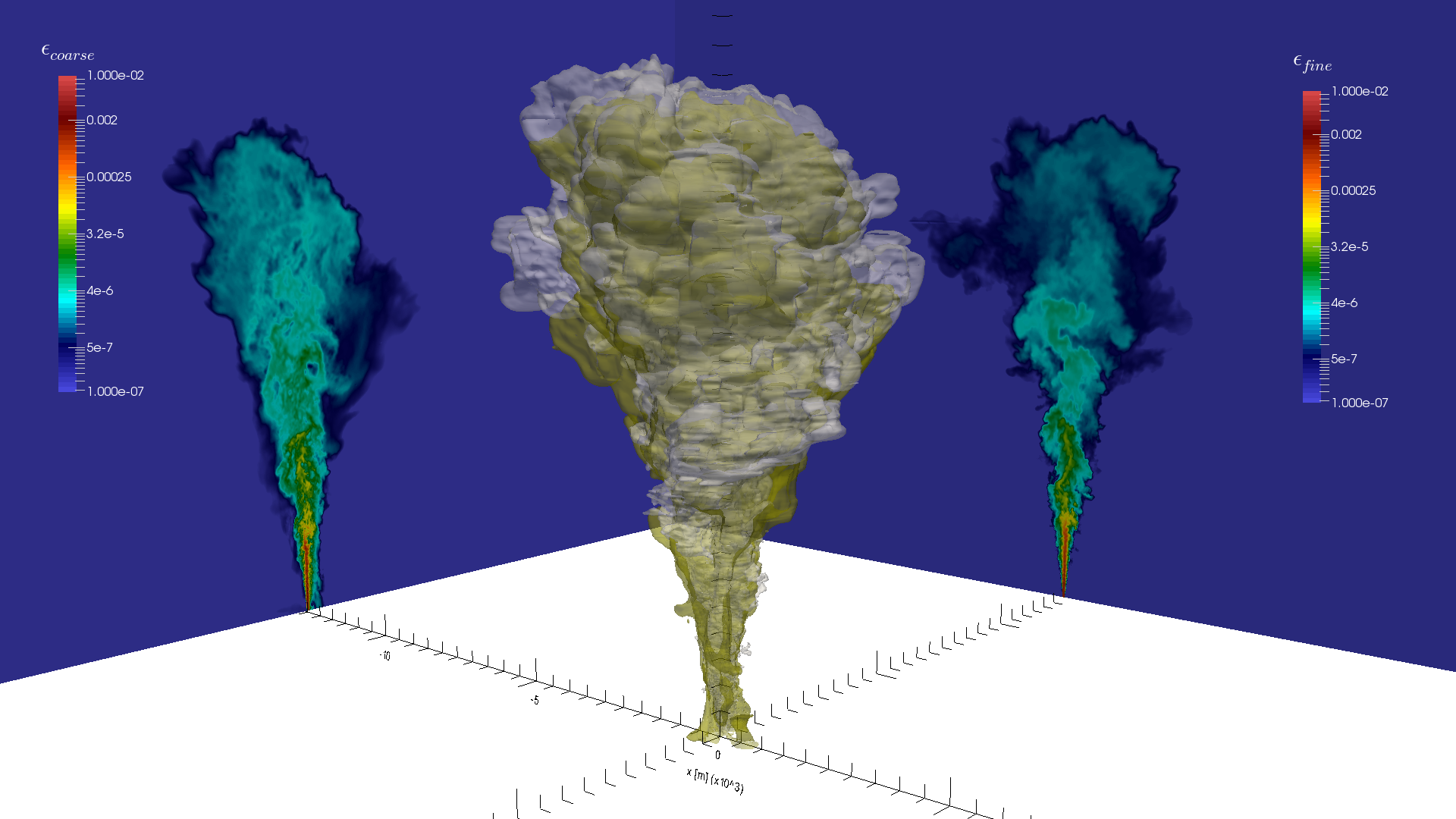}
\caption{Three-dimensional numerical simulation of a weak volcanic plume, 400 s after the beginning of the injection (inlet conditions as in Table \ref{tab:weakplume}). Isosurface and vertical sections of the fine (light white) and coarse (light sand) ash volume fractions. The two-dimensional plots represent the distribution of the volume concentration of coarse (left) and fine (right) particles across vertical orthogonal slices crossing the plume axis.}
\label{fig:WP3D}
\end{figure*}

\begin{figure*}
\centering
\includegraphics[width=0.9\textwidth]{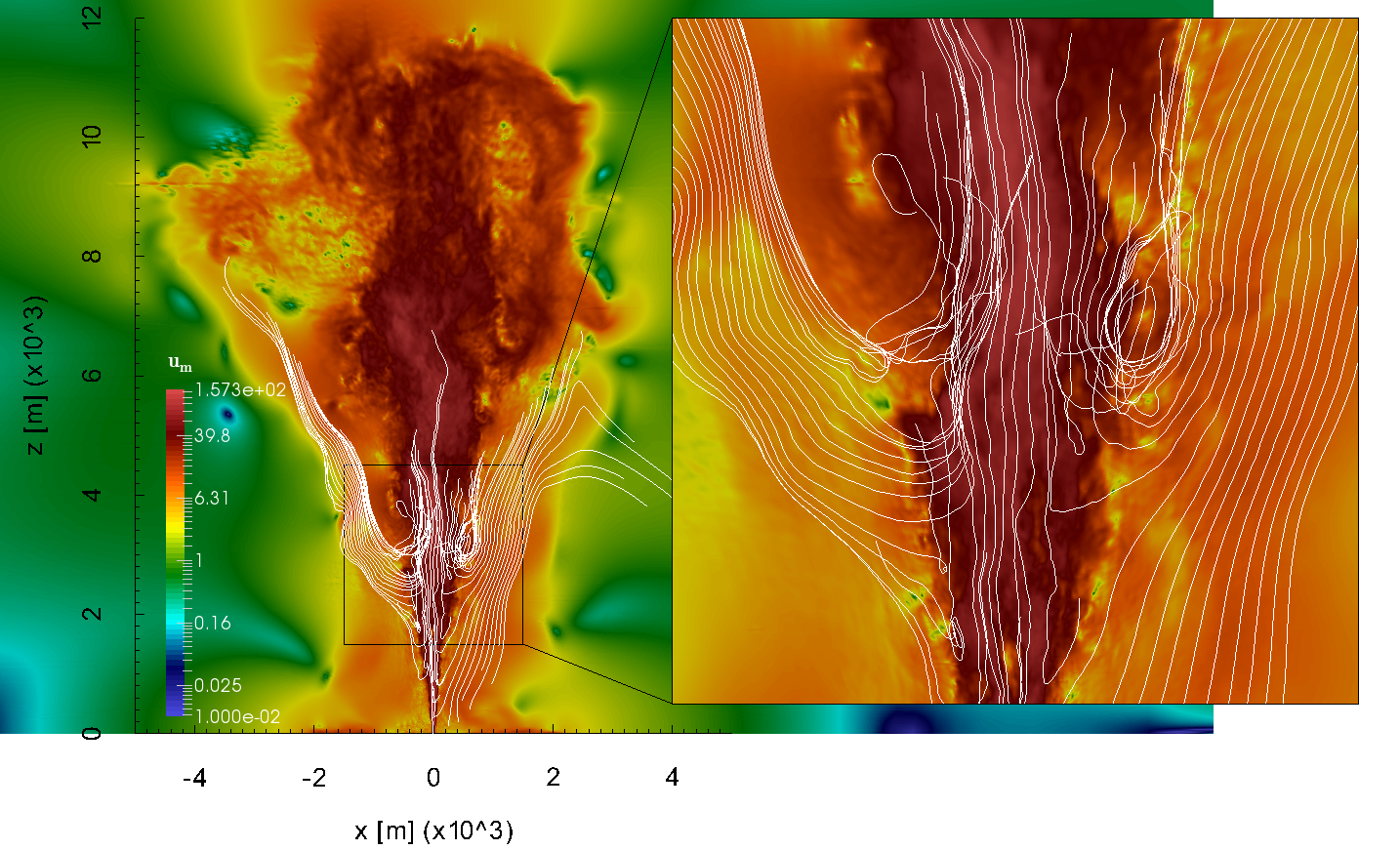}
\caption{Vertical section of the instantaneous value of the mixture velocity modulus (in logarithmic scale) at $t=400$ s and velocity streamlines.}
\label{fig:Ustream}
\end{figure*}

Figure \ref{fig:WP3D} shows the development of the volcanic plume at $t=400$ s. Because of the atmospheric stratification, the plume reaches a neutral buoyancy condition at about 10 km above the vent (i.e., 11.5 km above the sea level, still within the troposphere). Due to its inertia, the plume reaches its maximum plume height $H_{max}\approx 12$ km, higher than the neutral buoyancy level, before spreading radially to form the so-called volcanic umbrella.
%
The two orthogonal sections highlight the different spatial distribution of the volumetric fraction of fine (right) and coarse (left) ash particles, due to the different coupling regime with the gas phase. Coarse particles has indeed a larger settling velocity $\vec{w}_s=\tau_s \vec{g}$ which causes a more intense proximal fallout from the plume margins and a reduced transport by the umbrella.
This is highlighted by the plot of the streamlines of the mixture velocity along a vertical section (Fig.~\ref{fig:Ustream}), showing that the plume updraft is surrounded by a shell of settling coarse particles, which also inhibit air entrainment while promoting particle re-entrainment into the plume.

Besides settling, the large inertia of the coarse ash is responsible for the kinematic decoupling, leading to preferential concentration and clustering of particles at the margins of turbulent eddies.
To illustrate this phenomenon, in a non-homogeneous flow, the instantaneous preferential concentration is computed as the (normalized) ratio between the jth particle concentration and the concentration of a tracer (in our case, water vapor), 
i.e., 
\begin{equation}
\mathcal{C}_j= \frac{y_j}{y_{j,0}} \cdot \frac{y_\textup{tracer,0}}{y_\textup{tracer}}\,,
\label{eq:prefconc}
\end{equation} 
where the $0$ subscript corresponds to the value at the vent.

\begin{figure*}
\centering
\includegraphics[width=0.9\textwidth]{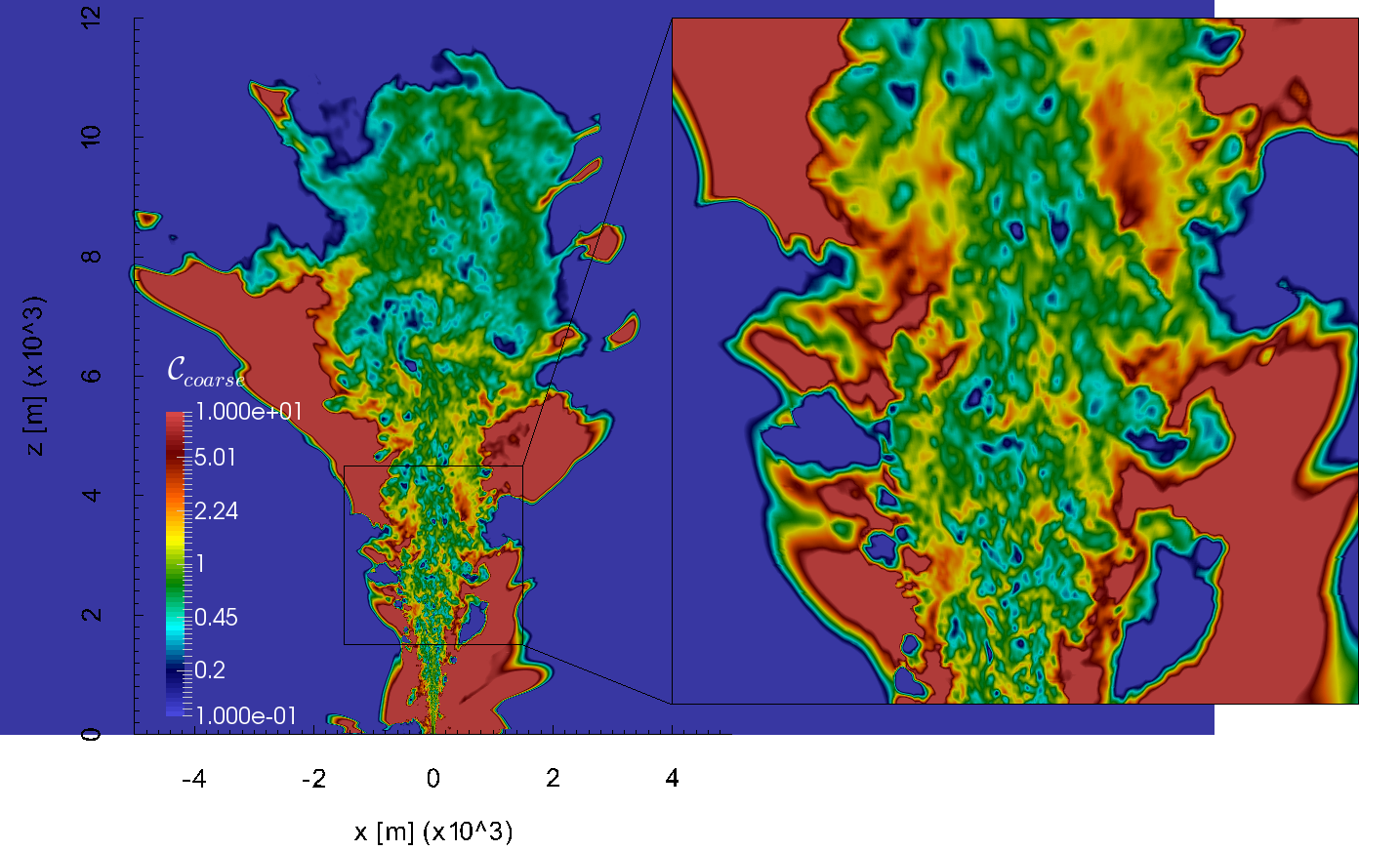}
\caption{Distribution of $\mathcal {C}_{\textup{coarse}}$ (Eq.~\ref{eq:prefconc}) 
for the coarsest particles across a vertical section at $t=400$ s.}
\label{fig:WP_preferential_coarse}
\end{figure*}

\begin{figure*}
\centering
\includegraphics[width=0.9\textwidth]{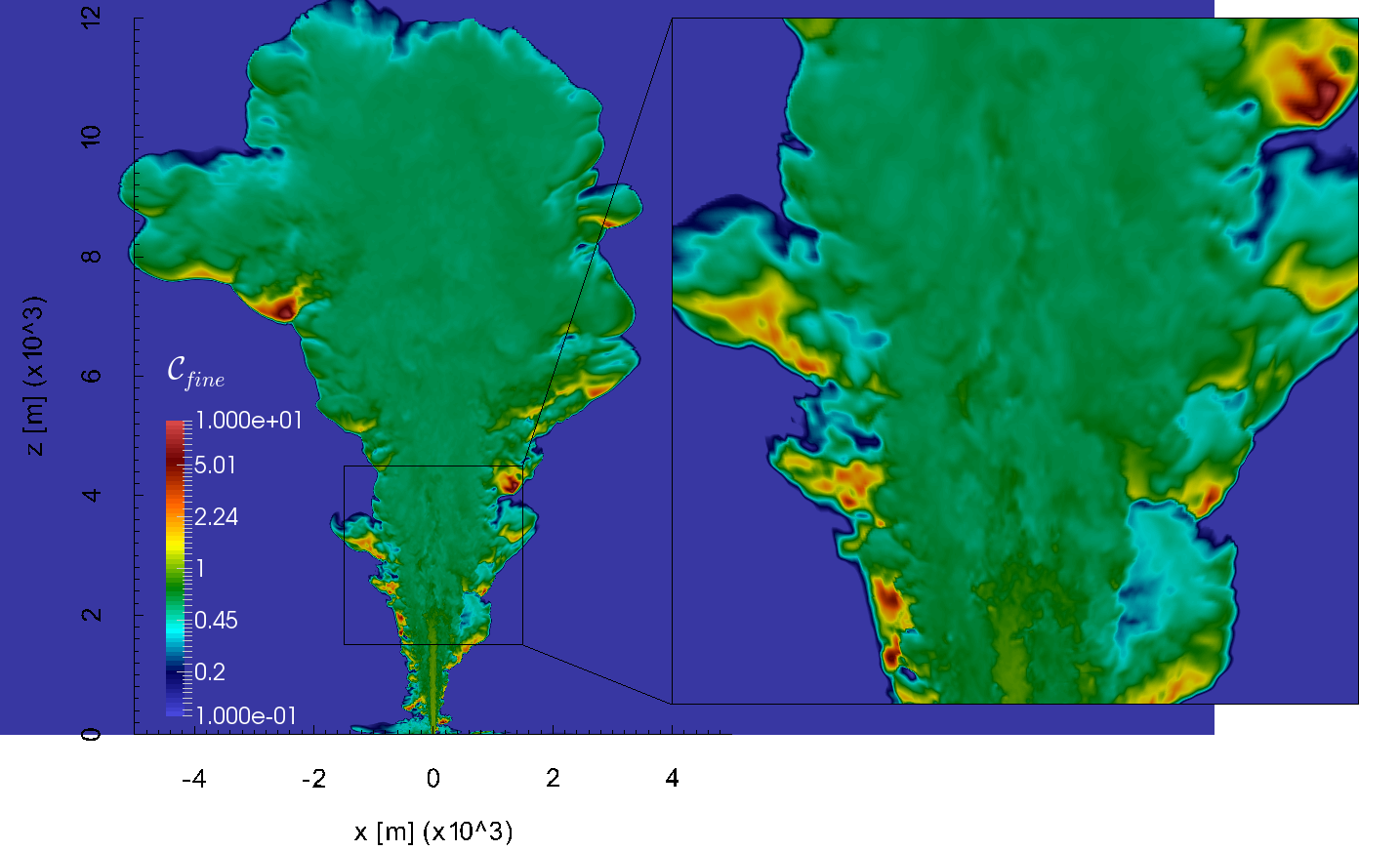}
\caption{Distribution of $\mathcal C_{\textup{fine}}$ (Eq.~\ref{eq:prefconc}) for the finest particles across a vertical section at $t=400$ s.}
\label{fig:WP_preferential_fine}
\end{figure*}

Fig. \ref{fig:WP_preferential_coarse} shows the distribution of $\mathcal{C}_j$ for the coarsest particles at $t=400$ s. The color scale is logarithmic and symmetric with respect to 1, which corresponds to the nil preferential concentration. For $\mathcal{C}_j<1$, the mixture is relatively depleted of particles (green to blue scale); for $\mathcal{C}_j>1$, particles are clustered (green to red scale), with mass fraction up to 5 times larger and 20 times smaller than the value it would have in absence of preferential concentration. This behavior is expected to affect the mixing and entrainment process.
It is also worth remarking that the more uniform red area beyond the plume margins corresponds to the region of settling particles below the umbrella region. On the other hand, the top of the plume is relatively depleted of coarse particles. The corresponding Figure \ref{fig:WP_preferential_fine} for fine particles confirms that these are tightly coupled to the gas phase and almost behave as tracers (value of $\mathcal{C}_{\textup{fine}}$ is everywhere around 1). These conclusions are coherent with the {\em a-priori} estimate of $\St_j$ we gave at the beginning of this section, based on the Taylor microscale time (Eq.~\eqref{eq:tau_xi}).


Finally, we present the results obtained by averaging the volcanic plume flow field over time (in a time-window \mbox{[300-720] s} where the plume has reached statistically stationary conditions) and over the azimuthal angle, in order to allow comparison with one-dimensional integral models \citep[e.g.,][]{Woods1988a} and discuss the effect of numerical resolution.
The averaging procedure is the generalization of that explained in Sect. \ref{validation} to the multiphase case~\citep[see][]{cerminara2015b}. The form of the results presented are similar to those obtained in Fig. \ref{fig:Zhou1D} for the laboratory plume test case.

\begin{figure*}[h]
\centering
\subfloat[][$\Delta r = D/32$]{\includegraphics[width=\columnwidth]{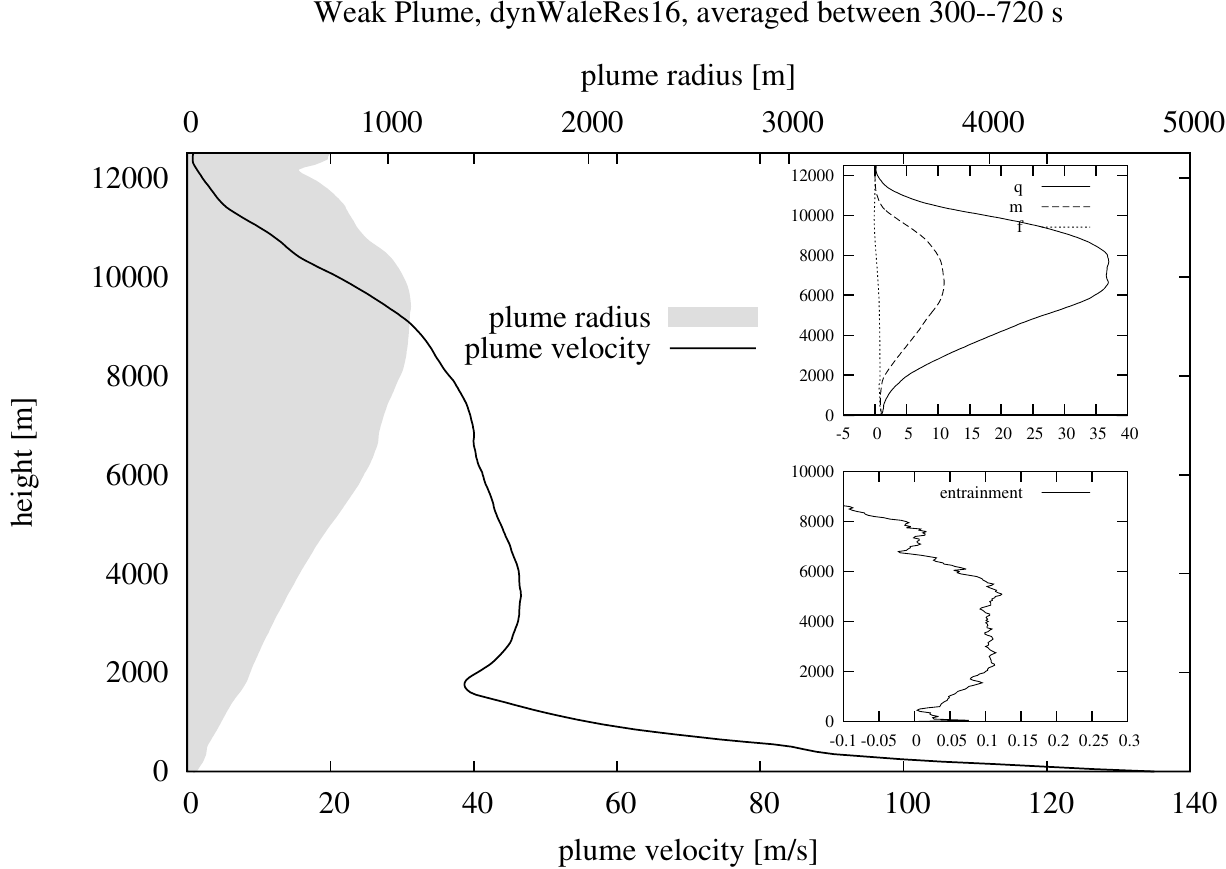}\label{fig:dynWaleRes16_meanPlume}}\quad
\subfloat[][$\Delta r = D/16$]{\includegraphics[width=\columnwidth]{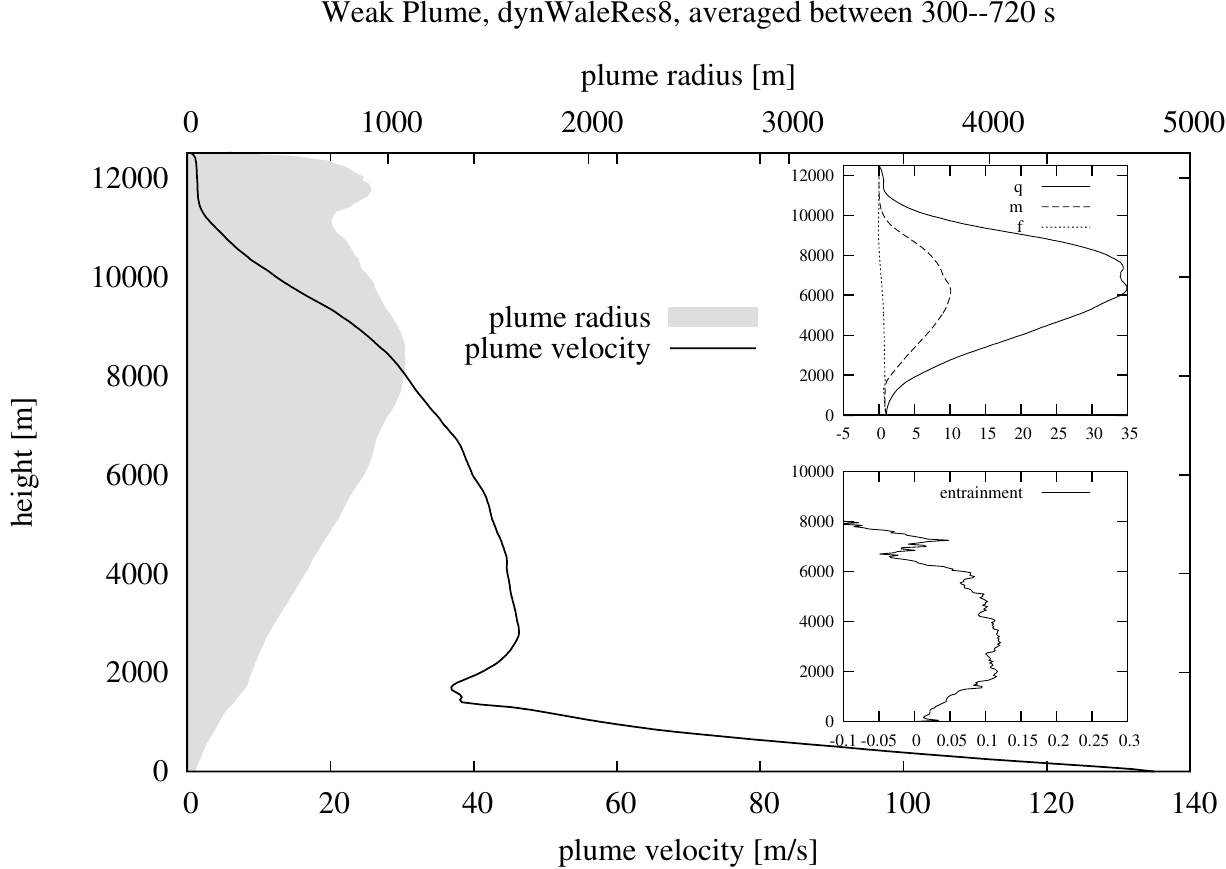}\label{fig:dynWaleRes8_meanPlume}}\\
\subfloat[][$\Delta r = D/8$]{\includegraphics[width=\columnwidth]{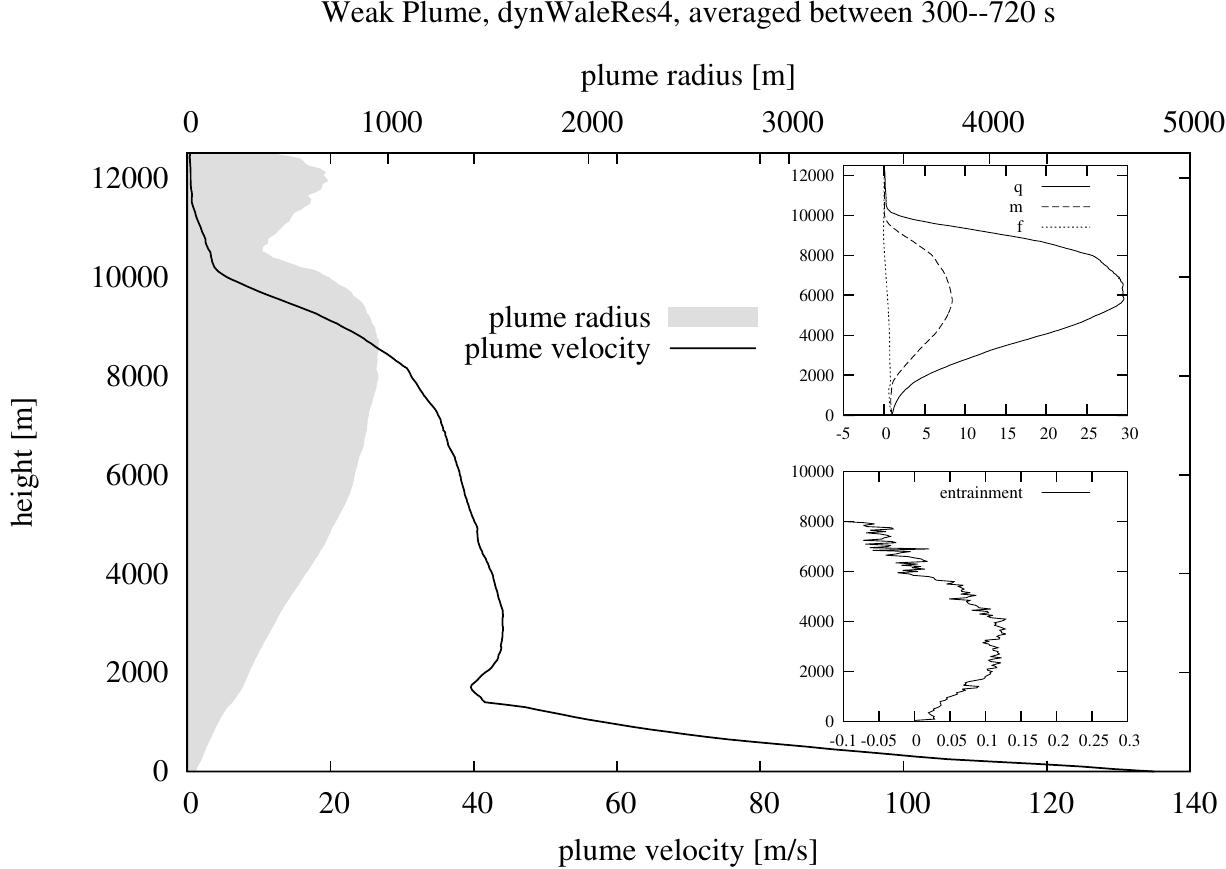}\label{fig:dynWaleRes4_meanPlume}}\quad
\subfloat[][$\Delta r = D/8$, dusty-gas]{\includegraphics[width=\columnwidth]{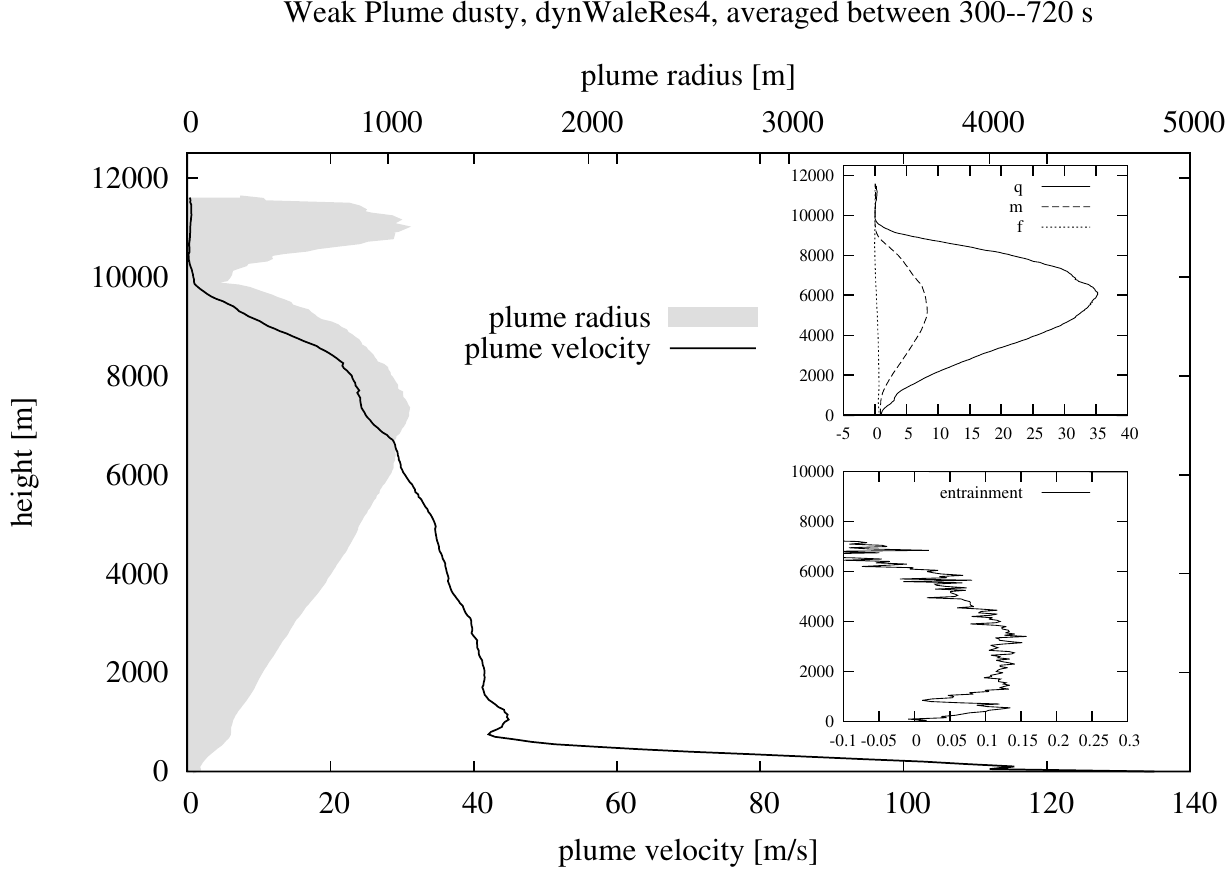}\label{fig:dynWaleRes4Dusty_meanPlume}}
\caption{Time-averaged plume radius and velocity. The insets display the non-dimensional mass, momentum and buoyancy fluxes (top) and the time-averaged entrainment coefficient. Panels a--c) ASHEE model at different resolutions; panel d) dusty-gas model.}
\label{fig:weakPlumeVs1D}
\end{figure*}

Figure \ref{fig:weakPlumeVs1D} presents the results of the averaging procedure for three multiphase flow simulations at different resolution (panels a--c). In particular, panel a) has the highest resolution (minimum radial cell size $\Delta r = 2b_0/32$ with $2b_0$ equal to the inlet diameter); panel b) has $\Delta r = 2b_0/16$; panel c) has $\Delta r = 2b_0/8$.
In panel d) we present results at the lowest-resolution obtained by imposing the full kinematic equilibrium between gas and particles, i.e., by adopting the dusty-gas model \citep{Marble1970a}.

Results demonstrate that the numerical model is quite robust and accurate so that even low-resolution simulations are able to capture the main features of the volcanic plume development. However, the maximum plume height systematically decreases from 12100 m (a), to 11300 m (b) to 11000 m (c) when we decrease the resolution. Analogously, the Neutral Buoyancy Level (NBL) decreases from 7800 m (a) to 7200 m (b) to 7100 m (c).
Although the lowest resolution run seems to underestimate the maximum plume height and the plume radius by about 10\%, the average velocity profile (and the vertical profiles of $q$, $m$ and $f$) is consistent in the three runs, showing a jet-plume transition at about 2000 m above the vent, also corresponding to the transition to a super-buoyancy region \citep{Woods2010}. The computed entrainment coefficient is also consistent and relatively independent on the grid resolution and shows a different behavior with respect to the laboratory case, associated with the effect of the density contrast. In this case, a maximum value of about $k\sim 0.1$ is obtained in the buoyant plume region between 2 and 5 km above the vent. 

Interestingly, we find that in three-dimensional simulations the entrainment decreases near the NBL and it become negative above that level. This happens because the mass exit from the plume region defined in Eq.~\eqref{eq:omegaPlume} moving from it to the umbrella cloud. In this way, the mass flow $q$ of the plume decreases above the NBL and a stationary solution can be achieved. This is not the case in integral plume models with positive entrainment coefficient, where the maximum plume height is reached as a singularity point with divergent mass flow and plume radius~\citep[cf.][]{Morton1959, Woods1988a}. We plan to study this behavior more thoroughly in future studies.

The dusty-gas model shows a significantly different behavior, with a larger plume radius, a slightly higher entrainment coefficient and a more marked jet-plume transition with no further acceleration (without a super buoyancy transition).
The plume height is slightly lower than the non-equilibrium case at the same resolution having maximum plume height and neutral buoyancy level of 9900 m and 6100 m, respectively.
Numerical simulations thus suggest that the effects of non-equilibrium gas-particle processes (preferential concentration and settling) on air entrainment and mixing are non-negligible. These effects are certainly overlooked in the volcanological literature and will be studied more thoroughly in future studies, by applying the present model to other realistic volcanological case studies.

\conclusions
We have developed a new, equilibrium-Eulerian model to numerically simulate compressible turbulent gas-particle flows. The model is suited to simulate relatively dilute mixtures (particle volume concentration $\epsilon \lesssim 10^{-3}$) and particles with Stokes number $St \lesssim 0.2$. It is appropriate to describe the dynamics of volcanic ash plumes, with kinematic decoupling between the gas and the particles, assumed in thermal equilibrium.

We have tested the model against controlled experiments to assess the reliability of the physical and numerical formulation and the adequacy of the model to simulate the main controlling phenomena in volcanic turbulent plumes, and in particular:
1) multiphase turbulence (including preferential concentration and density effects); 2) buoyancy and compressibility effects; 3) stratification and density non-homogeneity.

The model reproduces the main features of volcanic plumes, namely: 1) buoyancy reversal and jet-plume transition; 2) plume maximum height and spreading of the umbrella above the neutral buoyancy level; 3) turbulent mixing and air entrainment; 4) clustering of particles; 5) proximal fallout and re-entrainment of particles in the plume. Results demonstrate that the compressible equilibrium-Eulerian approach adopted in the ASHEE model is suited to simulate the three-dimensional dynamics of volcanic plumes, being able to correctly reproduce the non-equilibrium behavior of gas-particle mixtures with a limited computational cost. 

Finally, the adopted open-source computational infrastructure, based on OpenFOAM, will make the model easily portable and usable and will ease the maintenance and implementation of new modules, making ASHEE suitable for collaborative research in different volcanological contexts.
\label{conclusion}
\appendix
%
\begin{acknowledgements}
This work includes some results achieved in the PhD work by the first author (MC), carried out at Scuola Normale Superiore, Pisa, with a grant by Istituto Nazionale di Geofisica e Vulcanologia.
We thank Matteo Bernardini and Sergio Pirozzoli for useful discussion on decaying turbulence and for providing DNS data for model comparison and validation.
We acknowledge the CINECA Award N. 2011 for the availability of high performance computing resources and technical support on porting OpenFOAM on HPC architectures by I. Spisso and M. Culpo.
\end{acknowledgements}
\bibliographystyle{copernicus}
\bibliography{biblio_JouAbbr}
\end{document}